\documentclass[preprint]{aastex}

\slugcomment{Accepted by ApJS}

\shorttitle{3DMHD Simulations of Gravitational Collapse of a 15M$_\odot$ Star}

\shortauthors{Kuroda \& Umeda}

\begin{document}

\title{Three Dimensional Magneto Hydrodynamical Simulations of\\ Gravitational Collapse of a 15M$_\odot$ Star}

\author{Takami Kuroda and Hideyuki Umeda}

\affil{Department of Astronomy, School of Science,
   University of Tokyo, Bunkyo-ku, Tokyo, 113-0033, Japan;
   kuroda@astron.s.u-tokyo.ac.jp, umeda@astron.s.u-tokyo.ac.jp,
   }

\begin{abstract}
 We introduce our newly developed two different, three dimensional magneto hydrodynamical codes in detail.
One of our codes is written in the Newtonian limit (NMHD) and the other is in the fully general relativistic code (GRMHD).
Both codes employ adaptive mesh refinement and, in GRMHD, the metric is evolved with the
"Baumgarte-Shapiro-Shibata-Nakamura" formalism known as the most stable method at present.
We did several test problems and as for the first practical test, we calculated gravitational collapse of a $15M_\odot$ star.
Main features found from our calculations are;
(1) High velocity bipolar outflow is driven from the proto-neutronstar and moves through along the rotational
axis in strongly magnetized models;
(2) A one-armed spiral structure appears which is originated from the low-$|T/W|$ instability;
(3) By comparing GRMHD and NMHD models, the maximum density increases about $\sim30\%$ in GRMHD
models due to the stronger gravitational effect.
These features agree very well with previous studies and our codes are thus reliable to numerical simulation of gravitational collapse of
massive stars.
\end{abstract}


\section{Introduction}
\label{sec:Introduction}
There are lots of works searching for the explosion mechanisms of core-collapse supernovae (CCSNe), however we have still not obtained any conclusive results in these decades. 
Recent works, both observational and theoretical ones, show several indications that their explosions are commonly aspherical.
For instance, \citet{Maeda08} obtained late-time spectra for a lot of CCSNe and showed that the explosion morphologies of stars without
H envelope are close to bipolar configurations.
Furthermore, non-axisymmetric explosion is found from the observation of SN 2005bf by \citet{Tanaka09b}.
In their report, they presented an optical spectropolarimetric observation of Type Ib supernova 2005bf and claimed that SN 2005bf can be
explained as unipolar explosion and also the direction of launched unipolar blob is tilted from the symmetric axis.
Therefore, the asphericities might be key ingredients to understand the explosion mechanisms, especially for a subset class of CCSNe such as Type Ib SNe.
These asphericities found from the observations are thought to be products of hydrodynamical instabilities occurring in the vicinity
of the proto neutronstars (PNSs).
From the previous theoretical/numerical works, it is widely known that there are many types of hydrodynamical instabilities which would occur during 
CCSNe, such as: the Ledoux convection \citep{Keil96}: vortical-acoustic instability \citep{Blondin03}: magneto hydrodynamical instabilities, e.g.,
Magneto Rotational Instability (MRI) \citep{Balbus91,Akiyama03,Obergaulinger09} or Kelvin Helmholtz instability:
rotational instabilities, e.g., dynamical bar-mode instability \citep{Rampp98,Shibata03a}, secular instability \citep{Imamura03},
low-$|{\rm T/W}|$ instability \citep{Shibata03a,Watts05,Ott05}.
Which of these instabilities would occur depends on progenitor mass, the rotational/magnetic field velocity
configuration, the interaction between matters and neutrinos, etc.

Among these mechanisms, rotational instabilities are common byproducts of relatively fast-spinning progenitors and their subsequent collapses.
If ratio of rotational to gravitational potential energy $\beta \equiv | T/W |$, at core bounce, exceeds $\beta _ {\rm dyn}\sim0.27$ then the 
dynamical bar-mode instability appears \citep{Rampp98,Shibata03a}.
Simple estimate, in which we assume the angular momentum is almost conserved during core collapse, gives us a rough lower limit of the central angular velocity $\Omega_{\rm c}$, at pre-collapse stage
which exceeds $\sim 10$ rad/s.
Such rotational speed is at least $\sim10$ times faster than the results of recent stellar evolutional calculations \citep{Yoon05}.
Even if progenitor does not spin so rapidly at the beginning, the secular instability may be appeared if $\beta \ga \beta_{\rm sec}\sim0.14$
and also if some dissipative mechanisms exist such as the viscosity, the neutrino radiation or the gravitational radiation reaction \citep{Imamura03}.
Recently, another interesting rotational instability is reported to occur in CCSNe which is the low-$|{\rm T/W}|$ instability
\citep{Shibata03a,Watts05,Ott05}.
This is the resonance instability across the corotation point inside the differentially rotating area and occurs with more reasonable value $\beta\sim0.01$ or 
with relatively slow initial spin rate $\Omega_{\rm c}\sim 1$ rad/s.
Such a slow rotation is more realistic compared to aforementioned two mechanisms.

These rotational instabilities are intrinsically three dimensional, non-axisymmetric phenomena.
In the ideal MHD limit, which is a reasonable assumption in the CCSNe context, the non axisymmetric motion always convert the toroidal magnetic
field into the poloidal component by dragging the toroidal magnetic field.
Then, if the differential rotation exists, the converted poloidal magnetic field is again converted back into the toroidal ones.
Such a closed cycle never takes place in the axisymmetric motion and may play important roles to
amplify magnetic field via, e.g., the dynamo mechanism \citep{Thompson93} or the MRI.
Strongly amplified magnetic field ($\ga 10^{15-16}$G) launches high velocity outflow along the rotational axis \citep{Mikami08}
due to the magneto-spring or the magneto-centrifugal effects \citep{Wheeler02} and may leave highly magnetized ($\sim10^{15-16}$G at surface)
neutronstar which is a so-called "magnetar" \citep{Duncan92}.
Another feature of the magneto-rotational explosion is, through these magneto-rotational effects, the explosion morphology becomes
highly aspherical.
Since the explosion morphology shows stronger asphericity in case of energetic explosion such as Hypernova (HN) or
SN associated with gamma-ray burst (GRB) \citep{Maeda08}, the magneto rotational effects are considered to be more important to such hyper energetic CCSNe 
than the normal ones.

As just described, the non axisymmetric effects may play important roles and should be examined by three dimensional numerical simulations.
However, up to the present date, there are only a few numerical works about CCSNe with three dimensional MHD (see, e.g., 
\citet{Scheidegger08,Scheidegger09,Mikami08}).
In \citet{Scheidegger09}, they calculated a number of numerical models aiming for the gravitational wave signature during core collapse,
including two types of realistic EOSs, various magnetic-rotational configurations and a neutrino parametrization/leakage scheme.
In their work, they showed the low-$|T/W|$ instability appears when the progenitor rotates $2\pi$ rad/s or faster and alters the gravitational
wave radiation.
As for the explosion dynamics, they reported that bipolar outflow is driven by strong magnetic field if the initial central 
magnetic field strength is the order of $\ga\mathcal O(10^{12})$G \citep{Mikami08,Scheidegger09}.
This is because the magnetic field of the order of $\ga\mathcal O(10^{12})$G is easily amplified, a factor of $\sim\mathcal O(10^{3-4})$ \citep{Burrows07},
simply by the compression and the rotational winding effects during core collapse.
Magnetic pressure with $B\sim 10^{15-16}$G is comparable to matter pressure in the vicinity of the surface of PNS and thus drives bipolar outflow.
The amplification factor ($\sim\mathcal O(10^{3-4})$) hardly depends on the initial strength unless some other nonlinear amplification mechanisms work.
In the 3D context however the non axisymmetric motion may trigger the previously mentioned nonlinear amplification mechanisms (e.g., MRI) and may
alter the amplification factor or the time scale comparing to axisymmetric motion.
Therefore, especially when the initial magnetic field is much weaker than $\sim\mathcal O(10^{12})$G, the axisymmetric assumption may not be suitable for the CCSNe.
Additionally, since the strength of the order of $\sim\mathcal O(10^{12})$G is considered to be unrealistically strong for pre-collapse stage,
it should be examined how the initially weak (or more realistic) magnetic field, e.g., $\sim\mathcal O(10^{9})$G, is amplified and whether it affects the explosion dynamics or not.

Another challenge for numerical simulations of CCSNe is the treatment of general relativistic effects.
Since the gravity plays intrinsic role and also some very massive progenitors ($\ga25M_\odot$, see, \citet{Tanaka09a}) form black halls (BHs),
we cannot say any conclusive results about the explosion mechanisms of CCSNe without taking account of the general relativity (GR).
As same as the context of 3D MHD works of CCSNe, there are not so many works done by MHD simulations including GR effects (see, e.g.,
\citet{Shibata06,Cerda07}) and furthermore 3D GRMHD works of CCSNe have not been done yet.

In this paper, we describe our newly developed two 3D-MHD codes for CCSNe simulation.
Their features are as follows.
Both codes employ adaptive mesh refinement (AMR) technique and can cover wide dynamical ranges from the central compact object ($\sim10$km)
to beyond the several times of iron core radius ($\sim10000$km) or more.
Self gravity is included in the Newtonian approximated MHD code and the dynamical metric is included in GRMHD code.
High resolution shock capturing scheme is adopted in both codes and can handle the shock discontinuity
without numerical viscosity.
We use the staggered mesh algorithm and adopt the Constrained-Transport method to evolve the magnetic
field.
We have done many test problems and confirmed their abilities.

This paper is organized as follows. 
From Sec. \ref{sec:NMHDcode} to \ref{sec:AMR}, we describe our numerical codes, their detailed properties and adopted techniques.
In Sec.\ref{sec:NMHDtest} and \ref{sec:GRMHDtest}, we report several numerical tests of Newtonian approximated and general relativistic codes, respectively.
Section \ref{sec:collapse} shows our first core-collapse supernovae calculations and their initial setups.
We summarize in section \ref{sec:Summary and Discussions}.
We adopt cgs units for NMHD code and geometrical units for GRMHD code in which $G=c=1$.
In our practical calculations of a collapse of $15M_\odot$ star, in Sec. \ref{sec:collapse}, physical quantities are notated in cgs units.
Greek/Latin indices run through 0-3/1-3.


\section{NMHD Code}
\label{sec:NMHDcode}
Our NMHD code solves ideal NMHD equations written in conservative forms together with source terms of self gravity,
which are written in the following form:
\begin{equation}
\label{eq:NMHDEqs}
\frac{\partial {\bf Q}}{\partial t}+\nabla\cdot{\bf F}={\bf S}\\
\end{equation}
and Poisson's equation for gravitational potential
\begin{equation}
\label{eq:poisson}
\nabla^2\Phi=4\pi{\rm G}\rho
\end{equation}
Where, in Eq. (\ref{eq:NMHDEqs})
\begin{eqnarray}
\label{eq:NMHDcon}
{\bf Q}&=&
\left[
\begin{array}{c}
\rho \\
\rho{\bf u} \\
E \\
{\bf B} \\
\end{array}
\right] \\
\label{eq:NMHDflux}
{\bf F}&=&
\left[
\begin{array}{c}
\rho {\bf u} \\
\rho {\bf uu}+\left(P_{\rm tot}+\frac{\bf B\cdot B}{8\pi}\right){\bf I}-\frac{\bf BB}{4\pi} \\
\left(E+P_{\rm tot}+\frac{\bf B\cdot B}{8\pi}\right){\bf u}-\frac{\bf (u\cdot B)B}{4\pi} \\
{\bf uB-Bu} \\
\end{array}
\right]\\
\label{eq:NMHDsrc}
{\bf S}&=&
\left[
\begin{array}{c}
0 \\
-\rho\nabla\Phi\\
-\rho{\bf u}\cdot \nabla \Phi\\
0 \\
\end{array}
\right]
\end{eqnarray}
Eq. (\ref{eq:NMHDEqs}) represents the conservation of mass, momentum, energy and the Faraday's law.
$\rho$, ${\bf u}$, ${\bf B}$, $E$, $P_{\rm tot}$, $\Phi$ and ${\bf I}$ are rest-mass density, fluid velocity, magnetic field, energy density, total pressure, gravitational potential and $3\times3$ unit matrix, respectively.
The energy density and the total pressure are expressed as\\
\begin{equation}
\label{eq:Etot}
E=\rho \varepsilon+\frac{\rho {\bf u}^2}{2}+\frac{{\bf B}^2}{8\pi}
\end{equation}
\begin{equation}
\label{eq:Ptot}
P_{\rm tot}=p+\frac{{\bf B}^2}{8\pi}
\end{equation}
Here, $p$ and $\varepsilon$ are gas pressure and internal energy density, respectively, and are related via the equation of state, $p=p(\rho,\varepsilon)$.
We also define primitive variables ${\bf P}\tbond(\rho,\ {\bf u},\ \varepsilon,\ {\bf B})$ which is uniquely obtained from conservative variables {\bf Q} via EOS.
We employ staggered mesh and all variables except magnetic field are defined at cell center while, for instance, $B_{\rm x}$(i,j,k) is defined
in cell surface (i+1/2, j, k).
Hereafter we use ${\bf B'}={\bf B}/\sqrt{4\pi}$ and do not express "$\ '\ $", unless otherwise stated.

\begin{figure}[htpb]
\epsscale{0.3}
  \plotone{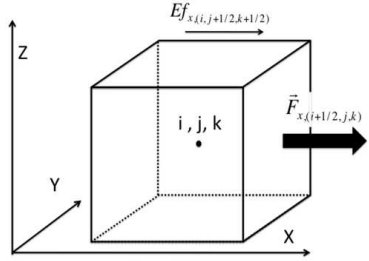}
  \caption{Positions where numerical flux and electric field are defined. For instances, $x$ directional numerical flux $\bf{F}_x$ is defined at
  (i+1/2, j, k) and x component of electric field $\bf{Ef}_x$ is defined at (i, j+1/2, k+1/2).}
\label{pic:f1}
\end{figure}

To solve time dependent equation (\ref{eq:NMHDEqs}), in our cartesian Eulerian grid, we adopt Roe-type upwind solver
in which numerical flux ${\bf \tilde F}$ is defined in cell surface (see Fig.\ref{pic:f1}).
Following \cite{Powell99}, ${\bf \tilde F}$ is constructed from the spectral decomposition of the system and expressed as
\begin{equation}
\label{eq:Froe}
{\bf \tilde F}=\left[({\bf F(P}_L)+{\bf F(P}_R))-\sum^{7}_{m=1}{\bf L}_m\cdot ({\bf Q}_R-{\bf Q}_L)|\lambda_m|{\bf R}_m\right]/2
\end{equation}
Here the index $L/R$ represents position immediate left/right of the cell boundary, i.e., when we evaluate
the numerical flux at (i+1/2, j, k), the position $L$ and $R$ stand for (i+1/2-0, j, k) and (i+1/2+0, j, k),
respectively.
${\bf L}_m$/${\bf R}_m$ and $\lambda_m$ correspond to left/right eigenvectors and eigenvalues of the system, respectively, and defined in cell surface.
Originally, in \citet{Powell99}, they decompose the system into eight spectral modes in which one mode carries 
the monopole moment of magnetic field.
Meanwhile, we adopt the constrained transport method for time evolution of magnetic field and we thus
consider only seven modes in our NMHD code (for explicit expressions of seven eigenvectors, see \citep{Ryu95}).
Eigenvalues $\lambda_m$ are defined as
\begin{equation}
\label{eq:Eigenroe}
\lambda_m=(u-c_f,\ u-c_{\mathcal A},\ u-c_s,\ u,\ u+c_s,\ u+c_{\mathcal A},\ u+c_f)
\end{equation}
Here $u$ is the fluid velocity normal to cell boundary; $c_f$ the fast magnetosonic speed; $c_s$ the slow magnetosonic speed; $c_{\mathcal A}$ the 
$Alf\acute ven$ speed. 
$c_{\mathcal A},\ c_f,\ c_s$ are expressed by
\begin{eqnarray}
\label{eq:Alfven}
c_{\mathcal A}&=&B_n/\sqrt{\rho} \\
\label{eq:fastslow}
c_{f,s}&=&\sqrt{\frac{a^2+{\bf B}^2/\rho \pm \sqrt{(a^2+{\bf B}^2/\rho)^2-4a^2c_{\mathcal A}^2}}{2} }
\end{eqnarray}
Where, $f/s$ takes $+/-$ in Eq. (\ref{eq:fastslow}), $B_n$ represents magnetic field component normal to cell boundary and $a$ is the speed of sound.
The speed of sound $a$ in a general form of EOS can be defined by
\begin{equation}
\label{eq:aNMHD}
a=\sqrt{\left(p/\rho^2-\frac{\partial\varepsilon}{\partial\rho}\Bigr|_{p}\right)\left(\frac{\partial\varepsilon}{\partial p}\Bigr|_{\rho}\right)^{-1}}
\end{equation}
Since we define the conservative variables other than magnetic field at cell center, we have to interpolate
to obtain those values at immediate left/right of cell boundary.
As for the interpolation, we adopt the monotonized central (MC) method \citep{VanLeer77} as
\begin{eqnarray}
{\bf P}_{i\pm1/2\mp0}&=&{\bf P}_i\pm{\rm MC}({\bf P}_{i+1}-{\bf P}_{i},{\bf P}_{i}-{\bf P}_{i-1})\nonumber \\
{\rm MC}(x,y)&=&
\left(
\begin{array}{ccc}
2\ {\rm sign}(x)\ {\rm min}(|x|,|y|,|x+y|/4) & {\rm for} & xy>0 \\
0 & {\rm otherwise} &
\end{array}
\right.
\end{eqnarray}
Additionally, we employ total variational diminishing (TVD) scheme which has second order convergence in space (see \citet{Ryu95}) and we briefly summarize how we implement it into our NMHD code below.
In TVD scheme, summation respect to the spectral modes in Eq. (\ref{eq:Froe}) is modified as
\begin{equation}
\label{eq:TVDsum}
\sum^{7}_{m=1}{\bf L}_m\cdot ({\bf Q}_R-{\bf Q}_L)|\lambda_m|{\bf R}_m 
=\sum^{7}_{m=1}\left( Q_l\left(\frac{\Delta t}{\Delta x}\lambda_{\rm m} +\gamma_{\rm m}\right)
LdQ_{\rm m,i}-(g_l+g_r)\right){\bf R}_m
\end{equation}
In Eq. (\ref{eq:TVDsum}) we consider $x$ directional numerical flux ${\bf \tilde F}_{x,i+1/2}$ defined at (i+1/2, j, k) and
\begin{eqnarray}
LdQ_{\rm m,i}&=&{\bf L}_m\cdot ({\bf Q}_{\rm i+1/2+0}-{\bf Q}_{\rm i+1/2-0})\\
g_{\rm i}&=&\frac{LdQ_{\rm m,i}}{2}\left( Q_l\left(\frac{\Delta t}{\Delta x}\lambda_{\rm m}\right)-\left(\frac{\Delta t}{\Delta x}\lambda_{\rm m}\right)^2 \right)\\
g_r&=&{\rm sign}(g_{\rm i+1})\ {\rm max}\left[0,\ {\rm min}(|g_{\rm i+1}|,\ g_{\rm i}{\rm sign}(g_{\rm i+1}) )\right]\\
g_l&=&{\rm sign}(g_{\rm i})\ {\rm max}\left[0,\ {\rm min}(|g_{\rm i}|,\ g_{\rm i-1}{\rm sign}(g_{\rm i}) )\right]\\
\gamma_{\rm m}&=&
\left(
\begin{array}{ccc}
(g_r-g_l)/LdQ_{\rm m,i} & {\rm for} & LdQ_{\rm m,i}\ne0 \\
0 & {\rm otherwise} & 
\end{array}
\right.\\
Q_l\left(x\right)&=&
\left(
\begin{array}{ccc}
\frac{x^2}{4\varepsilon}+\varepsilon & {\rm for} & |x|<2\varepsilon\\
|x| & {\rm for} & |x|\ge2\varepsilon
\end{array}
\right.\\
\varepsilon&=&
\left(
\begin{array}{ccc}
0.01 & {\rm for} & m=1,7\\
0.1 & {\rm for} & m=2,6\\
0.1 & {\rm for} & m=3,5\\
0 & {\rm for} & m=4
\end{array}
\right.
\end{eqnarray}
After we obtain the numerical fluxes in all sides of cell, the conservative variables other than magnetic field are updated through predictor and corrector steps \citep{Press92}.
In predictor step, ${\bf Q}^n$ is updated from time level $n$ to $n+1/2$ by $\Delta t/2$, i.e.,
\begin{equation}
\label{eq:predictor}
{\bf Q}^{n+1/2}={\bf Q}^{n}+0.5\Delta t\left(-\frac{{\bf \tilde F}_{x,i+1/2}^n-{\bf \tilde F}_{x,i-1/2}^n}{\Delta x}
-\frac{{\bf \tilde F}_{y,j+1/2}^n-{\bf \tilde F}_{y,j-1/2}^n}{\Delta y}
-\frac{{\bf \tilde F}_{z,k+1/2}^n-{\bf \tilde F}_{z,k-1/2}^n}{\Delta z}+{\bf S}_{\rm i,j,k}^n\right)
\end{equation}
and in corrector step from time level $n$ to $n+1$ by $\Delta t$ with using predicted values, i.e.,
\begin{equation}
\label{eq:corrector}
{\bf Q}^{n+1}={\bf Q}^{n}+\Delta t\left(-\frac{{\bf \tilde F}_{x,i+1/2}^{n+1/2}-{\bf \tilde F}_{x,i-1/2}^{n+1/2}}{\Delta x}
-\frac{{\bf \tilde F}_{y,j+1/2}^{n+1/2}-{\bf \tilde F}_{y,j-1/2}^{n+1/2}}{\Delta y}
-\frac{{\bf \tilde F}_{z,k+1/2}^{n+1/2}-{\bf \tilde F}_{z,k-1/2}^{n+1/2}}{\Delta z}+{\bf S}_{\rm i,j,k}^{n+1/2}\right)
\end{equation}
This method is second order convergence with respect to time.

On the other hand, as for the time evolution of magnetic field $\bf B$, we adopt the constrained transport (CT) scheme \citep{Balsara99} in which $\bf B$ is evolved as
\begin{equation}
\frac{\partial {\bf B}}{\partial t}=\nabla \times {\bf E}
\end{equation}
Electric field ${\bf E}$ for CT scheme is defined at the cell edge (see Fig.\ref{pic:f1}) and is evaluated from 
Roe-type numerical flux, Eq. (\ref{eq:Froe}), with appropriate interpolation.
We simply express the electric field as
\begin{equation}
\label{eq:electricfield}
E_{x,i,j+1/2,k+1/2}=\left({\bf \tilde F}_{y,i,j+1/2,k}^6+{\bf \tilde F}_{y,i,j+1/2,k+1}^6
-{\bf \tilde F}_{z,i,j,k+1/2}^7-{\bf \tilde F}_{z,i,j+1,k+1/2}^7\right)/4
\end{equation}
Here the upper suffixes in the right hand side denote components of the numerical flux.
$y$ and $z$ components of the electric field are obtained straightforwardly by permutation as
$x\rightarrow y,\ y\rightarrow z,\ z\rightarrow x$ and
$x\rightarrow z,\ y\rightarrow x,\ z\rightarrow y$, respectively.

In self gravitating system, the source term contains gravitational potential which is obtained by solving
Poisson equation (our method to solve Poisson equation with AMR framework is described in Sec. \ref{sec:Poisson Solver under the AMR Framework}).
Since solving Poisson equation is very time consuming task, we solve it only once in one hydrodynamical
time step \citep{Truelove98} after predictor step is completed.
Practically, in predictor step, we extrapolate gravitational potential $\Phi^n$ at $n$th time level via
$\Phi^n=(3\Phi^{n-1/2}-\Phi^{n-3/2})/2$ and then predict variables at $(n+1/2)$th time level through Eq. (\ref{eq:predictor}).
After the predictor step is completed, we solve Poisson equation by using $\rho^{n+1/2}$.
In corrector step, we fully evolve the $n$th time level to $(n+1)$th variables by using $\bf Q^{\rm n+1/2}$ and $\Phi^{n+1/2}$.
Even though we extrapolate gravitational potential at $n$th time level, our numerical results
do not show any large error in conservation of energy and numerical convergence is also achieved
(described later in Sec. \ref{sec:Energy and Angular Momentum Conservations}).

On a final note, we mention about a numerical instability which is characteristic to Roe-type scheme.
Even though Roe-type numerical flux is less numerically dissipative and has good shock capturing ability,
one problem arises which is a so-called "odd-even decoupling".
This instability appears when the shock normal is directed parallel to the grid alignment.
To avoid this instability, we adopt "carbuncle cure" in our NMHD code by following \citet{Hanawa08}.

\section{GRMHD Code}
\label{sec:GRMHDcode}
Formalism of our GRMHD code is based mainly on \citet{Shibata05}.
It can be divided into two parts, one is MHD part and the other is Einstein's equation part.
MHD part describes time evolution of matter on the background of spacetime metric and the metric is evolved according to
Einstein's equation through the so-called "Baumgarte-Shapiro-Shibata-Nakamura (BSSN)" formalism \citep[see, e.g.,][]{Shibata95,Baumgarte99,Yo02}.

Before going to brief summary of our method, we describe our fundamental variables of MHD and metric parts.
We set fundamental variables of MHD part as rest-mass density $\rho$, specific internal energy $\varepsilon$, 4-velocity $u^\mu$
and magnetic field $b^\mu$ measured by a comoving observer.
For metric part, the 3-metric $\gamma_{ij}$ and the extrinsic curvature $K_{ij}$ are the fundamental ones adopting the 3+1 formulation
of "Arnowitt-Deser-Misner (ADM)" formalism \citep{Arnowitt62}.
Then the line element of the spacetime can be expressed as
\begin{equation}
\label{eq:line}
ds^2=-\alpha^2dt^2+\gamma_{ij}(dx^i+\beta^idt)(dx^j+\beta^jdt)
\end{equation}
Here $\alpha,\ \beta^i$ are the lapse function and the shift vector, respectively, and determined by arbitrary chosen gauge condition (Sec.  \ref{sec:gauge}).
Hypersurface of constant $t$ is foliated in the spacetime so that a unit normal vector $n^\mu(n_\mu)$ to this hypersurface becomes
\begin{equation}
\label{eq:normalVec}
n^\mu=(1,-\beta^i)/\alpha \ \ \ \ \& \ \ \ \  n_{\mu}=(-\alpha,0)
\end{equation}

Fundamental metrics $\gamma_{ij}$, $K_{ij}$ are converted to 5 variables in BSSN formalism which are;
the conformal exponent $\phi={\rm ln}(\gamma)/12$, here $\gamma$ is the determinant of 3-metric $\gamma_{ij}$;
the conformal 3-metric $\tilde\gamma_{ij}=e^{-4\phi}\gamma_{ij}$; the trace of the extrinsic curvature $K={\rm tr}(K_{ij})$; the tracefree extrinsic curvature $
\tilde A_{ij}=e^{-4\phi}(K_{ij}-\gamma_{ij}K/3)$; and three auxiliary variables $F_i=\delta^{jk}\tilde\gamma_{ij,k} $, here "$,k$'' represents partial derivative with respect to $k$ direction.
Hereafter, $\tilde D_i$ and $D_i$ denote the covariant derivatives with respect to $\tilde \gamma_{ij}$ and $\gamma_{ij}$, respectively.

Stress energy tensor for ideal magneto-hydrodynamical fluid is expressed as
\begin{equation}
\label{eq:SEtensor}
T_{\mu\nu}=(\rho h+b^2)u_\mu u_\nu+(p+b^2/2)g_{\mu \nu}-b_{\mu}b_{\nu}
\end{equation}
Here $h(=1+\varepsilon+p/\rho)$ is enthalpy and we define magnetic pressure as $b^2/2=b^\mu b_\mu/2$.
With these settings, we define other useful quantities which are; 3-velocity $v^i$ observed by Eulerian observer at rest; and magnetic field $B^\mu$ observed in the fluid flame, i.e., $u^\mu=n^\mu$.
\begin{equation}
\label{eq:Vi}
v^i=u^i/u^t
\end{equation}
\begin{equation}
\label{eq:B}
B^\mu=(0,e^{6\phi}(Wb^i-\alpha b^t u^i))
\end{equation}
Here $W=\alpha u^t$ is the Lorentz factor. We also define primitive/conservative variables  $\bf P/Q$ as
\begin{eqnarray}
\label{eq:GRpri}
{\bf P}&=& (\rho, u_i, \varepsilon, B^i) \\
\label{eq:GRcon}
{\bf Q}&=&
\left[
\begin{array}{cc}
\rho_{\ast} \\
S_i \\
\tau \\
B^i  \\
\end{array}
\right]=
\left[
\begin{array}{cc}
\rho e^{6\phi} W \\
e^{6\phi}((\rho h+b^2)W u_i-\alpha b^tb_i) \\
e^{6\phi}((\rho h+b^2)W^2-(p+b^2/2)-(\alpha b^t)^2)-\rho_{\ast} \\
B^i  \\
\end{array}
\right]
\end{eqnarray}

\subsection{Magneto Hydrodynamical Equations}
\label{sec:GRMHDeq}
Basic equations of magneto-hydrodynamical part in general relativistic form are written as the following conservative-like equations.
\begin{equation}
\label{eq:GRmass}
\partial_t \rho_{\ast}+\partial_i(\rho_\ast v^i)=0
\end{equation}
\begin{eqnarray}
\label{eq:GRmomentum}
\partial_t S_i+\partial_j(S_i v^j+\alpha e^{6\phi}P_{\rm tot}\delta_i^j-B^jb_i/u^t)=\nonumber\\
-S_0\partial_i \alpha+S_k\partial_i \beta^k+2\alpha e^{6\phi}S_k^k\partial_i \phi-\alpha e^{2\phi} ({S}_{jk}-P_{\rm{tot}} \gamma_{jk}) \partial_i 
\tilde{\gamma}^{jk}/2
\end{eqnarray}
\begin{eqnarray}
\label{eq:GRenergy}
\partial_t \tau+\partial_i (S_0v^i+e^{6\phi}P_{\rm tot}(v^i+\beta^i)-\alpha b^tB^i /u^t-\rho_\ast v^i)=\nonumber\\
\alpha e^{6\phi} K S_k^k /3+\alpha e^{2\phi} ({S}_{ij}-P_{\rm{tot}} \gamma_{ij})\tilde{A^{ij}}-S_iD^i\alpha
\end{eqnarray}
\begin{equation}
\label{eq:GRfaraday}
\partial_t B^i+\partial_j(B^iv^j-v^iB^j)=0
\end{equation}
Where $S_0=\tau+\rho_{\ast}$ and these equations can be expressed in the form of $\partial_t {\bf Q}
+\partial_i {\bf F}^i={\bf S}$.

Our procedure to evolve the MHD conservative variables $\bf Q$ is similar to our NMHD code except we use the HLL (Harten-Lax-van Leer) flux \citep{Harten1983} and not Roe-type one.
HLL-flux is less numerically expensive compared to Roe-flux, since we only have to consider the two fastest left and right going wave speed without 
considering eigenvectors like expressed in Eq. (\ref{eq:Froe}), however HLL-flux has sufficient capability to follow shocks and is suitable for our aims.
The fastest left/right going wave speed are evaluated from 4th order eigenvalue problem \citep[see, Eqs. (58), and (63-65), in][]{Anton06}.
We solve this problem by iterative Newton method with given sound speed.
Following \citet{Shibata05}, the sound speed $a$ is defined as
\begin{equation}
a=\sqrt{\frac{1}{h}\left[ \frac{\partial P}{\partial \rho} \Biggr| _\varepsilon +\frac{P}{\rho^2}\frac{\partial P}{\partial \varepsilon} \Biggr| _\rho \right]}
\end{equation}
By solving 4th order eigenvalue problem the fastest left/right going wave speed $\lambda_1 / \lambda_7$ (correspond to $\lambda_1 / \lambda_7$ in
Eq. (\ref{eq:Eigenroe})) can be obtained.
We evaluate $\lambda_1$ and $\lambda_7$ at immediate left and right of the cell boundary by adopting several types of reconstruction schemes such as monotonized central (MC) or piecewise linear method (PLM).
In this paper, we adopt only MC method which has second order convergence with respect to space.
Then the fastest left/right going wave speed $\lambda_- / \lambda_+$ at cell boundary are defined as
\begin{eqnarray}
\label{eq:EigenGR}
\lambda_-={\rm max}(0,\ \lambda_{1,L},\ \lambda_{1,R}) \\
\lambda_+={\rm max}(0,\ \lambda_{7,L},\ \lambda_{7,R})
\end{eqnarray}
With these wave speed, we define HLL flux, by following \citet{Anton06}, as
\begin{eqnarray}
\label{eq:FHLL}
{\bf F}_{HLL}=\frac{\tilde\lambda_+{\bf F(P}_L)-\tilde\lambda_-{\bf F(P}_R)+\tilde\lambda_-\tilde\lambda_+({\bf Q}_R-{\bf Q}_L)}
{\tilde\lambda_+-\tilde\lambda_-}
\end{eqnarray}
Here $\tilde\lambda=\lambda/\alpha$ and ${\bf F(P)}$ is an appropriate flux vector in Eqs. (\ref{eq:GRmass}-\ref{eq:GRfaraday}).
Solenoidal constraint of magnetic field is satisfied by CT scheme as the same procedure as NMHD.

Once we update conservative variables $\bf Q$, we have to obtain primitive variables $\bf P$ by solving following three coupled equations with iterative Newton method.
\begin{eqnarray}
\tau&=&\tau(\rho,\varepsilon,W) \\
S^iS_i&=&S^2(\rho,\varepsilon,W) \\
\rho_\ast&=&\rho_\ast(\rho,W)
\label{eq:pccp}
\end{eqnarray}
We employ the same recovering procedure as proposed in \citet{Cerda08} with adopting "safe-guess values" when the iteration does not converge.

\subsection{The BSSN Equations}
\label{sec:BSSN}
Next we describe our method to evolve metric part. As previously mentioned we evolve BSSN variables ($\phi,\ \tilde\gamma_{ij},\ K,\ \tilde A_{ij},\ F_i$)
according to following equations (see, e.g., \citet{Shibata95,Baumgarte99,Yo02}), .
\begin{eqnarray}
\label{eq:BSSN1}
(\partial_t-\mathcal{L}_\beta)\tilde\gamma_{ij}&=&-2\alpha\tilde A_{ij} \\
\label{eq:BSSN2}
(\partial_t-\mathcal{L}_\beta)\phi&=&-\alpha K/6 \\
\label{eq:BSSN3}
(\partial_t-\mathcal{L}_\beta)\tilde A_{ij}&=&e^{-4\phi}\left[ \alpha (R_{ij} -8\pi S_{ij})-D_iD_j \alpha\right]^{\rm trf}+\alpha(K\tilde A_{ij}-2\tilde A_{ik}
\tilde \gamma ^{kl}\tilde A_{jl}) \\
\label{eq:BSSN4}
(\partial_t-\mathcal{L}_\beta)K&=&-\Delta \alpha +\alpha (\tilde A_{ij}\tilde A^{ij}+K^2/3)+4\pi \alpha (S_0 e^{-6\phi}+\gamma^{ij}S_{ij}) \\
\label{eq:BSSN5}
(\partial_t-\beta^k\partial_k)F_i&=&-16\pi\alpha e^{-6\phi}S_i  \nonumber \\
&+&2\alpha\left[f^{jk}\tilde A_{ij,k}+{f^{jk}}_{,k}\tilde A_{ij}-\tilde A^{jk}h_{jk,i}/2+6\phi_{,j}\tilde A^j_i-2K_{,i}/3\right] \nonumber \\
&+&\delta^{jk}\left[-2\alpha_{,k}\tilde A_{ij}+{\beta^l}_{,k}h_{ij,l}+(\tilde\gamma_{il}{\beta^l}_{,j}+\tilde\gamma_{jl}{\beta^l}_{,i}-2\tilde\gamma_{ij}{\beta^l}_{,l}/3)_{,k}\right]
\end{eqnarray}
In these Eqs. (\ref{eq:BSSN1}-\ref{eq:BSSN5}), $\mathcal{L}_\beta$ is $Lie$ derivative with respect to $\beta^i$; "trf'' denotes trace-free operator; $\Delta=D^iD_i$;
$f^{ij}=\tilde\gamma^{ij}-\delta^{ij}$ and $h_{ij}=\tilde\gamma_{ij}-\delta_{ij}$.
$R_{ij}$ is the Ricci tensor and consisted of two parts in the form of
\begin{equation}
\label{eq:Ricci}
R_{ij}=\tilde R_{ij}+R^\phi_{ij}
\end{equation}
For explicit forms of the Ricci tensor and several notes when calculating the Ricci scalar, see \citet{Shibata02}.
We evolve these BSSN variables by the second order scheme in space \citep[e.g., Appendix of][]{Shibata99}
and by the iterative Crank-Nicholson scheme with three steps in time.
Many recent numerical simulations in full general relativity adopt fourth order scheme in space such as \citet{Zlochower05} or \citet{Etienne08}.
However, such higher order scheme is necessary especially when the metric is highly distorted such as around the BH.
Our numerical simulations with a relatively low mass star ($15M_\odot$) do not show any BH formation and we thus consider second order scheme is acceptable at this time.

For the metric, there are several mathematical and physical constraints.
As for the mathematical constraints, ${\rm det}(\tilde\gamma_{ij})=1$ and ${\rm tr}(\tilde A_{ij})=0$ should be satisfied. We enforce following two artificial procedures
\begin{eqnarray}
\tilde\gamma_{ij} &\rightarrow &\tilde\gamma_{ij}/{\rm det}(\tilde\gamma_{ij}) \\
\tilde A_{ij}&\rightarrow &\tilde A_{ij}- \tilde\gamma^{kl}\tilde A_{kl}\tilde\gamma_{ij}/3
\end{eqnarray}
after each update to maintain numerical stability.
Physical constraints are the Hamiltonian and momentum constraints.
\begin{eqnarray}
\label{eq:HamiltonianCon}
\mathcal{H}&=&\tilde D^i\tilde D_i e^{\phi}-\frac{e^{\phi}\tilde R}{8}+2\pi S_0e^{-\phi}+\frac{e^{5\phi}}{8}\left(\tilde A_{ij}\tilde A^{ij}-\frac{2}{3}K^2\right)=0 \\
\label{eq:MomentumCon}
\mathcal{M}_i&=&\tilde D_j\left(e^{6\phi}\tilde {A^j}_i\right)-\frac{2}{3}e^{6\phi}\tilde D_iK-8\pi S_i=0
\end{eqnarray}
We do not enforce any artificial modifications to satisfy these constraints, though monitor these values just as our code check.
However, we enforce Hamiltonian constraint every time we refine/coarsen the AMR blocks by solving above Poisson like non-linear equation(\ref{eq:HamiltonianCon}).
We monitor $C_\mathcal{H}$ defined by
\begin{eqnarray}
  \label{eq:L1normHamiltonianCon}
  C_\mathcal{H}=\frac{1}{M_{\rm bar}}\int{\frac{\rho_\ast\mathcal{H}}
    {\left[|\tilde D^i\tilde D_i e^{\phi}|+|\frac{e^{\phi}\tilde R}{8}|+|2\pi S_0e^{-\phi}|+|\frac{e^{5\phi}}{8}\left(\tilde A_{ij}\tilde A^{ij}-\frac{2}{3}K^2\right)|\right]}dx^3}
\end{eqnarray}
to check the Hamiltonian constraint and accuracy of our code.
Here, $M_{\rm bar}$ is proper rest mass and defined by Eq. (\ref{eq:Mbar}).

\subsection{Gauge Conditions}
\label{sec:gauge}
Hyper-surface at constant time $t$ can be foliated in spacetime arbitrary, but is usually chosen so as to keep time evolution numerically most stable.
As for the time slicing condition which determines lapse ($\alpha$), we adopt several choices in our GRMHD code such as "the approximate maximal slicing" or "harmonic slicing" or "1+log" slicing conditions.
Since the approximate maximal slicing condition requires to solve poisson like non-linear equation every time step and very time consuming method,
we usually adopt 1+log gauge condition given by
\begin{equation}
\label{eq:1+log}
\partial_t\alpha=\beta^i\partial_i\alpha-2\alpha K
\end{equation}
We have implemented "dynamical gauge condition" for shift ($\beta^i$), following \citet{Shibata03b}, which is given by solving
\begin{equation}
\label{eq:dynamicalgauge}
\partial_t\beta^i=\tilde\gamma^{ij}\left(F_j+\Delta t\partial_t F_j\right)
\end{equation}
Where, $\Delta t$ is the numerical time step.
By imposing these gauge conditions, we do not suffer from time consuming Poisson like equations at every time step,
while we do not encounter any numerical instabilities throughout our calculations of core collapse of massive star.

\subsection{Diagnostics in GRMHD}
\label{sec:diagnostics}
In GRMHD code, global quantities such as total baryon rest mass $M_{\rm bar}$; ADM mass $M_{\rm ADM}$; total angular momentum along the rotational (z) axis
$J_{\rm z}$; internal energy $E_{\rm int}$; magnetic energy $E_{\rm mag}$; kinetic energy $T_{\rm kin}$; rotational kinetic energy $T_{\rm rot}$ are defined as expressed below
\citep[see, e.g.,][]{Duez03,Kiuchi08}.
\begin{eqnarray}
\label{eq:Mbar}
M_{\rm bar}&=&\int {\rho_{\ast}dx^3}\\
\label{eq:Madm}
M_{\rm ADM}&=&\int \left[S_0e^{-\phi}+\frac{e^{5\phi}}{16\pi}\left(\tilde A_{ij}\tilde A^{ij}-\frac{2}{3}K^2-\tilde{\gamma}^{ij}\tilde{R}_{ij}e^{-4\phi}\right)
\right]dx^3\\
\label{eq:Jtot}
J_z&=&\int{ \left[\frac{\tilde A^x_y-\tilde A^y_x}{8\pi}+xS_y-yS_x+\frac{xK_{,y}-yK_{,x}}{12\pi}-\frac{x\tilde \gamma^{ij}_{,y}\tilde A_{ij}-
y\tilde \gamma^{ij}_{,x}\tilde A_{ij}}{16\pi}\right]e^{6\phi}dx^3}\\
\label{eq:Eint}
E_{\rm int}&=&\int \rho_\ast \varepsilon dx^3\\
\label{eq:Tkin}
T_{\rm kin}&=&\int \rho_\ast hu_i v^i dx^3\\
\label{eq:Trot}
T_{\rm rot}&=&\int \rho_\ast hu_\phi v^\phi dx^3\\
\label{eq:Emag}
E_{\rm mag}&=&\int \frac{b^2}{2} W e^{6\phi} dx^3
\end{eqnarray}
Then the gravitational potential energy $E_{\rm grv}$ is defined by $E_{\rm grv}=-(M_{\rm bar}-M_{\rm ADM}+E_{\rm int}+T_{\rm kin}+E_{\rm mag})$.
Because of our formulae for MHD part and because of our AMR scheme (see, Sec. \ref{sec:Boundary of AMR Blocks}), $M_{\rm bar}$ is conserved with high accuracy.
On the other hand, conservation of $M_{\rm ADM}$ which is guaranteed from the Einstein's equation
in the absence of gravitational radiation is violated due to the accumulation of numerical errors in our CCSNe
simulations.
Then, several \% fluctuation appears (see, Fig.\ref{pic:f40} in Sec. \ref{sec:GRMHDtest}) which is approximately 2-3 orders larger compared to that of $M_{\rm bar}$ in our CCSNe simulations.
Consequently even though our initial condition satisfies $E_{\rm grv}<0$ (i.e., gravitationally trapped system), there sometimes appear that $E_{\rm grv}>0$ during calculation.
Therefore we estimate the gravitational potential energy as $E_{\rm grv}\sim -(E_{\rm int}+T_{\rm kin}+E_{\rm mag})$ and apply it to such as the rotational to gravitational energy $\beta_{\rm rot}=T_{\rm rot}/|E_{\rm grv}|$.

\section{Adaptive Mesh Refinement}
\label{sec:AMR}
One of difficulties in computational astrophysics is that we have to handle wide dynamical range in a limited computational resource.
For instance, in the context of CCSN simulation, the PNS is a size of $ \sim$10 km and on the other hand radius of the iron core is the order of $\ga O(1000)$km.
If we cover such a vast range (several times of the iron core) with a uniform resolution, e.g., $\sim 500$m to resolve interior of the PNS, it becomes impossible to calculate with our limited computational resource.
We thus raise resolution in the vicinity of proto-neutronstar and, at the same time, lower resolution far from the centre.
To realize such situation, we incorporate the AMR technique \citep[e.g.,][]{Berger89,Powell99} into our codes.
\subsection{AMR Structure}
In our codes, computational domain is divided into "$blocks$" (hereafter, AMR block) and every AMR block consists of $8\times8\times8$ cubic cells and of 2 additional cells as ghost zones in every side of block.
Every AMR block belongs to a refinement level "$l$" and if the refinement level $l$ is raised by one,
the AMR block is divided into 8 blocks with halved cell width.
On the other hand, if all $2\times2\times2$ neighboring AMR blocks are assigned to lower their refinement level by one, 8 blocks merge into one block with twice cell width.

Our codes are fully parallelized and adopt Message Passing Interface (MPI) for communication between
different nodes.
Then it requires load balancing in AMR frame work.
Our method for this purpose is like this.
Three dimensional structure of AMR blocks are projected on one dimensional structure connected by {\it "Hilbert}" space filling curve \citep{Hilbert1891}.
Along the $Hilbert$ curve, AMR blocks are numbered sequentially.
Then AMR blocks projected on one dimension are allocated to all computational nodes in a straight forward manner.
Using such a projection scheme, e.g., connecting by $Hilbert$ curve, enables us to minimize the data transfer between different computational nodes.
This is because surface area of a "chunk" of AMR blocks allocated in one node by above method
is minimized as much as possible and, thus, we can minimize time to spare for the data communication.
\begin{figure}[htpb]
\begin{tabular}{cc}
\begin{minipage}{0.5\hsize}
\begin{flushright}
\includegraphics[width=30mm,angle=-90]{f2.eps}
\end{flushright}
\end{minipage}
\begin{minipage}{0.5\hsize}
\begin{flushleft}
\includegraphics[width=30mm,angle=-90]{f3.eps}
\end{flushleft}
\end{minipage}
\end{tabular}
\begin{center}
\includegraphics[width=50mm]{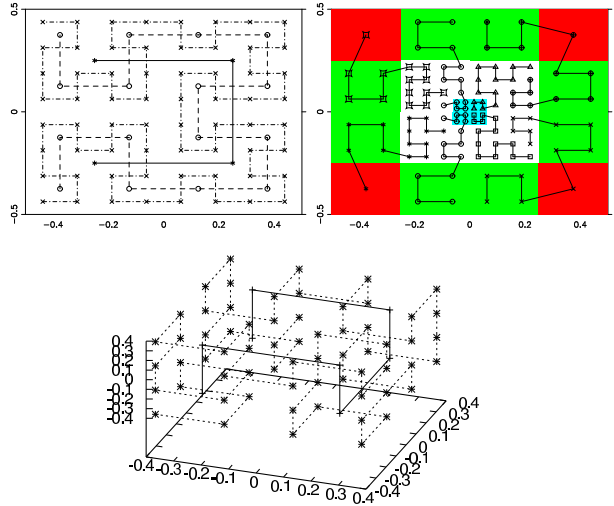}
\end{center}
\caption{$Top$-$left$: $Hilbert$ curves filling two dimensional area covered by uniform AMR blocks with three different levels.
$Top$-$right$: $Hilbert$ curves filling two dimensional area covered by nonuniform AMR blocks with four
different AMR levels.
In this panel, AMR blocks are allocated to 8 nodes and blocks allocated to one node are connected through one sequential line.
$Bottom$: Three dimensional extension of $top$-$left$ panel with two different levels.
}
\label{pic:Filling_curve.epsAMR_conf.eps}
\end{figure}
In $top$-$left$ panel of Fig.\ref{pic:Filling_curve.epsAMR_conf.eps}, we display an example of $Hilbert$
curve in two dimension.
In this figure, asterisks(circles, crosses) denote centers of cells with width 0.5(0.25, 0.125) and lines are 
$Hilbert$ curves.
As seen in this panel, two dimensional structure of AMR blocks is projected on one dimension.
In $right$-$top$ panel, we again display $Hilbert$ curve which fills two dimensional computational domain covered by blocks of four different AMR levels.
Background colors represent AMR levels and blocks connected one curve are allocated to one computational node.
Therefore, in this panel, all AMR blocks are allocated to 8 nodes with maintaining load balancing.
In $bottom$ panel, we also display three dimensional extension of $Hilbert$ curves with two different AMR levels for reference.

By adopting such method, our AMR structure has flexibility to refine or coarsen AMR blocks locally.

\subsection{Poisson Solver under the AMR Framework}
\label{sec:Poisson Solver under the AMR Framework}
\begin{figure}[htbp]
  \begin{center}
    \includegraphics[width=65mm]{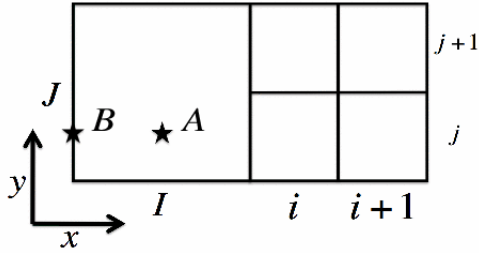}
  \end{center}
  \caption{Schematic figure of interpolation of ghost zone value.
    Four small boxes and one large box belong to AMR level $l$ and $l-1$, respectively.}
  \label{pic:buffer.eps}
\end{figure}

In NMHD code, we have to solve Poisson equation in the form of $Ax=B$ for the self gravity and; in GRMHD code,
Poisson like non linear equation in the form of $Ax=B(x)$ for the initial Hamiltonian and momentum constraints.
Here $A$ is a given $(N_{\rm block}\times8\times8\times8)\times(N_{\rm block}\times8\times8\times8)$ matrix,
$B$ is a given $(N_{\rm block}\times8\times8\times8)$ vector and $x$ is a solution we seek which is $(N_{\rm block}\times8\times8\times8)$ vector.
$N_{\rm block}$ is a total number of AMR blocks.
In GRMHD, $B(x)$ is a vector containing non-linear term of $x$.
We adopt iterating method, the so called "BiConjugate Gradient Stabilized (BiCGSTAB)" \citep{vanderVorst92}
method to solve such huge simultaneous equations.
Our strategy for solving this equation under our AMR structure is; (1) we set an AMR level $l$ which is 0 at initial;
(2) for all AMR boxes whose AMR levels are larger or equal to $l$, we project their physical quantities, such as the density, to boxes of AMR level $l$
by the coarsening procedure and construct $A\ \&\ B$; (3) we then solve equation by BiCGSTAB on the uniform mesh with appropriate
boundary conditions at the interface of AMR level $l$ and $l-1$ and also at the outer boundary; (4) increment $l$ by one and repeat these procedures
from (1) again.
Note that, we have to pay special attention at the interface of different AMR level, i.e., $l$ and $l-1$.
At here, we have to connect both the solutions and their first derivation smoothly, otherwise there appear some non-physical divergence.
To avoid this, we adopt quadratic and bilinear interpolation methods following \citet{Matsumoto07} to evaluate ghost zone values of AMR level $l$, e.g., $\Phi(i-1,\ j)$ and $\Phi(i-2,\ j)$ in Fig.\ref{pic:buffer.eps}.
Here, we summarize our interpolation method in two dimension.
Three dimensional extension can be done in a straightforward manner.
If we seek gravitational potential $\Phi(i-1,\ j)$, we first have to obtain $\Phi(A)$ at $(I,\ J-1/4)$ which is derived via
\begin{eqnarray}
\label{eq:PhiA}
\Phi(A)=\Phi(I,J)-\frac{1}{2}{\rm MC}(\Phi(I,J+1)-\Phi(I,J),\Phi(I,J)-\Phi(I,J-1))
\end{eqnarray}
Here, MC$(x,y)$ is the monotonized central method.
Then $\Phi(i-1,\ j)$ is derived via quadratic interpolation
\begin{eqnarray}
\label{eq:quadratic}
\Phi(i-1,j)=\frac{10\Phi(i,j)-8\Phi(A)+3\Phi(i+1,j)}{15}
\end{eqnarray}
$\Phi(B)$ which is required to evaluate $\Phi(i-2,j)$ is obtained by following equation
\begin{eqnarray}
\Phi(B)=\frac{\Phi(I,J)+\Phi(I-1,J)}{2}-\frac{1}{2}{\rm MC}(\frac{\Phi(I,J+1)+\Phi(I-1,J+1)}{2}-\frac{\Phi(I,J)+\Phi(I-1,J)}{2}\nonumber \\
,\frac{\Phi(I,J)+\Phi(I-1,J)}{2}-\frac{\Phi(I,J-1)+\Phi(I-1,J-1)}{2})
\end{eqnarray}
Three dimensional extension of this smoothening method is done by replacing, e.g, Eq.(\ref{eq:PhiA}), with bilinear interpolation
\begin{eqnarray}
\Phi(A)=\Phi(I,J,K)&-&\frac{1}{2}{\rm MC}(\Phi(I,J+1,K)-\Phi(I,J,K),\Phi(I,J,K)-\Phi(I,J-1,K))\nonumber \\
&-&\frac{1}{2}{\rm MC}(\Phi(I,J,K+1)-\Phi(I,J,K),\Phi(I,J,K)-\Phi(I,J,K-1))
\end{eqnarray}
On the other hand, lower level ghost zone value $\Phi(I+1,J)$ is derived by $"restriction"$ procedure
and this is simply averaging $\Phi$ over the $2\times2$ (or $2\times2\times2$ in 3D) adjacent cells via
\begin{eqnarray}
\Phi(I+1,J)=\frac{1}{4}\sum_{n,m=0,1}{\Phi(i+n,j+m)}
\end{eqnarray}

\subsection{Boundary of AMR Blocks}
\label{sec:Boundary of AMR Blocks}
\begin{figure}[htpb]
\begin{center}
\includegraphics[width=60mm]{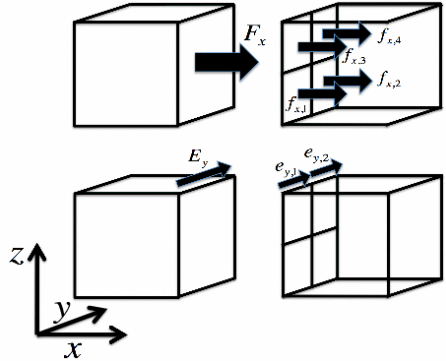}
\end{center}
\caption{Schematic picture of refluxing of the numerical flux $F_x(f_x)$ and the electric field $E_y(e_y)$.
Left two and right boxes belong to AMR level $l-1$ and $l$, respectively. 
}
\label{pic:refluxing.eps}
\end{figure}

To guarantee the conservation law and the solenoidal constraint of magnetic field, we have to reflux the numerical flux, Eqs. (\ref{eq:Froe}) and (\ref{eq:FHLL}), and the electric field, Eq.(\ref{eq:electricfield}), at where AMR boxes of different levels are contacting.
In Fig.\ref{pic:refluxing.eps}, we display schematic picture of refluxing procedures.
For instance, the numerical flux $F_x$ belonging to AMR level $l-1$ and defined at cell boundary is replaced by summation of $f_{x,i}(i=1,2,3,4)$ which belong to AMR level $l$.
Similarly, as for the electric filed, the electric field $E_y$ defined at cell edge is replaced by summation of $e_{y,i}(i=1,2)$.
These procedures ensure the conservation and the solenoidal constraint below the round off error.

Additionally, as for the ghost zones, we have to obtain physical variables every after the time updating and
this procedure is sorted into three cases.

(1)For AMR box whose neighbor has the same AMR level, we simply copy all physical variables.

(2)For AMR box whose neighbor has lower level, ghost zone variables are interpolated and are sent from lower to higher level box.
For this interpolation, we use the same method as used in our Poisson solver (described in Sec.
\ref{sec:Poisson Solver under the AMR Framework}) other than the magnetic field $\bf B$.
As for the magnetic field, we have to interpolate while maintaining the solenoidal constraint and
this is done by adopting the same method proposed by \citet{Balsara01}.

(3)If AMR level of the neighbor is higher, physical variables are evaluated by "$restriction$" procedure.
In our codes, this restriction procedure is simply averaging the variables of $2\times2\times2$ adjacent cells
which is the same method as used in our Poisson solver (Sec. \ref{sec:Poisson Solver under the AMR Framework}).

\subsection{The BSSN Evolution Under the AMR Framework}
\label{sec:The BSSN Evolution under the AMR Framework}
During time evolution of the BSSN variables, we have to derive the spatial derivatives of metrics not only along one direction, e.g., $\partial_x$, but also
the cross derivatives, e.g., $\partial_{xy}$ to obtain such as the Ricci tensor.
Therefore, if there exist some discontinuity in the spatial derivatives across the AMR refinement boundary, spurious oscillations of the BSSN variables
appear near the refinement boundary.
Since the time marching is simultaneous across all the AMR boxes in our codes, there is no time lag between different AMR levels.
However, we should carefully interpolate the buffer zone's metrics especially for AMR boxes whose neighbors have lower AMR levels than theirs.
This situation is the same as that appeared in our Poisson solver (Sec. \ref{sec:Poisson Solver under the AMR Framework}) and we adopt the same strategy to obtain the buffer zone's variables.
In addition we have to evaluate the metrics along the edge of AMR block (e.g., $(i,j,k)=(9,9,1\sim8)$) for the cross derivatives and this evaluation can be done in a similar manner as that in the normal buffer zone's case.
For instance, if we seek a metric $X$ at $(i,j,k)=(9,9,1)$, $X(A)$ which corresponds to $\Phi(A)$ in Eq. (\ref{eq:PhiA}) is replaced by
\begin{equation}
X(A)=X(1,1,1)-\frac{1}{2}{\rm MC}(X(1,1,2)-X(1,1,1),X(1,1,1)-X(1,1,0))
\end{equation}
Then, in a straight forward manner of Eq. (\ref{eq:quadratic}), $X(9,9,1)$ can be derived by
\begin{eqnarray}
X(9,9,1)=\frac{10X(8,8,1)-8X(A)+3X(7,7,1)}{15}
\end{eqnarray}

However, we cannot completely suppress the spurious oscillations in the refinement boundary and, in such case, adding numerical dissipation is sometimes
useful \citep{Schnetter04}.
Even though we do not add the dissipation at present codes, we do not suffer from growth of the noises and the code crash.
Several tests of the BSSN evolution with AMR structure are summarized in Sec.\ref{sec:TestDynamical}.

\section{Tests for NMHD Code}
\label{sec:NMHDtest}
In this section, we introduce several test problems done by our newly developed 3DMHD code in the Newtonian approximation.
We verified the accuracies of our new AMR-NMHD code through several test suites.
\subsection{One Dimensional Shock Tube Test}
\label{sec:1DTubeNMHD}
\begin{figure}[htpb]
\epsscale{0.3}
\plotone{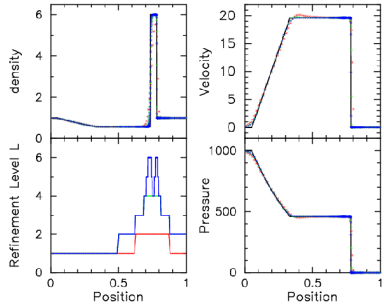}
\caption{Results of test 1) at $t=0.012$. $Black$ solid line is analytical solution and $blue$, $green$ and $red$ dots/lines are our numerical results.
Each color represents the maximum refinement level $L_{\rm AMR,max}$. We set $L_{\rm AMR,max}=2,4,6$ for $red$, $green$ and $blue$, respectively.}
\label{pic:f30}
\end{figure}

\begin{figure}[htbp]
\begin{center}
\includegraphics[scale=0.3]{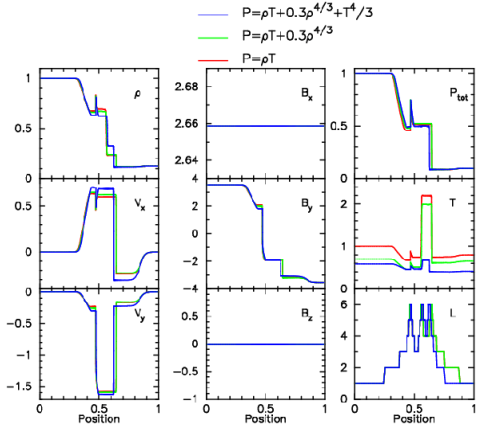}
\caption{Results of test 2) at $t=0.1$. Each line corresponds to different EOS shown at the top.
}
\label{pic:f31}
\end{center}
\end{figure}

We calculated several 1-D hydrodynamical shock tube tests and compared to the exact solutions obtained from the code HE-E1RPEXACT of the library $NUMERICA$\citep{ToroNumerica}.
We show one test in Fig.\ref{pic:f30} in which we assumed ideal gas with adiabatic index $\gamma=1.4$.
The mesh width of exact solution is 1/1000, meanwhile, the "effective" mesh width of our numerical run's are 1/2048, 1/512, 1/128 for $blue$, $green$ and 
$red$ dots, respectively.\\
test 1)
\begin{eqnarray}
(\rho\ ,\ V_{\rm x}\ ,\ P_{\rm tot})=
\left\{\begin{array}{cccc}
(1.0, & 0.0, & 1000.0) & x\le0.5 \nonumber \\
(1.0, & 0.0, & 0.01) & x>0.5 \nonumber
\end{array} \right.
\end{eqnarray}
For MHD shock tube test, we show the same test described in \citep{Brio88} with three different types of the EOSs in Fig.\ref{pic:f31}.\\
test 2)
\begin{eqnarray}
(\rho\ ,\ &{\bf V}&,\ B_{\rm x},\ B_{\rm y},\ B_{\rm z},\ P_{\rm tot})=\nonumber \\
&&\left\{\begin{array}{ccccccc}
(1.0,& {\bf 0}, & 0.75\sqrt{4\pi},& \sqrt{4\pi},& 0.0, & 1.0) & x\le0.5 \nonumber \\
(0.125, & {\bf0},& 0.75\sqrt{4\pi}, & -\sqrt{4\pi}, & 0.0, & 0.1) & x>0.5 \nonumber
\end{array} \right.
\end{eqnarray}
Initial total pressure $P_{\rm tot}$, which excludes the magnetic pressure, is divided into three parts such as the gas $(P_{\rm gas})$, the degenerate $(P_{\rm dege})$ and the radiation $(P_{\rm rad})$ pressure term.\\
\begin{eqnarray}
&a):& P_{\rm tot}=P_{\rm gas}=\rho T\nonumber \\
&b):& P_{\rm tot}=P_{\rm gas}+P_{\rm dege}=\rho T+0.3\rho^{4/3}\nonumber \\
&c):& P_{\rm tot}=P_{\rm gas}+P_{\rm dege}+P_{\rm rad}=\rho T+0.3\rho^{4/3}+T^4/3\nonumber
\end{eqnarray}
Internal energy has also three parts corresponding to the gas $(\varepsilon_{\rm gas})$, the Fermi $(\varepsilon_{\rm dege})$ and the radiation $(\varepsilon_{\rm rad})$ energy
as defined by the following.
\begin{eqnarray*}
\varepsilon_{\rm tot}(\rho,T) & = & \varepsilon_{\rm gas}+\varepsilon_{\rm dege}+
\varepsilon_{\rm rad}\\
 & = & \frac{P_{gas}}{(\gamma-1)\rho}+\int_0^\rho \frac{P_{\rm dege}}{\rho^2} d\rho
 +\frac{3P_{\rm rad}}{\rho}
\end{eqnarray*}
Here $\gamma=2.0$ and EOS a) corresponds to the original model reported in \citet{Brio88}.
From Fig.\ref{pic:f30}-\ref{pic:f31}, we see that our code captures the discontinuities accurately and also no numerical instabilities are seen.

\subsection{Poisson Solver}
\label{sec:Tests of Poisson Solver}
\begin{figure}[htpb]
\begin{center}
\includegraphics[width=55mm,angle=-90.]{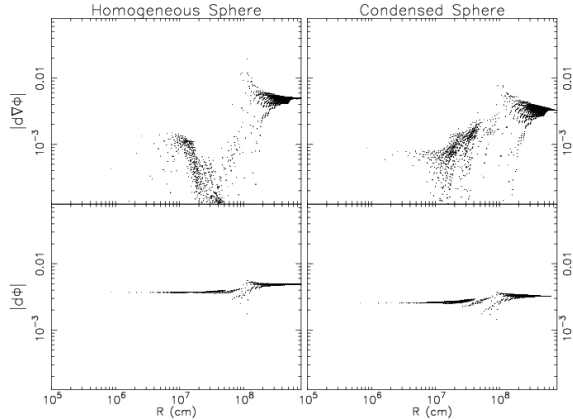}
\caption{Results of test 3) ($left$) and 4) ($right$).
See text for the definitions of $|{\rm d}\phi |$ and  $|{\rm d}\nabla\phi |$.
Five different levels of AMR boxes are drawn.}
\label{pic:f32}
\end{center}
\end{figure}

As for the tests of our Poisson solver, we set two types of spherically symmetric density distribution like below which have analytical solutions and compare our results with analytical ones.\\
test 3)
\hspace{1cm}Homogeneous sphere of radius R and density $\rho_0$\\
test 4)
\hspace{1cm}Centrally condensed sphere with density distribution $\rho(r)$
\begin{equation}
\label{eq:test_dens_dis}
\rho(r) = \left\{
\begin{array}{ll}
\frac{\rho_0}{1+(r/r_{\rm c})^2} & \mbox{r $<$ R}  \\
0 & \mbox{ r $>$ R}
\end{array}
\right.
\end{equation}
These tests are the same tests done in \citet{Stone92} and have analytical solutions, thus we can easily check
the accuracy of our Poisson solver\citep[for analytical formulae, see][]{Stone92}.
We set $R=10^8$cm, $\rho_0=10^{12}$g $\rm cm^{-3}$ and $r_{\rm c}=3\times10^7$cm for both tests 3) \& 4).
Fig.\ref{pic:f32} shows our numerical results with comparing analytical ones.
$Upper$ and $lower$ two panels show $|{\rm d}\nabla\phi |\tbond|(\nabla\phi_{\rm ana}-\nabla\phi)/\nabla\phi_{\rm ana} |$
and $|{\rm d}\phi |\tbond|(\phi_{\rm ana}-\phi)/\phi_{\rm ana} |$, respectively.
Here, $\phi$ is the numerical result of the Poisson equation and $\phi_{\rm ana}$ is the analytical one.
Deviations of our numerical results from the analytical ones are $\sim0.1$\% and we also find neither kink nor jump of both $\phi$ and
$\nabla\phi$ at the interface of different AMR level boxes.
Therefore, we consider our Poisson solver under AMR structure works with sufficient accuracy.

\subsection{Energy and Angular Momentum Conservations}
\label{sec:Energy and Angular Momentum Conservations}
\begin{figure}[htpb]
\begin{center}
\includegraphics[width=60mm]{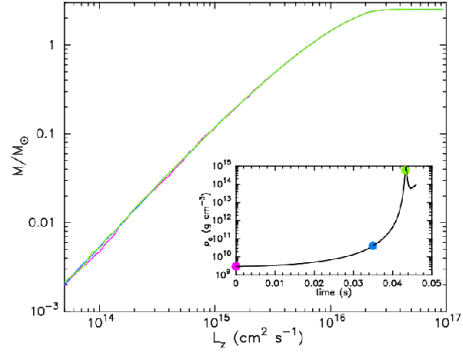}
\end{center}
  \caption{The angular momentum conservation. Abscissa and vertical axes represent the specific angular momentum and the total mass in solar 
  mass unit which is the summation of fluid elements having less or equal to the corresponding specific angular momentum on the abscissa axis, 
  respectively.
  In the inner mini-panel, the time evolution of the central density is displayed.
  Each Color of line represents the elapsed time from the initiation of collapse and corresponds to the same color-coded circle in the mini-panel.
 }
\label{pic:f33}
\end{figure}

\begin{figure}[htpb]
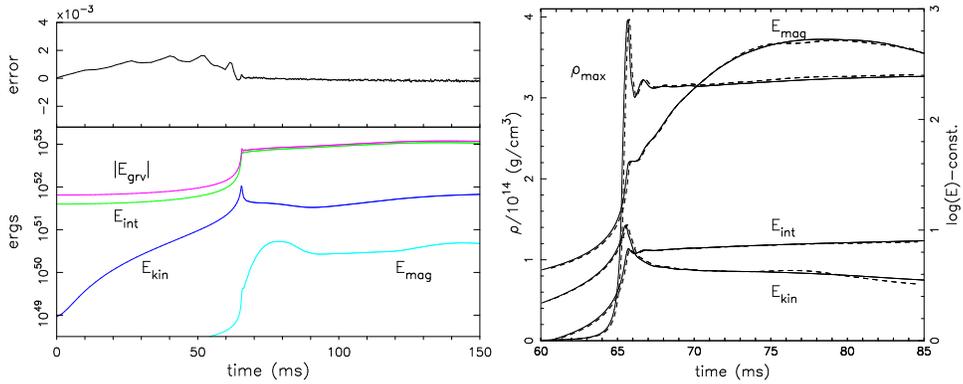

\begin{tabular}{cc}
\begin{minipage}{0.5\hsize}
\begin{flushright}
\includegraphics[width=50mm,angle=-90.]{f11.eps}
\end{flushright}
\end{minipage}
\begin{minipage}{0.5\hsize}
\begin{flushleft}
\includegraphics[width=50mm,angle=-90.]{f12.eps}
\end{flushleft}
\end{minipage}
\end{tabular}
  \caption{
  $Left$: The error in energy conservation and time evolutions of various energy components.
  $\rm E_{grv}$, $\rm E_{int}$, $\rm E_{kin}$ and $\rm E_{mag}$ represent gravitational, internal, kinetic and magnetic energy, respectively.
  Relative error is defined by a deviation of $\rm{(E_{grv}+E_{int}+E_{kin}+E_{mag})/|E_{grv}|}$ from its initial value and is shown in upper part.
  Energy conservation is maintained within $\sim$0.2\% error through our time evolution.
  $Right$: Magnified view around the time of core bounce to see the numerical convergence with respect to the
  grid resolution.
  $Solid$ and $dashed$ curves correspond to models with minimum resolution $\Delta x_{\rm min}=600$m and
  $\Delta x_{\rm min}=480$m, respectively.
  The internal, kinetic and magnetic energies are normalized by $10^{52}$, $10^{51}$ and $10^{48}$,
  respectively.
  }
\label{pic:f34}
\end{figure}

In this subsection, we check our NMHD code's accuracy against the energy and the angular momentum
conservations.
Since we consider one of energy source of the formation of bipolar outflow is the extracted angular momentum,
we have to carefully trace the time evolution of angular momentum.
As for the test of angular momentum transfer, we follow the collapse of a non-magnetized and rotating
25M$_{\odot}$ star with the adiabatic gas with index $\gamma=1.4$.
If no magnetic field exists and the fluid is adiabatic gas, the angular momentum is not transported.
Result is shown in Fig.\ref{pic:f33}.
Abscissa and vertical axes represent the specific angular momentum and the total mass in solar mass unit which is the summation of fluid elements having less or equal to the corresponding specific angular momentum on the abscissa axis, respectively.
If the angular momentum conservation is maintained, the curves do not change its form in time
and we thus see the angular momentum conservation is well maintained from this figure.
As for the energy conservation test, we calculate collapse of a rotating and magnetized 15M$_{\odot}$ star
(corresponds to model "NB12R020Sf" described later in Sec.\ref{sec:collapse}).
In Fig.\ref{pic:f34}, we display time evolutions of various energy components and the error in energy
conservation in $left$ panel and the magnified view around the time of core bounce with different numerical
resolutions to see the numerical convergence in $right$ panel.
As for the numerical convergence test, the computational domain is chosen as $(x,y,z)=[-5000,5000]$km for
$solid$ curves and $(x,y,z)=[-4000,4000]$km for $dashed$ ones.
Then the minimum grid widths become $\Delta x_{\rm min}=600(solid)$ and $480(dashed)$ m.
From Fig.\ref{pic:f34}, we see the energy conservation is well maintained and also see the numerical
convergence is achieved within the range of our adopted numerical resolutions.

\section{Tests for GRMHD Code}
\label{sec:GRMHDtest}
In this section, we introduce several test problems done by our newly developed 3DGRMHD code.
\subsection{One Dimensional Shock Tube Test}
\label{sec:1DTubeGRMHD}
\begin{figure}[htpb]
\begin{center}
\includegraphics[width=40mm,angle=-90.]{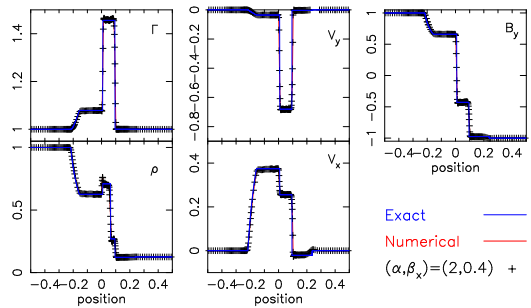}
\caption{The relativistic version of Brio \& Wu MHD shock tube test.
Exact solutions are obtained from \citet{Giacomazzo06} and drawn with blue lines.
$\Gamma$ is the Lorentz factor and other notations are the same as in Fig.\ref{pic:f31}.
Numerical results ($red$ lines) are models with the gauge condition $(\alpha,\ \beta_x)=(1.0,\ 0.0)$ and black crosses are the one with $(\alpha,\ \beta_x)=(2.0,\ 0.4)$.
Levels of AMR blocks are from 0 to 3 and the highest resolution is $\Delta \rm x=1/512$.
Time slice is chosen at $t=0.25$ for $\alpha=1$ models and $t=0.125$ for $\alpha=2$ model.
Results are shifted to $x\rightarrow x+0.4\times t$ for non-zero shift gauge model.}
\label{pic:f35}
\end{center}
\end{figure}

As for the basic test, we calculated the relativistic Brio \& Wu MHD shock tube test \citep[see,][]{Brio88,Komissarov99} with fixed flat metric and the results are
shown in Fig.\ref{pic:f35}.
In this figure, we show two models with different gauge conditions and compare with the exact solutions obtained from \citet{Giacomazzo06}.
One is $(\alpha,\beta_x)=(1.0,0.0)$ and represented by $red$ lines and the other is $(\alpha,\beta_x)=(2.0,0.4)$ and plotted by crosses.
Time slice is taken at $t=0.25$ and $t=0.125$ for $\alpha=1$ and 2, respectively.
In addition, for non-zero shift gauge ($\beta_x=0.4$) model, results are shifted to $x\rightarrow x+0.4\times t$ to let it coincides with other results.
We can see our GRMHD code can handle the non-zero gauge conditions and shocks.
\subsection{Bondi Accretion}
\begin{figure}[htpb]
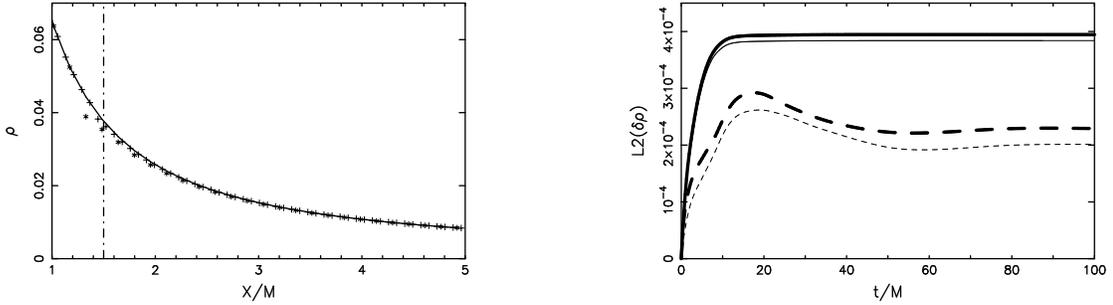

\begin{center}
\begin{tabular}{cc}
\begin{minipage}{0.5\hsize}
\begin{center}
\includegraphics[width=40mm,angle=-90.]{f14.eps}
\end{center}
\end{minipage}
\begin{minipage}{0.5\hsize}
\begin{center}
\includegraphics[width=40mm,angle=-90.]{f15.eps}
\end{center}
\end{minipage}
\end{tabular}
  \caption{$Left$: Density plots of magnetized Bondi accretion test with different resolutions taken at $t=100M$.
  $Solid$ line represents analytical solution and the vertical $dash$-$dotted$ line is where we excise the computational domain.
  $Asterisks$ and $crosses$ represent our results with different numerical resolutions.
  They correspond to $N_{\rm block}=16^3(Asterisk)$ and $N_{\rm block}=8^3(cross)$, respectively.
  $Right$: L2 norms of the errors in density from analytical value.
  $Long$-$dashed$ lines are non magnetized accretion models and solid lines are magnetized models with
  $B^{\rm r}|_{\rm Horizon}=1.0$ (or $b^2/\rho|_{\rm Horizon}=2.446$).
  Thick lines are $N_{\rm block}=8^3$ models, while thin lines are $N_{\rm block}=16^3$ models.
  Note that the errors of $N_{\rm block}=8^3$ models are multiplied by $(1/2)^2$ to show the second order convergence.
  }
\label{pic:Bondi}
\end{center}
\end{figure}

In this subsection, we test our GRMHD code in a strongly curved and fixed spacetime.
As for the test, we evolve the Bondi accretion flow with/without magnetic field and compare our results to
analytical solution which are obtained according to \citet{Hawley84}.
It is known that the radial magnetic field does not influence the Bondi accretion \citep{DeVilliers03} and, thus,
we add the initial magnetic field via
\begin{equation}
{\bf B}=\nabla\times\left[\frac{B_0}{r(r+z)}(-y,\ x,\ 0)\right]
\end{equation}
In this test, we adopt Kerr-Schild coordinate
\begin{eqnarray}
\alpha&=&\frac{1}{\sqrt{1+2/r}} \\
\beta^i&=&\left(\frac{2}{2+r}\right)\frac{x^i}{r} \\
\gamma_{ii}&=&\left(1+2/r,\ r^2,\ r^2 {\rm sin}^2\theta\right)
\end{eqnarray}
Here, the metric $\gamma_{ii}$ is written in spherical polar coordinate.
Event horizon locates $r=2$ and outer boundary is set at $|x,y,z|=10$.
We excise the computational domain $|x,y,z|\le1.5$ and simply connect the non- and excised region with first order extrapolation.
We run four models with two different resolutions ($N_{\rm block}=8^3,\ 16^3$) and $B^r_{\rm Horizon}=1$
(or $b^2/\rho|_{\rm Horizon}=2.446$) and $B^r_{\rm Horizon}=0$.
In Fig.\ref{pic:Bondi}, we show the rest density profiles of magnetized models at $t=100M$ in $left$ 
panel and L2 norm of the errors in density in $right$ panel.
The errors in $N_{\rm block}=8^3$ models are multiplied by $(1/2)^2$ to show the second order convergence.
From this test, we see that our GRMHD code which employs HLL flux with MC limiter actually reproduces the second order convergence and also that our code can treat the strongly curved spacetime.

\subsection{${\rm{div}}({\bf{B}})=0$ Constraint}
\begin{figure}[htpb]
\begin{center}
\includegraphics[scale=0.4,angle=-90.,clip]{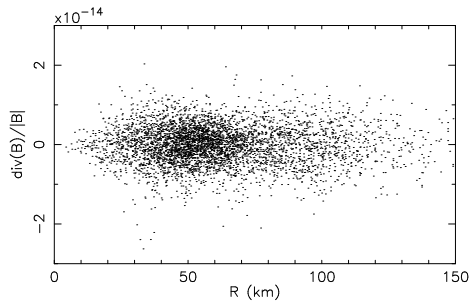}
  \caption{${\rm{div}}({\bf{B}})/B$ is plotted of model "GB12R020Sf". In this figure, three different AMR levels are plotted in the central $[0,100]\times[0,100]\times[0,100]$km$^3$ region.
  }
\label{pic:f36}
\end{center}
\end{figure}

To check whether our two codes satisfy the solenoidal constraint for the magnetic field, we have checked the value ${\rm{div}}({\bf{B}})/B$.
Fig.\ref{pic:f36} display ${\rm{div}}({\bf{B}})/B$ in the central $[0,100]\times[0,100]\times[0,100]$km$^3$ region from one representative model
of GRMHD models.
In this region, three different AMR level boxes exist.
At this moment, it almost reaches core bounce time and three times refinement procedures have been done since the initiation of calculation.
From this figure, we see that solenoidal constraint is well maintained below the round-off error which tells us that both the refinement procedures and 
the electric field refluxing work well.

\subsection{Test Problems With Dynamical Background}
\label{sec:TestDynamical}
In this subsection, we test our dynamical metric solver written by the BSSN formalism with and without matters.
\subsubsection{Linearized Teukolsky Wave}
\label{sec:Linearized Teukolsky Wave}
\begin{figure}[htpb]
\begin{center}
\includegraphics[scale=0.4,angle=-90.,clip]{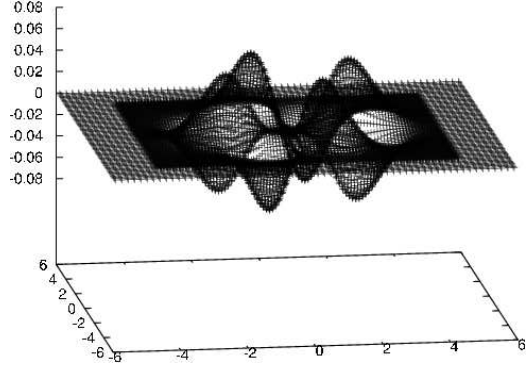}
\caption{Our AMR structure and initial value of the extrinsic curvature $K_{12}$ in the equatorial plane
of linearized "Teukolsky wave test".
The inner region of $|x,y,z|\la5$ is covered by higher resolution meshes.
}
\label{pic:f37}
\end{center}
\end{figure}

\begin{figure}[htpb]
\begin{center}
\includegraphics[scale=0.3,angle=-90.]{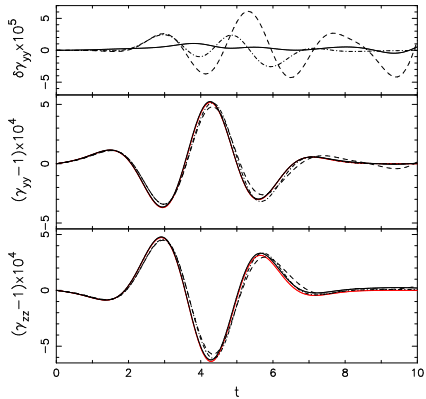}
\caption{Results of linearized "Teukolsky wave test".
$Bottom$ and $middle$ panels represent $\gamma_{zz}-1$ and $\gamma_{yy}-1$, extracted at (x, y, z)=(4.2, 0, 0), respectively.
$Red$ lines are analytical solutions and $solid$, $dash$-$dotted$ and $dashed$ lines are results of
maximum AMR level 3, 2 and 1, respectively.
In model with maximum AMR level 3, the minimum grid resolution is $\Delta x_{\rm min}=0.078$.
$Top$ panel represents deviations of $\gamma_{yy}$ from analytical value with three different grid resolutions.
}
\label{pic:f38}
\end{center}
\end{figure}

First test is to follow the linearized gravitational waves, the so called "Teukolsky wave" \citep{Teukolsky82}, in a vacuum space.
Following \citet{Shibata95}, we adopt the same mode $l=|m|=2$ and the initial wave amplitude is set to $C=0.01$.
For time slicing gauge condition, we adopt "1 + log" condition (see, Sec. \ref{sec:gauge}).
In this test, we also check the influence of the boundary where different AMR level boxes are contacting.
In Fig.\ref{pic:f37}, the initial condition (a component of the extrinsic curvature, $K_{12}$) and AMR structure are displayed.
As seen in this figure, the inner region of $|x,y,z|\le5$ is covered by maximum AMR level meshes which
we vary as $L_{\rm AMR}=1$, 2 and 3 to see convergence of the error.
We extract the metric variables $\gamma_{yy}$ and $\gamma_{zz}$ at (x, y, z)=(4.2, 0, 0) and
compared them with analytical ones in Fig.\ref{pic:f38}.
$Red$ curves are analytical solutions (see, e.g., \citet{Nakamura87}) and $black$ curves are numerical results.
Numerical resolutions are $\Delta x_{\rm min}=0.078(solid)$, 0.156($dash$-$dotted$) and 0.312($dashed$).
From $top$ panel of Fig.\ref{pic:f38}, we see that the errors decrease with increasing numerical resolution.

\subsubsection{Rotating Neutron Star}
\label{sec:Rotating Neutron Star}
\begin{figure}[htbp]
\begin{center}
 \includegraphics[angle=-90.,width=50mm,clip]{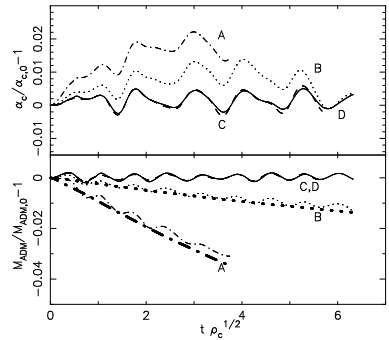}
  \caption{Time evolutions of the central lapse ($top$) and the ADM mass ($bottom$) normalized by their initial values.
  A ($dash$-$dotted$) and B ($dotted$) are fully (both matters and metrics) evolved models,
  while, in C ($dash$) and D ($solid$), only metrics are evolved.
  In the $bottom$ panel, we also plot the time evolutions of baryon mass with thick lines.
  Grid numbers are $64^3$ for A and C and $128^3$ for B and D.
  }
\label{pic:f39}
\end{center}
\end{figure}

Next test is an evolution of rigidly rotating neutronstar in equilibrium state.
Initial parameters are the central rest density $\rho_{\rm c}=10^{-3}$ and the central angular velocity $\Omega_{\rm c}=10^{-2}$.
We use the polytropic EOS $P=K\rho^\Gamma$ where $K=10$ and $\Gamma=5/3$ and assume rigid rotation.
With these parameters, the central lapse is $\alpha=0.701$ and the ADM/baryon masses are $M_{\rm ADM}=1.49$/$M_{\rm bar}=1.55$.
Outer boundary is taken at $|x,y,z|=L_{\rm out}=17$ and the equatorial radius of the NS is $\sim13.1$.
In Fig.\ref{pic:f39}, we display 4 models with two different numerical resolutions $\Delta x_{\rm min}=L_{\rm out}/32$ (models A and C), $\Delta x_{\rm min}=L_{\rm out}/64$ (models B and D).
Models C and D are evolved with fixed matter distribution and only metrics are evolved, while, in models A and B, both matters and metrics are evolved.
$Upper$ and $lower$ panels display time evolutions of deviations of the central lapse and the ADM mass from their initial values, respectively.
In the $lower$ panel, we also display the time evolutions of baryon mass for models A and B with thick lines.
From this figure, we see that models C and D keep their initial configurations within 1$\%$,
while fully evolved models A and B show gradual decrease(increase) of ADM mass(central lapse).
In this test, our treatment of the low density region outside of the NS is like this.
We set the floor density value as $\rho_{\rm floor}=10^{-9}\times\rho_{\rm c,max}$ and, in every time step, for
all cells whose density is smaller than $10\times\rho_{\rm floor}$, we assume them as vacuum.
Then their density and velocity are reset to $\rho_{\rm floor}$ and 0, respectively.
We consider that this treatment is too simple and, e.g., the conservation of baryon mass is violated as seen
in thick lines.
However, in the context of CCSNe, we currently do not have to treat vacuum space and, additionally,
models A and B show numerical convergence with respect to grid resolution,
we consider our GR code works with sufficient level for our aims.

\subsubsection{Box Refinement and Numerical Convergence}
\label{sec:Box Refinement}
In the end, we mention about the influence of AMR refinement procedures during the collapse
and about the numerical convergence.
In GRMHD models, we refine the AMR blocks as the central density grows to save the computational time.
Typically the total number of AMR blocks increases from $\sim\mathcal O(10^2)$ at initial to $\sim\mathcal O(10^{3-4})$ at core bounce.
We check whether this refinement procedure breaks such as the Hamiltonian constraint or the ADM mass conservation.

\begin{figure}[htbp]
\begin{center}
 \includegraphics[angle=-90.,width=50mm]{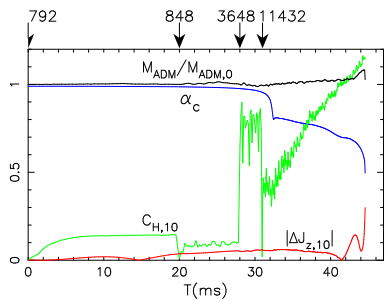}
  \caption{The Hamiltonian constraint $C_\mathcal H(\times10)$ ($green$), the central lapse $\alpha$ ($blue$),
  the deviation of the total angular momentum $|\Delta J_z|(\times10)$ ($red$) and the ADM mass normalized by its initial value ($black$)
  are plotted against time.
  Refinement procedures and enforcing the Hamiltonian constraint are done four times at when arrows point in the top.
  Initial ADM mass is $4.337M_\odot$ and the total number of AMR blocks is increased as 792, 848, 3648, and 11432.}
\label{pic:f40}
\end{center}
\end{figure}

In Fig.\ref{pic:f40}, we plot the central lapse $\alpha$, the error of the Hamiltonian constraint $C_{\mathcal H}(\times10)$ (see, Eq. (\ref{eq:L1normHamiltonianCon})) and the ADM mass normalized by its initial value
of magnetorotational collapse of a 100$M_\odot$ star with zero metallicity \citep{Umeda08}.
Such a highly massive star is considered to form BH and a good objection to test our GRMHD code.
Initial conditions and the adopted EOS are the same as those used in model "GB12R020Sf" (see, Sec.\ref{sec:collapse}) and the minimum cell width is $\Delta x=600$m.
We also plot the deviation of the total angular momentum along the rotational axis from its initial value $|\Delta J_z|=|(J_z-J_{z,0})/J_{z,0}|$, here $J_z$ is
defined by Eq. (\ref{eq:Jtot}) and $J_{z,0}$ is the value at $t=0$ ms.
In this test, the total number of AMR blocks increases from 792 to 11432 through three times refinement procedures until the time of core bounce.
The ADM mass and the total angular momentum should be almost constant until the core bounce and for a short while after it.
This is because it takes $5000$km/$c\sim10$ms till the gravitational radiation, emitted at core bounce around the center, reaches the outer boundary 5000km.
From this test, we find that the total angular momentum are conserved within $\sim1\%$
until the time of core bounce.
In addition, the time evolution of the central lapse is smooth and no influence of the refinement procedures is seen.
As for the Hamiltonian constraint, $C_{\mathcal H}$ is kept within several percentage until the time of core bounce except $t=28\sim31$ ms
and after a short while from the core bounce.
We consider that the sudden increase of $C_{\mathcal H}$ during $t=28\sim31$ ms is due to the low resolution and thus the next refinement at $t\sim31$ ms suppresses the error.
After the core bounce at $t\sim32$ ms, $C_{\mathcal H}$ increases gradually due to the collapse but not rapidly and finally the calculation is crushed.
We found the apparent horizon \citep{Shibata97} is formed at the end of the calculation and we thus
consider the BH is born.

\begin{figure}[htbp]
\begin{center}
 \includegraphics[angle=-90.,width=60mm]{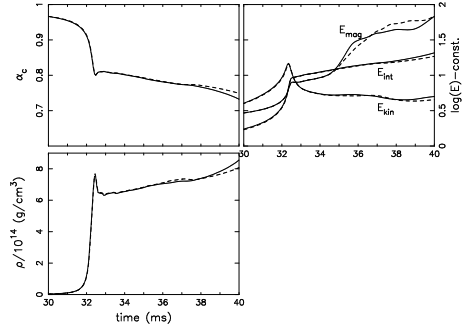}
  \caption{Time evolutions of the central lapse ($top$-$left$), various total energies ($top$-$right$)
  and the maximum density ($bottom$-$left$) with different numerical resolutions.
  The minimum cell widths are set as $\Delta x=600$m for $solid$ lines and as $\Delta x=480$m for $dashed$ lines.
  The internal, kinetic and magnetic energies are normalized by $10^{52}$, $10^{51}$ and $10^{48}$, respectively.}
\label{pic:GRconverge}
\end{center}
\end{figure}

In Fig. \ref{pic:GRconverge}, we display magnified views of Fig. \ref{pic:f40} around the time of core bounce
with two different numerical resolutions.
We put the outer boundary at 5000 km for lower resolution model and 4000 km for higher one and
maintain the AMR structure almost the same in both models.
Then the cell widths of the higher resolution model are 0.8 times smaller than those of lower resolution model.
In Fig. \ref{pic:GRconverge}, $dashed$ and $solid$ lines correspond to higher (minimum cell width is $\Delta x=480$m) and lower ($\Delta x=600$m) resolution model, respectively.
From this figure, we see that the good numerical convergence is achieved.

\section{Collapse of a $\bf 15M_\odot$ Star}
\label{sec:collapse}
In this section, we calculate magneto rotational collapse of a $15{\rm M}_\odot$ progenitor star as our practical test using GR/NMHD codes and check their abilities.
We follow the gravitational collapse with varying the initial magnetic field, the stiffness of the EOS and the initial central angular velocity.
We first describe our initial setups and then show results after subsection \ref{sec:Results}.

\subsection{Equation of State}
\label{EOS}
We adopt a parametric type EOS \citep[e.g.,][]{Takahara88} in this study.
It is divided into two parts, "cold" and "thermal" part. The cold part is expressed as
\begin{eqnarray}
\label{eq:Pcold}
P_{\rm c}=
\left\{\begin{array}{cc}
K_1\rho^{\Gamma_1} &  0\le\rho<\rho_1\\
K_2\rho^{\Gamma_2} &  \rho_1\le\rho<\rho_2\\
K_3\rho^{\Gamma_3} &  \rho_2\le\rho<\rho_3\\
K_4\rho^{\Gamma_4} &  \rho_3\le\rho<\rho_4
\end{array} \right.
\end{eqnarray}
where, we fix $K_1=2.78\times10^{14}$ in cgs units and $K_{2,3,4}$ are determined from the continuity of $P_c$ at $\rho_{1,2,3}$.
$\rho_3$ corresponds to the nuclear density $\rho_{\rm nuc}$.
Polytropic indexes represent physical processes occurring  during core collapse such as the electron-capture, onset of the neutrino-trap and the nuclear 
repulsive force.
We adopt two types of polytropic indexes and, for convenience, we call "Soft" and "Stiff" EOS as summarized in Table \ref{tb:Pcold}.
Stiff EOS corresponds to the models reported in \citet{Mikami08} and Soft EOS adopts smaller polytropic index compared to the Stiff case in the range of $\rho_1\le\rho<\rho_2$.
That density region corresponds to the electron capture regime.
\begin{deluxetable}{cccccccc}
\tablecolumns{8} 
\tablewidth{0pc} 
\tablecaption{Parameters of cold part EOS} 
\tablehead{ 
\colhead{}    & \multicolumn{3}{c}{Soft EOS} &   \colhead{}   & 
\multicolumn{3}{c}{Stiff EOS} \\ 
\cline{2-4} \cline{6-8} \\ 
\colhead{i} & \colhead{$\rho_i ({\rm g\ \rm cm^{-3}})$}   & \colhead{$K_i$}    & \colhead{$\Gamma_i$} & 
& \colhead{$\rho_i ({\rm g\ \rm cm^{-3}})$}   &\colhead{$K_i$}    & \colhead{$\Gamma_i$}}
\startdata 
1 & $4\times10^9$ & $2.78\times10^{14}$ & 4/3 & & $4\times10^9$ & $2.78\times10^{14}$ & 4/3 \\ 
2 & $1\times10^{12}$ & $7.25\times10^{14}$ & 1.29 & & $1\times10^{12}$ &$4.66\times10^{14}$ & 1.31 \\ 
3 & $2\times10^{14}$ & $3.17\times10^{14}$ & 1.32 & & $2.8\times10^{14}$ & $2.45\times10^{14}$ & 4/3 \\ 
4 & $1\times10^{16}$ & $4.22\times10^{-3}$ & 2.5 & & $1\times10^{16}$ &$3.42\times10^{-3}$ & 2.5 \\
\enddata 
\label{tb:Pcold}
\end{deluxetable}

The thermal part is expressed by
\begin{equation}
\label{eq:Ptherm}
P_{\rm t}=(\Gamma_{\rm t}-1)\rho\varepsilon_{\rm t}
\end{equation}
and we fix the index of the thermal part as $\Gamma_{\rm t}=1.3$ in this study.
Then the total pressure $p$ and internal energy $\varepsilon$ contributed from both thermal and cold part are written by
\begin{eqnarray}
\label{eq:Pcold+Ptherm}
p&=&P_{\rm c}+P_{\rm t} \\
\varepsilon&=&\varepsilon_{\rm t}+\varepsilon_{\rm c} \nonumber \\
&=&\varepsilon_{\rm t}+\int_0^\rho{\frac{P_{\rm c}(\rho')}{\rho'^2}d\rho'}
\end{eqnarray}

\subsection{Grid Setup}
\label{sec:grid}
In NMHD models, we did not change their initial AMR structures and fixed them.
On the other hand, we turn on the switch of AMR and refine AMR boxes in the vicinity of center in GRMHD models.
This is because, the time step $\Delta T$ to keep the Courant-Friedrichs-Lewy condition (CFL condition) is determined from the maximum
wave speed which usually becomes that of dynamical background and not hydrodynamical wave speed (i.e., the fast magneto sonic) in GRMHD models.
The wave speed of the dynamical background is nearly the speed of light $c$ and hardly change throughout the time evolution and also not depend on the
hydrodynamical properties such as maximum density.
We set $\Delta T$ as $0.3\Delta{\rm x}_{\rm min}/c$ in GRMHD models to keep the CFL condition where $\Delta{\rm x}_{\rm min}$ is the minimum cell width in 
the computational domain.
If we set maximum $L_{\rm AMR}$ as the maximum allowed one ($L_{\rm AMR,max}=8\ {\rm or}\ 9$ in this study) from the beginning of calculation, it takes too much time until 
corebounce since the time step depends solely on $\Delta{\rm x}_{\rm min}$.
Therefore in GRMHD models, we set maximum $L_{\rm AMR}$ as 5 at the beginning and then increment it as the collapse proceeds to save computational
time.
We define the criterion to increment the maximum AMR level (i.e., refine the AMR boxes in the vicinity of center) that every time the central density exceeds 
$10^{11},\ 10^{12},\ 10^{13}$g $\rm cm^{-3}$.
Influences of box refinements are described in Sec.  \ref{sec:Box Refinement}.

\begin{figure}[htpb]
\begin{center}
\includegraphics[width=40mm,angle=-90.]{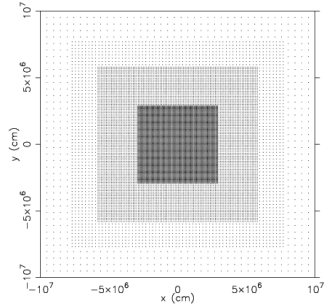}
\end{center}
\caption{Schematic figure of grid setup in standard resolution run.
Each point represents center of cell.
The central region of $|x,y,z|\la30$ and 60 km is covered by $L_{\rm AMR}=8$ and 7, respectively.
In high resolution run, the central region of $|x,y,z|\la30$, 60 and 120 km is covered by $L_{\rm AMR}=9$, 8 and 7, respectively.
}
\label{pic:AMRstructure.eps}
\end{figure}

The outer boundary is taken at 5000 km from the origin and $8\times8\times8$ AMR boxes with 0 refinement level ($L_{\rm AMR}=0$) cover the whole computational domain which is $(x,y,z)=[-5000,5000]$km.
We set the maximum allowed AMR refinement level to 8 in standard model and to 9 for high resolution model, thus the highest resolution is $\Delta {\rm x}
\sim 600$m in standard and $\Delta {\rm x}\sim 300$m in high resolution run.
In standard models, the central region of $(x,y,z)=[-60,60]$km is covered by $L_{\rm AMR}=7$ 
and $|x,y,z|\la30$km is covered by $L_{\rm AMR}=8$ (see, Fig.\ref{pic:AMRstructure.eps}).
On the other hand in high resolution run, the central regions of $|x,y,z|\la120$, 60, 30 km are covered by $L_{\rm AMR}=7$, 8, 9, respectively.
Since the most dynamical and active region is above the surface of central core ($r\ga20$km) and inside the prompt shock ($r\la200$km),
we increase resolutions as much as possible in this region.
Then the total numbers of AMR blocks become $\sim6000$ in standard run and $\sim42000$ in high resolution run.

\subsection{Initial Setup}
\label{sec:initialmodels}
Our progenitor is a $15M_\odot$ star with the metallicity Z=0.02 and the pre-collapse model is taken from \citet{Umeda08} which calculates stellar evolution 
with spherical symmetry.
We add rotation and magnetic field to the spherical progenitor model.
Since little is known about the rotational law and magnetic field configuration
in the central iron core and its surroundings at the pre-collapse stage, we assume the rotational law such as
\begin{eqnarray}
\label{eq:rotlawNMHD}
\Omega(\varpi)=\Omega_0\frac{{\varpi_0}^2}{{\varpi_0}^2+{\varpi}^2} \ \ \ \ \ \ \ \ \ \ \ \ \ \ \ &(\rm{for\ NMHD})& \\
\label{eq:rotlawGRMHD}
u^t u_\phi ={\varpi_0}^2(\Omega_0-\Omega) \ \ \ \ \ \ \ \ \ \ \ \ \ \ \ &(\rm{for\ GRMHD})&
\end{eqnarray}
Where $\varpi=\sqrt{x^2+y^2}$, $u_\phi=\sqrt{u_x^2+u_y^2}$ and $\varpi_0$ is the parameter which is set as $10^8$cm in this study.
These configurations are commonly used rotational law in which the core, within $\sim \varpi_0$, rotates rigidly with angular velocity $\Omega_0$ and differentially beyond that.
Both equations represent same rotational profile, since in the Newtonian limit of Eq. (\ref{eq:rotlawGRMHD}) becomes Eq. (\ref{eq:rotlawNMHD}) by replacing
$u^t\rightarrow 1$ and $u_\phi\rightarrow \gamma_{\phi\phi}v^\phi=\varpi^2\Omega$, where $\gamma_{\phi\phi}$ is a component of cylindrical flat metric.

As for the initial magnetic field configuration, to ensure divergence-free constraint, we adopt the following form of vector potential \citep[see, e.g.,][]{Takiwaki09}.
\begin{eqnarray}
\label{eq:vectorpotential}
\left(A_r,A_\theta,A_\phi\right)=\left(0,0,\frac{B_0}{2}\frac{R_0^3}{r^3+R_0^3}\varpi\right)
\end{eqnarray}
Where $B_0\ \&\ R_0$ are the parameters and $r=\sqrt{x^2+y^2+z^2}$.
$R_0$ is fixed as $10^8$cm in  this study.
This vector potential represents almost uniform magnetic field within $\sim R_0$ and dipole-like magnetic field configuration beyond $\sim R_0$
with the central magnetic field strength $\sim B_0$.
We calculated several models with various initial magnetic field strength $B_0$, in Eq. (\ref{eq:vectorpotential}), and central angular velocity $\Omega _0$,
in Eq. (\ref{eq:rotlawNMHD}-\ref{eq:rotlawGRMHD}).
We also add random perturbation  to trigger asymmetric motion in the form of velocity with 5\% amplitude when the maximum density exceeds $10^{13}$g $\rm cm^{-3}$.
Such perturbation is added in the central sphere of radius $10^8$cm via
\begin{equation}
\label{eq:perturbation}
{\bf u}={\bf u}\left(1+\sigma\frac{1}{1+(r/R_0)^2}\right)
\end{equation}
In this equation, $R_0$ is a parameter fixed with $R_0=10^8$cm and $-0.05\le\sigma\le0.05$ is a random number.
Here, we have to comment about the momentum constraint, Eq.(\ref{eq:MomentumCon}), in GRMHD model,
since the momentum constraint would be violated by adding the perturbation.
In our practical simulations, the momentum constraint $\mathcal{M}_i$ is not kept strictly 0 during time evolution due to the numerical error.
In our GRMHD model "GB12R020Sf", for instance, $C_{\mathcal{M}_{x}}$ defined by Eq.(\ref{eq:L1normMomentumCon}) is $C_{\mathcal{M}_{x}}\sim1.53\times10^{-2}$ before the perturbation is added.
\begin{eqnarray}
  \label{eq:L1normMomentumCon}
  C_{\mathcal{M}_{i}}=\frac{1}{M_{\rm bar}}\int\rho_\ast\mathcal{M}_idx^3
\end{eqnarray}
By adding the perturbation, this value increases to $C_{\mathcal{M}_{x}}\sim1.55\times10^{-2}$.
On the other hand, the error after core bounce is around $C_{\mathcal{M}_{x}}\sim4.5\times10^{-2}$ in
both models with and without the perturbation and we thus consider that the violation of the momentum constraint by the perturbation is negligibly small within the range of our numerical accuracy.
\begin{deluxetable}{cccccccccc}
\tabletypesize{\scriptsize}
\tablecolumns{10}
\tablewidth{0pc} 
\tablecaption{Initial parameters and adopted EOS}
\tablehead{ 
\colhead{model name} & \colhead{$B_0$(G)}  & \colhead{$\Omega_0$(rad/s)}  &$\beta_{\rm mag}$ &$\beta_{\rm rot}(\%)$ & 
$E_{\rm grv}$(ergs) & $\alpha_{\rm c}$ & $M_{\rm bar}/M_\odot$ & $M_{\rm ADM}/M_\odot$ &  \colhead{EOS} }
\startdata
\cutinhead{NMHD}
NB00R02St & 0E0 & 2 & 0E0 & 0.135 & -6.46E51 & \nodata & 2.0483 &\nodata & Stiff  \\ 
NB12R02St & 4.8E12 & 2 & 2.95E-4 & 0.135 & -6.46E51 &\nodata &2.0483 & \nodata& Stiff  \\ 
\cline{1-10}
NB00R02Sf & 0E0 & 2 & 0E0 & 0.135 & -6.46E51&\nodata &2.0483& \nodata& Soft  \\ 
NB12R02Sf & 4.8E12 & 2 & 2.95E-4 & 0.135 &-6.46E51 &\nodata & 2.0483&\nodata & Soft  \\ 
NB12R06Sf & 4.8E12 & 6 & 2.95E-4 & 1.22 & -6.46E51&\nodata &2.0483 &\nodata & Soft  \\ 
NB09R02Sf\tablenotemark{1}& 1E9 & 2 & 1.15E-11 & 0.135 &-6.46E51 &\nodata&2.0483&\nodata & Soft  \\ 
\cline{1-10}
\cutinhead{GRMHD}
GB00R02Sf & 0E0 & 2 & 0E0 & 0.132 & -6.62E51& 0.994 & 2.0688 & 2.0673 & Soft  \\ 
GB12R02Sf & 4.8E12 & 2 & 2.00E-4 & 0.132 &-6.62E51 & 0.994 &2.0688 &2.0673 & Soft  \\ 
\enddata 
\tablenotetext{1}{This model is calculated with high resolution.}
\tablecomments{Each column denotes model name and initial parameters. From left; model name; magnetic field $B_0$; central angular velocity
$\Omega_0$; $\beta_{\rm mag}=E_{\rm mag}/|E_{\rm grv}|$; $\beta_{\rm rot}=T_{\rm rot}/|E_{\rm grv}|$; gravitational energy $E_{\rm grv}$;
central lapse; baryon rest mass $M_{\rm bar}/M_\odot$; ADM mass $M_{\rm ADM}/M_\odot$; and adopted EOS.}
\label{tb:model}
\end{deluxetable}

Model names and adopted parameters are summarized in Table \ref{tb:model}.
The first character "N" and "G" of model names indicate that the calculation is done by NMHD and GRMHD code, respectively.
Numbers after "B" and "R" represent the exponent of magnetic field strength and the central angular velocity, respectively.
The last characters St/Sf represents Stiff/Soft EOS adopted.
As for the initial magnetic field strength, we adopted 0, $10^{9}$ and $4.8\times10^{12}$G so that we can easily check the roles of magnetic field.

The initial magnetic field is first amplified mainly through the compression and the rotational winding effects during collapse from previous many studies \citep[e.g.,][]{Shibata06,Burrows07,Takiwaki09}.
For instance, the compression mechanism amplifies the magnetic field about $\sim10^3$ times
\citep{Burrows07} and then $B_0\sim \mathcal{O}(10^{12})$G is amplified to $\sim \mathcal{O}(10^{15-16})$G which is equivalent strength to that of magnetar \citep{Duncan92}.
\citet{Heger05} studied stellar evolution with including magnetic field and rotation and they reported the
strength of magnetic field is of the order of $\sim10(9)$G or weaker and the poloidal magnetic field is much weaker, approximately $10^{-4}$, than the toroidal component.
Thus our initial condition with purely poloidal and extremely strong magnetic field might be unrealistic one,
however we employ such initial condition to see the effects of magnetic field easily and also to compare our
results with other previous studies.
In addition to the very strong initial magnetic field models, we calculated one model (NB09R02Sf) with initially weak magnetic field $B_0=10^{9}$G in a high resolution run.
Since, there are possibly several non linear magnetic field amplification mechanisms in the vicinity of PNS
such as MRI or dynamo mechanism which are intrinsically 3D phenomena,
we examine how the initially weak magnetic field is amplified through this model.

As for the central angular velocity $\Omega_0$ we adopt 2 or 6 (rad/s).
In \citet{Hirschi04}, they calculated the evolutions of various rotating stars with changing initial stellar mass and metallicity.
Their results are that $25M_\odot$ star with solar metallicity ($Z=0.02$) has $\Omega_0\sim 1.0\ {\rm s}^{-1}$ at the end of silicon burning stage.
They also reported that lower initial metallicity raises the final angular velocity due to lower mass loss rate.
\citet{Yoon05} did similar works to \citet{Hirschi04}, though they included magnetic effects.
Their results showed magnetic torques lower the local specific angular momentum approximately 
one order of magnitude compared to non-magnetic field cases.
However these two works are done by one dimensional calculations and are not still conclusive results.
Thus, even though our adopted parameters are comparatively faster than the theoretical works those values may be reasonable.

\subsubsection{Initial Setup for GRMHD}
In GRMHD calculations, we have additional setups to be done which are constraining the Hamiltonian and momentum equations
(\ref{eq:HamiltonianCon}) and (\ref{eq:MomentumCon}).
To solve Eqs. (\ref{eq:HamiltonianCon}-\ref{eq:MomentumCon}), we assume conformally flat metric for initial condition,
in which $\tilde\gamma_{ij}=\delta_{ij}$ (therefore, $F_i=0$) and $K=0$.
Then, following \citep{Shibata02}, constraint equations become the following 4 equations to obtain rest of BSSN variables $\phi$ and $\tilde A_{ij}$ and
gauge variables $\alpha$ and $\beta_i$.
\begin{eqnarray}
\label{eq:initialHamiltonian}
\Delta_{\rm flat}e^\phi=-2\pi S_0e^{-\phi}-\frac{1}{8}\delta^{ik}\delta^{jl} \tilde A_{ij}\tilde A_{kl}e^{5\phi}\\
\label{eq:initialMS}
\Delta_{\rm flat}(\alpha e^\phi)=2\pi \alpha e^{5\phi}(S_0 e^{-6\phi}+2\gamma^{ij}S_{ij})+\frac{7\alpha e^{5\phi}}{8}\delta^{ik}\delta^{jl} \tilde A_{ij}\tilde A_{kl}\\
\label{eq:initialAMD}
\delta_{ij}\Delta_{\rm flat}\beta^j+\frac{1}{3}{\beta^k}_{,ki}-2\tilde A_{ik}\delta^{kj}\left(\alpha_{,j}-\frac{6\alpha}{e^\phi}{e^\phi}_{,j}\right)
=16\pi\alpha S_i e^{-6\phi}\\
\label{eq:initialAij}
\tilde A_{ij}=\frac{1}{2\alpha}\left(\delta_{ik}{\beta^k}_{,j}+\delta_{jk}{\beta^k}_{,i}-\frac{2}{3}\delta_{ij}{\beta^k}_{,k}\right)
\end{eqnarray}
Where $\Delta_{\rm flat}$ is the Laplacian in flat space.
We solve above 4 equations by iterative method with initial guess as $\phi=0$, $\tilde A_{ij}=0$, $\alpha=1$, $\beta^i=0$.
Then first, we evaluate conservative variables $\bf Q$ from given metrics and primitive variables
$\bf P$ from a progenitor model.
Second, we solve above each equation.
We iterate these two procedures until sufficient convergence is achieved.
After these procedures, the Hamiltonian constraint $C_\mathcal H$ at initial is kept below arbitrary chosen small number (in our calculations $10^{-8}$ is adopted).

\subsection{Results}
\label{sec:Results}
\subsubsection{Global Dynamics}
\label{sec:GlobalDynamics}
\begin{figure}[htbp]
  \begin{center}
    \includegraphics[width=35mm,angle=-90]{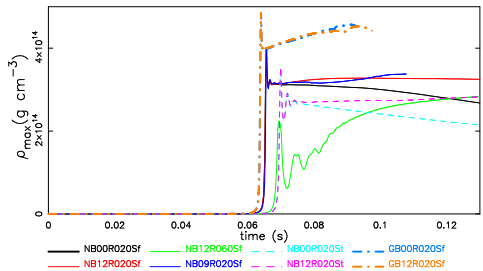}
    \caption{Time evolutions of the maximum density in different models.
    Model names are listed in the bottom.}
    \label{pic:f2}
  \end{center}
\end{figure}

\begin{figure}[htbp]
\begin{center}
\includegraphics[width=50mm,angle=-90]{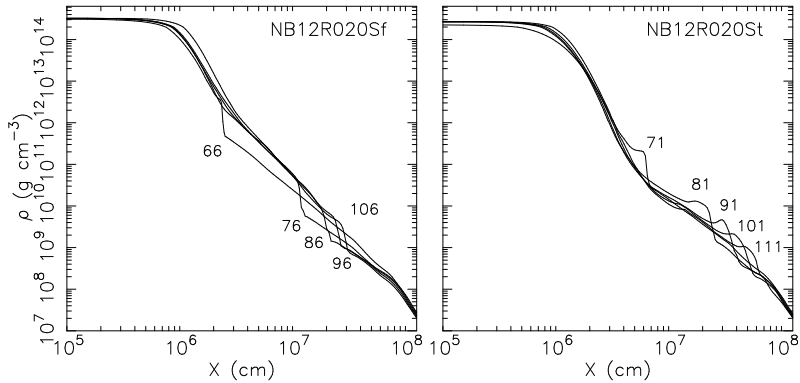}
\end{center}
\caption{Density profiles along the $x$ axis in model NB12R020Sf($left$) and NB12R020St($right$).
Numbers denote time in ms.}
\label{pic:RhoX}
\end{figure}

We first show the time evolutions of the maximum density $\rho_{\rm max}$ in Fig.\ref{pic:f2}, which tells us rough overview of how the core collapse proceeds.
In this study our calculations are done in full three-dimension and the central density does not always become the maximum one, thus we do not
use central value.
However all of our models show that the maximum density exists nearly the center.
In Fig.\ref{pic:f2}, each line corresponds to different model and the model names are shown in the bottom part.
From this figure, we can find $\rho_{\rm max}$ is increased $\sim$ 30\% in GRMHD models when compared with corresponding
NMHD models (e.g., "GB12R020Sf" vs "NB12R020Sf").
We however do not see any rapid increase of $\rho_{\rm max}$ which possibly indicates collapse toward BH formation.
Another feature is that the central density gradually increases after core bounce when the magnetic field exists, on the other hand
the central density decreases in non-magnetized models (e.g., $black$ vs $red$ solid curves).
This difference means that the magnetic torque works strongly and the maximum density increases due to the angular momentum transfer.
In rapidly rotating model "NB12R060Sf", the maximum density at core bounce marginally exceeds the nuclear density
($\rho_{\rm nuc}=2\times10^{14}$g $\rm cm^{-3}$) and it is thus not rotational supported core bounce but due to the nuclear repulsive force which is the same as the rest of models.
In this model, $\rho_{\rm max}$ eventually relaxes to similar value to other models after several oscillations.
We also summarize physical properties at core bounce in Table \ref{tb:corebounce}.
\begin{deluxetable}{ccccc}
\tabletypesize{\scriptsize}
\tablecolumns{5}
\tablewidth{0pc} 
\tablecaption{Physical properties at core bounce}
\tablehead{ 
\colhead{model name} &  $\rho_{\rm max}/10^{14}$(g $\rm cm^{-3}$)  & $\beta_{\rm mag}$ &$\beta_{\rm rot}(\%)$ & $\alpha_{\rm c}$}
\startdata
\cutinhead{NMHD}
NB00R02St  & 3.54 &0 & 3.62 & -  \\ 
NB12R02St  & 3.55 & 6.17E-4 & 3.58 & - \\
\cline{1-5}
NB00R02Sf  & 3.95 & 0 & 3.00 & - \\
NB12R02Sf  & 3.96 & 5.88E-4 & 2.82 & - \\
NB12R06Sf  & 2.25 & 165E-1 & 13.9 & - \\
NB09R02Sf  & 3.98 & 2.88E-11 & 3.04 &- \\
\cutinhead{GRMHD}
GB00R02Sf & 4.87 & 0 & 2.25 & 0.839 \\
GB12R02Sf & 4.87 & 2.31E-4 & 2.20 & 0.839 \\
\enddata 
\tablecomments{From left; maximum density $\rho_{\rm max}$; $\beta_{\rm mag}=E_{\rm mag}/|E_{\rm grv}|$; $\beta_{\rm rot}=T_{\rm rot}/|E_{\rm grv}|$;
central lapse $\alpha_{\rm c}$ at core bounce.}
\label{tb:corebounce}
\end{deluxetable} 
Fig. \ref{pic:RhoX} displays time evolution of density profiles along the $x$ axis in model NB12R020Sf($left$)
and NB12R020St($right$).
Numbers beside each line denote time in ms.
Prompt shock which is formed at core bounce moves outward and then stays around $x\sim200$-300km
in model NB12R020Sf, on the other hand the shock moves further out in stiff EOS model NB12R020St.
Model NB12R020St adopts same EOS as that of \citet{Mikami08} and our result seen in the prompt shock
propagation agrees well with their results.

\begin{figure}[htpb]
  \begin{center}
    \includegraphics[width=50mm,angle=-90]{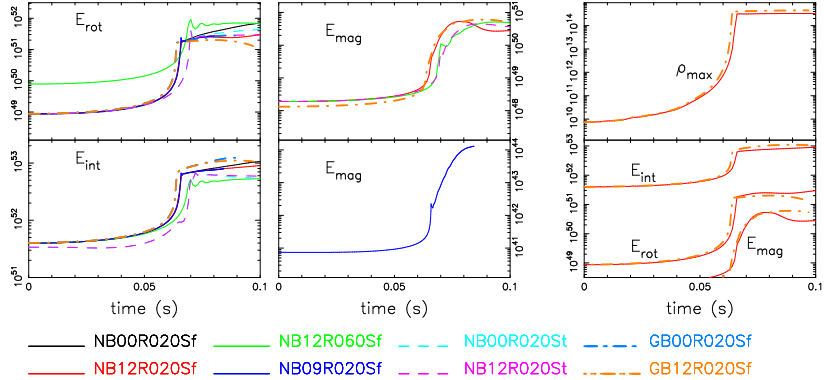}
    \caption{$Left$ four panels; Evolutions of rotational, internal and magnetic energy.
    Initially weak magnetic field model "NB09R020Sf" is plotted with different range in the lower middle panel.
    $Right$ two panels; Comparison between GRMHD(GB12R020Sf) and NMHD(NB12R020Sf) models.
    $Upper$ and $lower$ panels represent the maximum density and each energy component, respectively.
    }
    \label{pic:f3}
  \end{center}
\end{figure}

In Fig.\ref{pic:f3}, we display time evolutions of the rotational, internal and magnetic energies in $left$ four panels and we also plot comparison between
GRMHD(GB12R020Sf) and NMHD(NB12R020Sf) models in $right$ two panels.
The magnetic energy of model "NB09R020Sf" is too small compared to other strong field models and we display it separately with different range in
$bottom$-$right$ panel.
We see that the rotational and the internal energies are kept almost constant or gradual increase after core bounce, on the other hand the magnetic energy
increases rapidly, approximately $\sim2$ orders, after core bounce.
In strongly magnetized models ($upper$-$middle$ panel), the final magnetic energies saturate around $\sim5\times10^{50}$ergs.
When we compare GRMHD and NMHD models, shown in $right$ two panels, the evolution tracks look similar
except $E_{\rm rot}$ and $E_{\rm mag}$ after $t\ga80$ms.
We consider the difference is originated from the bipolar-outflow and will be described in Sec.
\ref{sec:FormationofOutflow}.

Finally we compare our results shown here with those reported by other groups.
\citet{Obergaulinger06} reported magnetorotational collapse in axisymmetry with various initial rotation, magnetic field and EOS and also with including general relativistic effects by replacing spherical Newtonian potential with "Tolman-Oppenheimer-Volkoff" potential.
They showed that maximum rest mass density is increased several 10 \% after core bounce when they
compare GR and Newtonian models and also that the magnetic field works to raise the maximum rest mass
density.
These features agree to ours since $\sim30$\% rise in the maximum density in our GRMHD model can be seen.
Additionally, if we compare our results with previous three-dimensional NMHD work reported by \citet{Mikami08},
similar time evolutions are obtained such as time evolution of various energy components and also the shock propagation
(as seen in Fig. \ref{pic:RhoX}).
Thus, we consider the results shown here are common and robust features.

\subsubsection{Formation of Outflow}
\label{sec:FormationofOutflow}
\begin{figure}[htbp]
\begin{center}
\includegraphics[width=90mm,angle=-90]{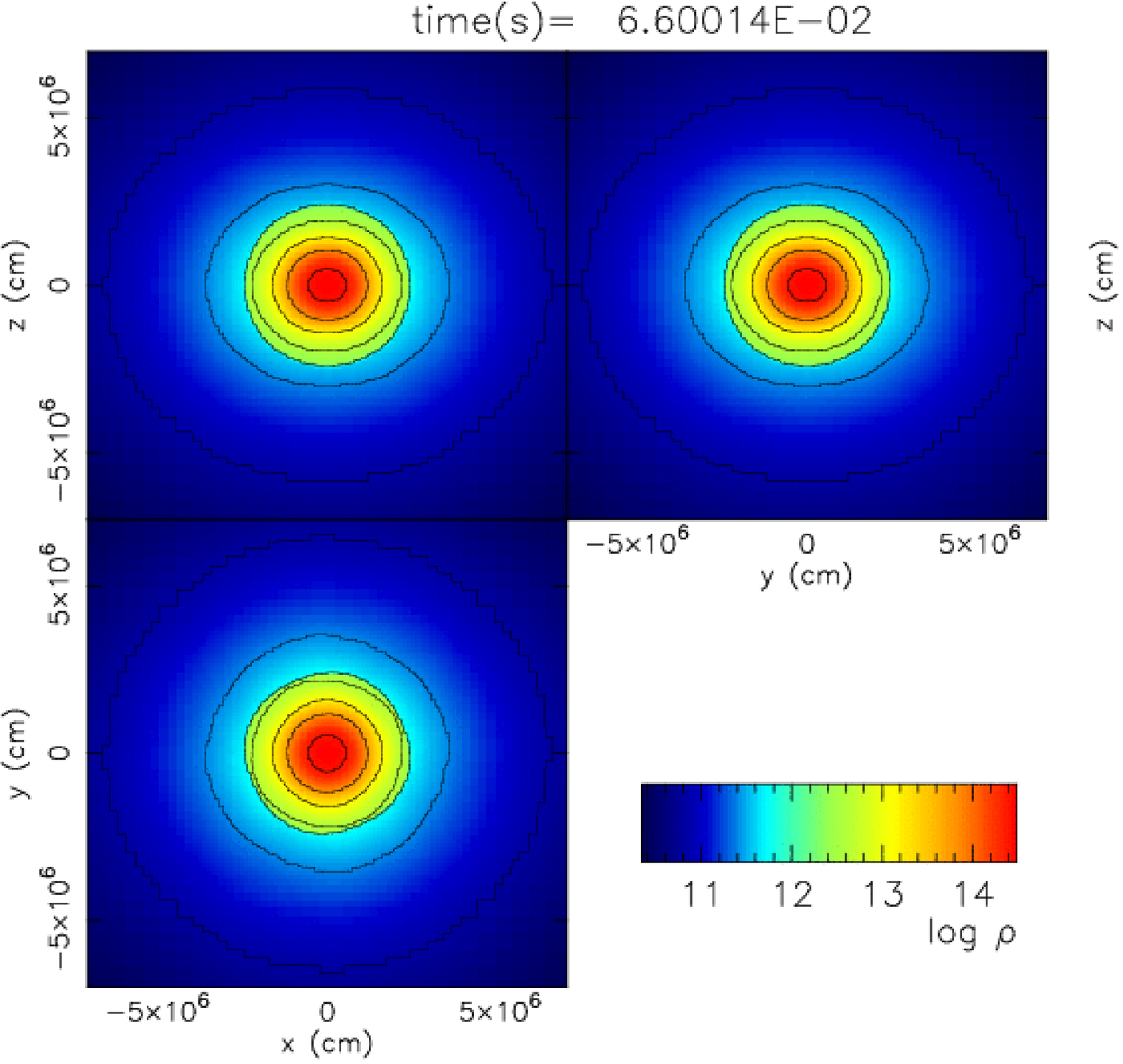}
\end{center}
\begin{center}
\includegraphics[width=90mm,angle=-90]{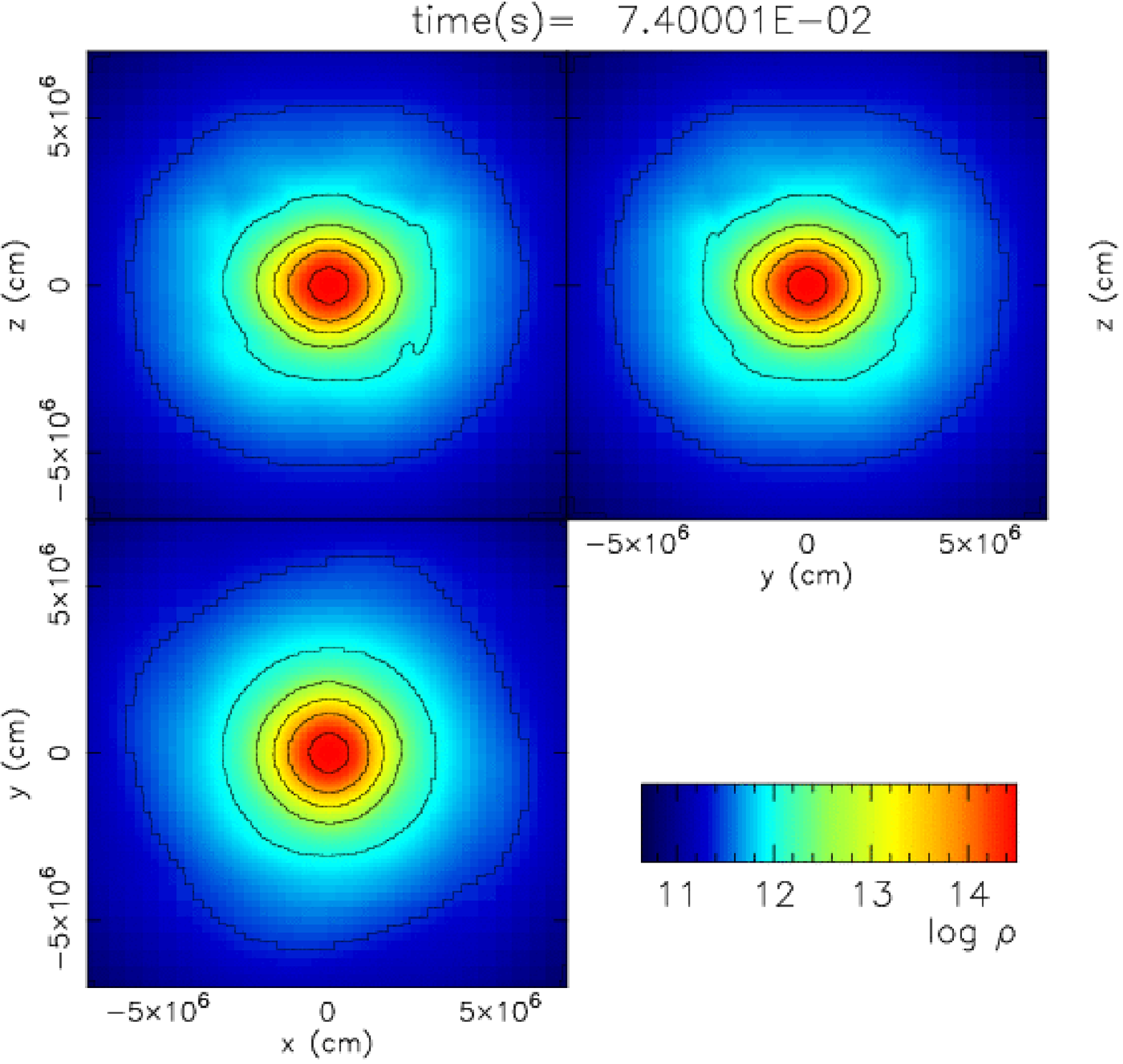}
\end{center}
\caption{Logarithmic scale of rest mass density in model NB12R020Sf at different time slices are depicted. Time slices are chosen at t=66ms (nearly the time of core bounce) and t=74ms.
}
\label{pic:rho5rho7}
\end{figure}

\begin{figure}[htbp]
\begin{center}
\includegraphics[width=90mm,angle=-90]{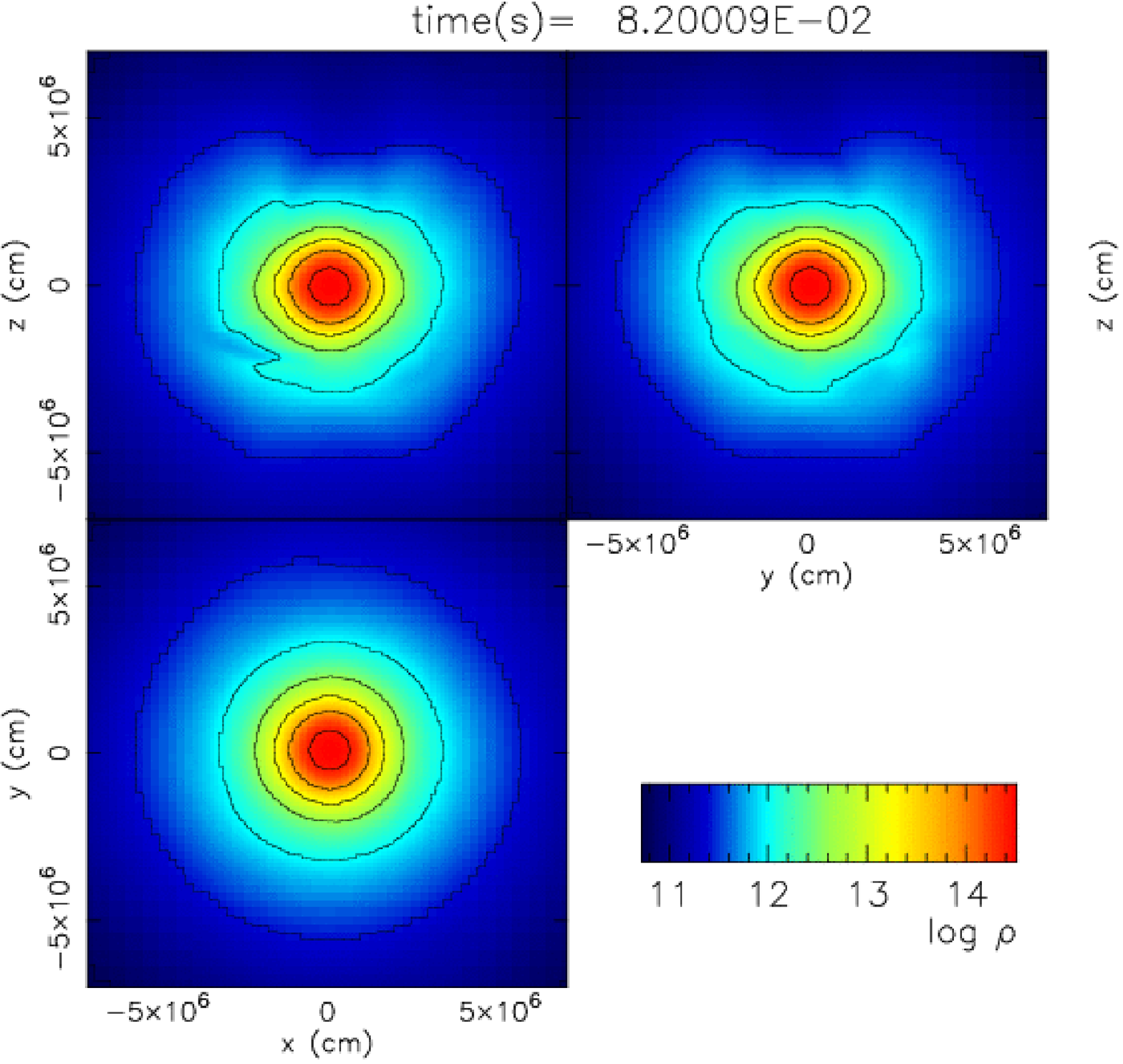}
\end{center}
\begin{center}
\includegraphics[width=90mm,angle=-90]{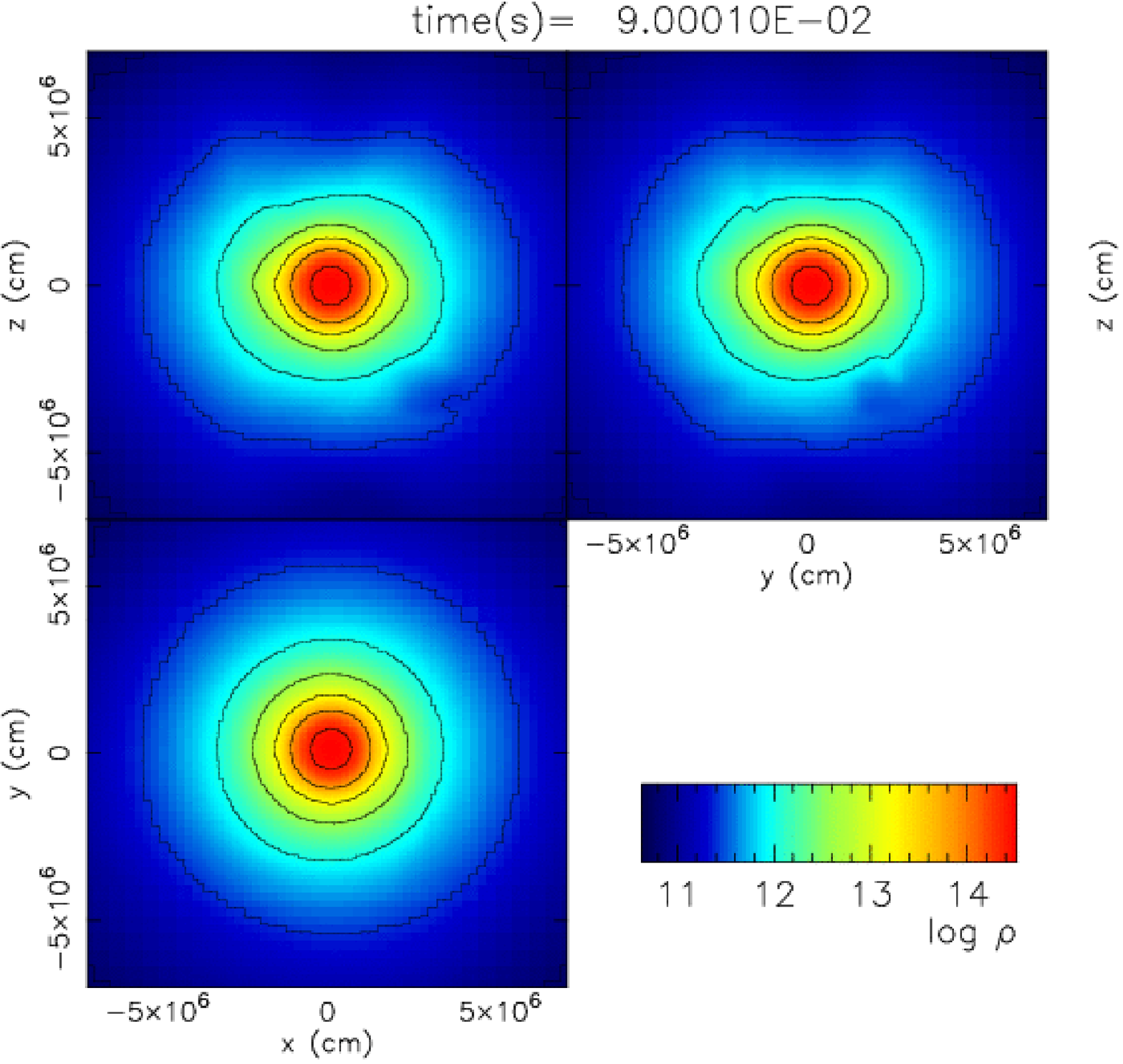}
\end{center}
\caption{Same as Fig. \ref{pic:rho5rho7} but for t=82ms and t=90ms.}
\label{pic:rho8rho9}
\end{figure}

\begin{figure}[htbp]
\begin{center}
\includegraphics[width=90mm,angle=-90]{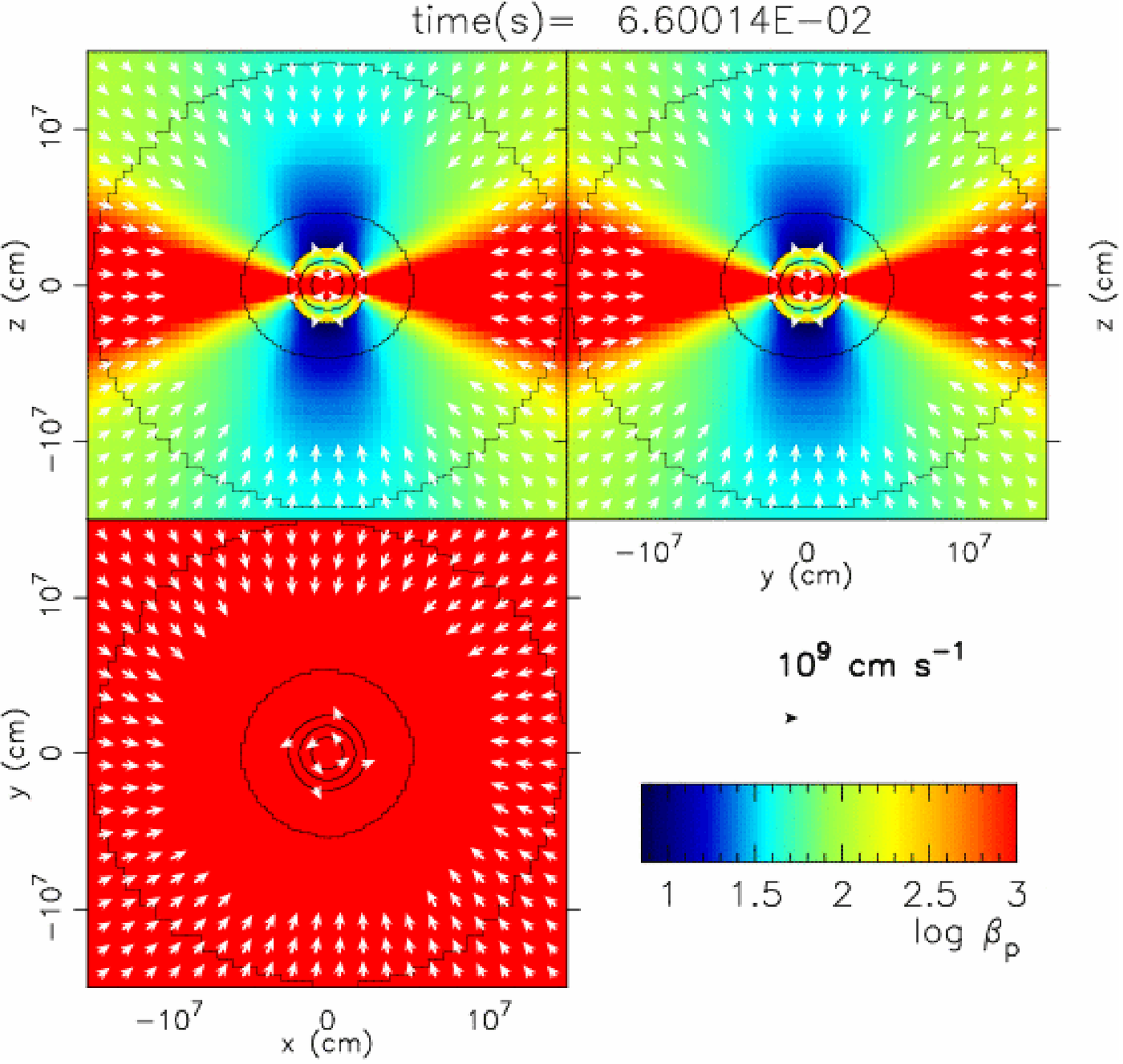}
\end{center}
\begin{center}
\includegraphics[width=90mm,angle=-90]{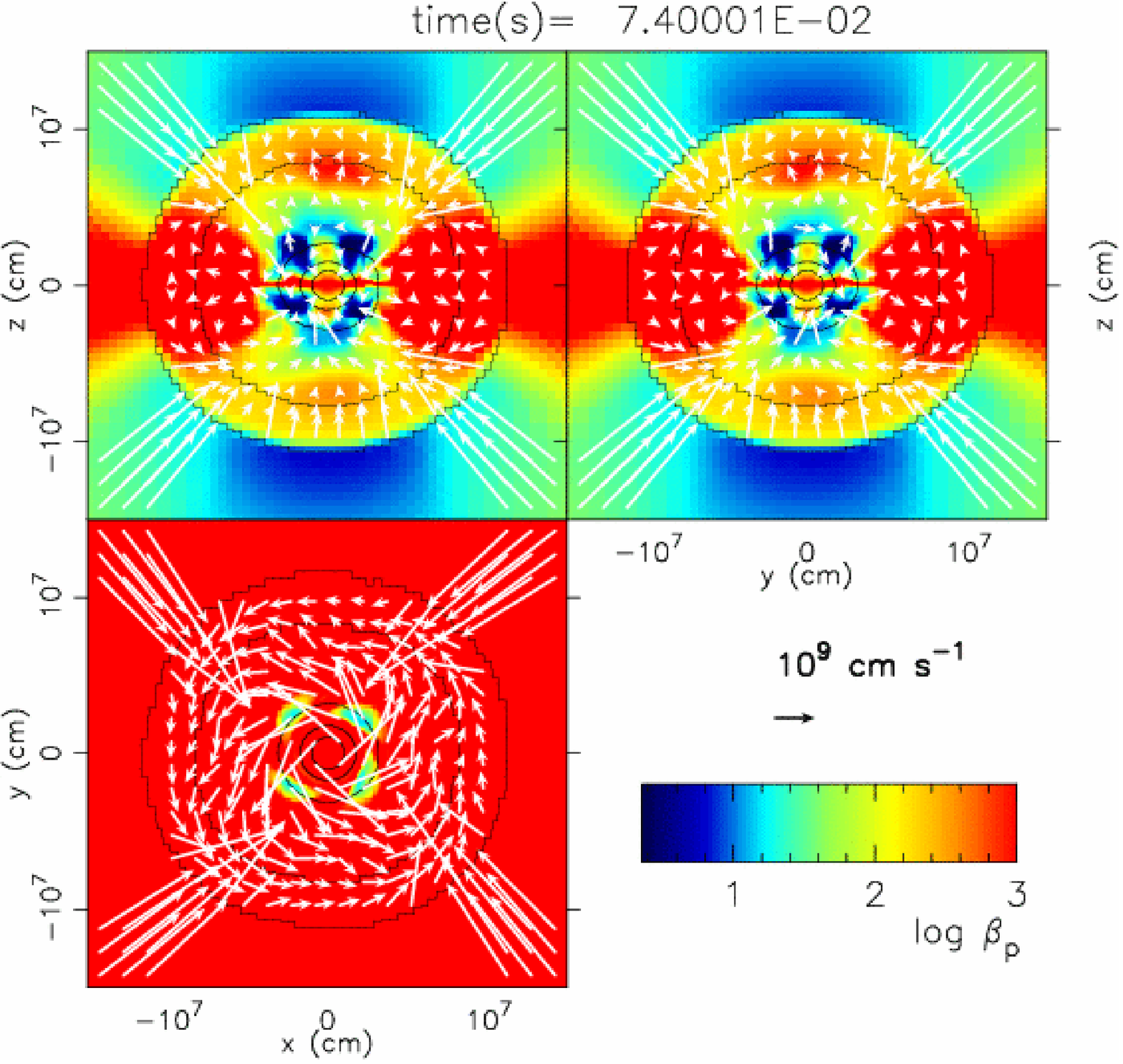}
\end{center}
\caption{Logarithmic scale of plasma beta in model NB12R020Sf at different time slices are depicted. Time slices are chosen at t=66ms (nearly the time of core bounce) and t=74ms.
Arrows represent flow velocity and black curves represent iso-density contour.
In the figure, we cut-off $\log\beta_{\rm p}$ and the flow velocity higher than 3 and $3\times10^9$ cm s$^{-1}$, respectively.
}
\label{pic:f5f7}
\end{figure}

\begin{figure}[htbp]
\begin{center}
\includegraphics[width=90mm,angle=-90]{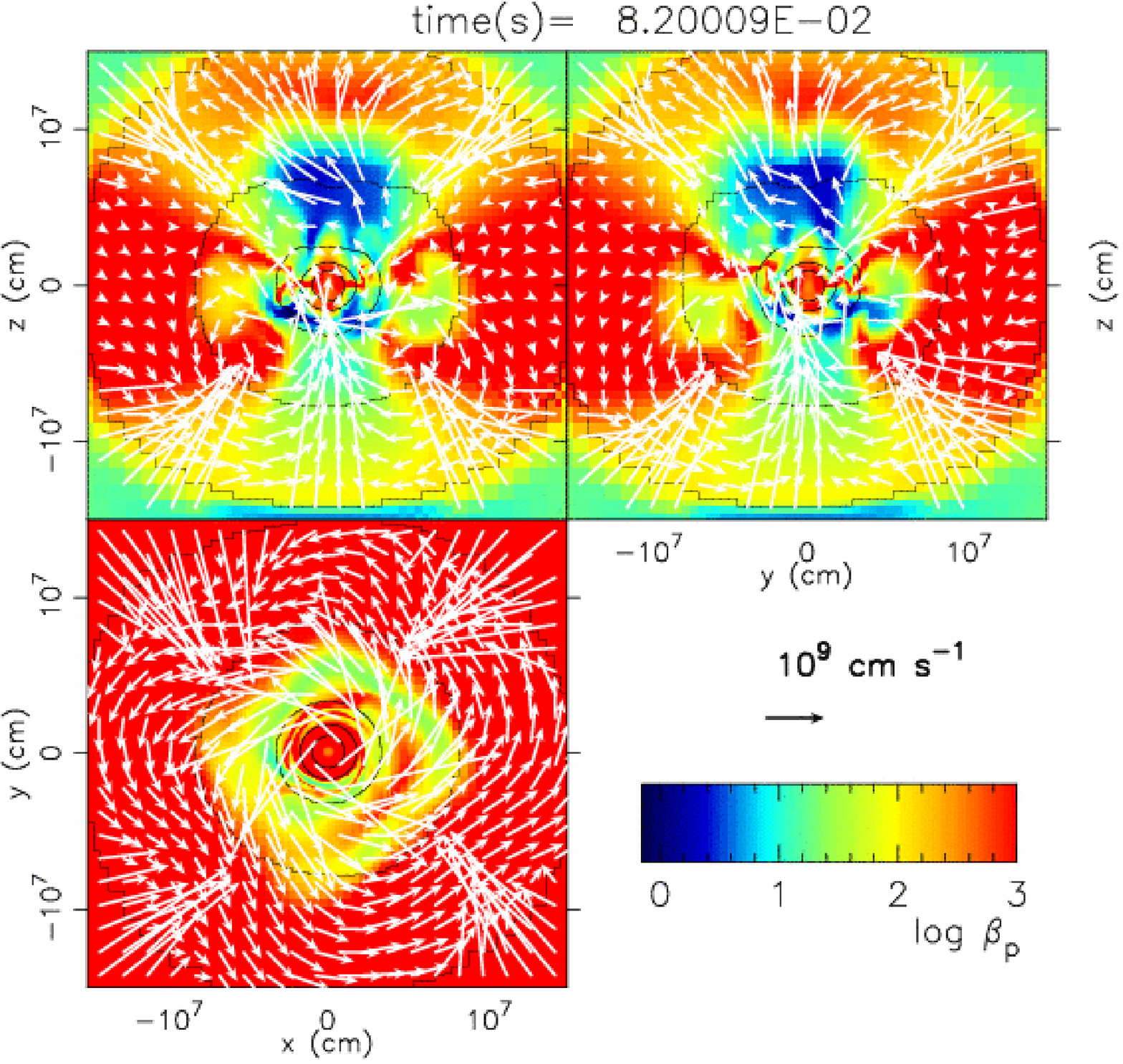}
\end{center}
\begin{center}
\includegraphics[width=90mm,angle=-90]{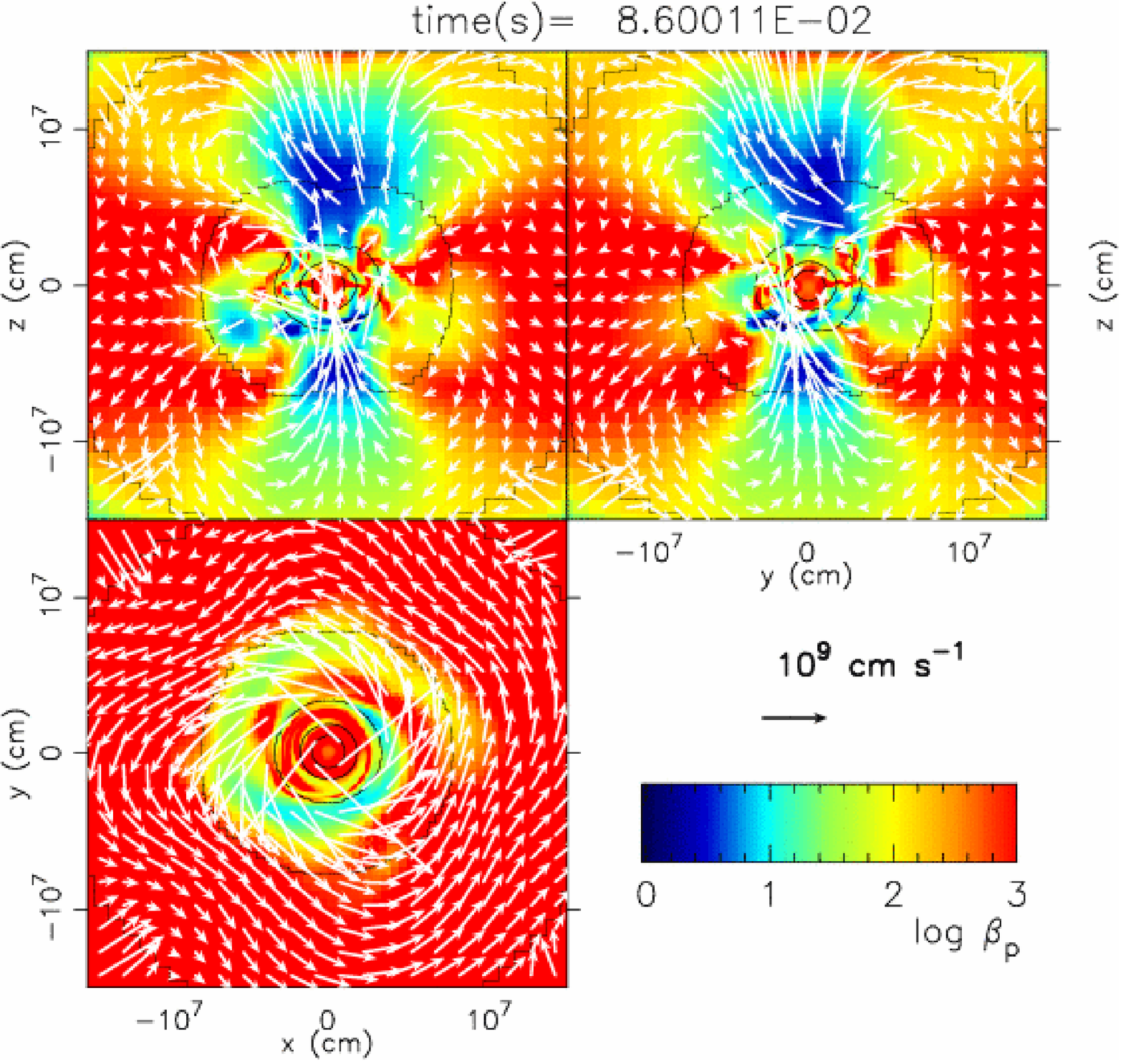}
\end{center}
\caption{Same as Fig. \ref{pic:f5f7} but for t=82ms and t=86ms.}
\label{pic:f8f9}
\end{figure}

\begin{figure}[htbp]
\begin{tabular}{cc}
\begin{minipage}{0.5\hsize}
\begin{flushright}
\includegraphics[width=50mm]{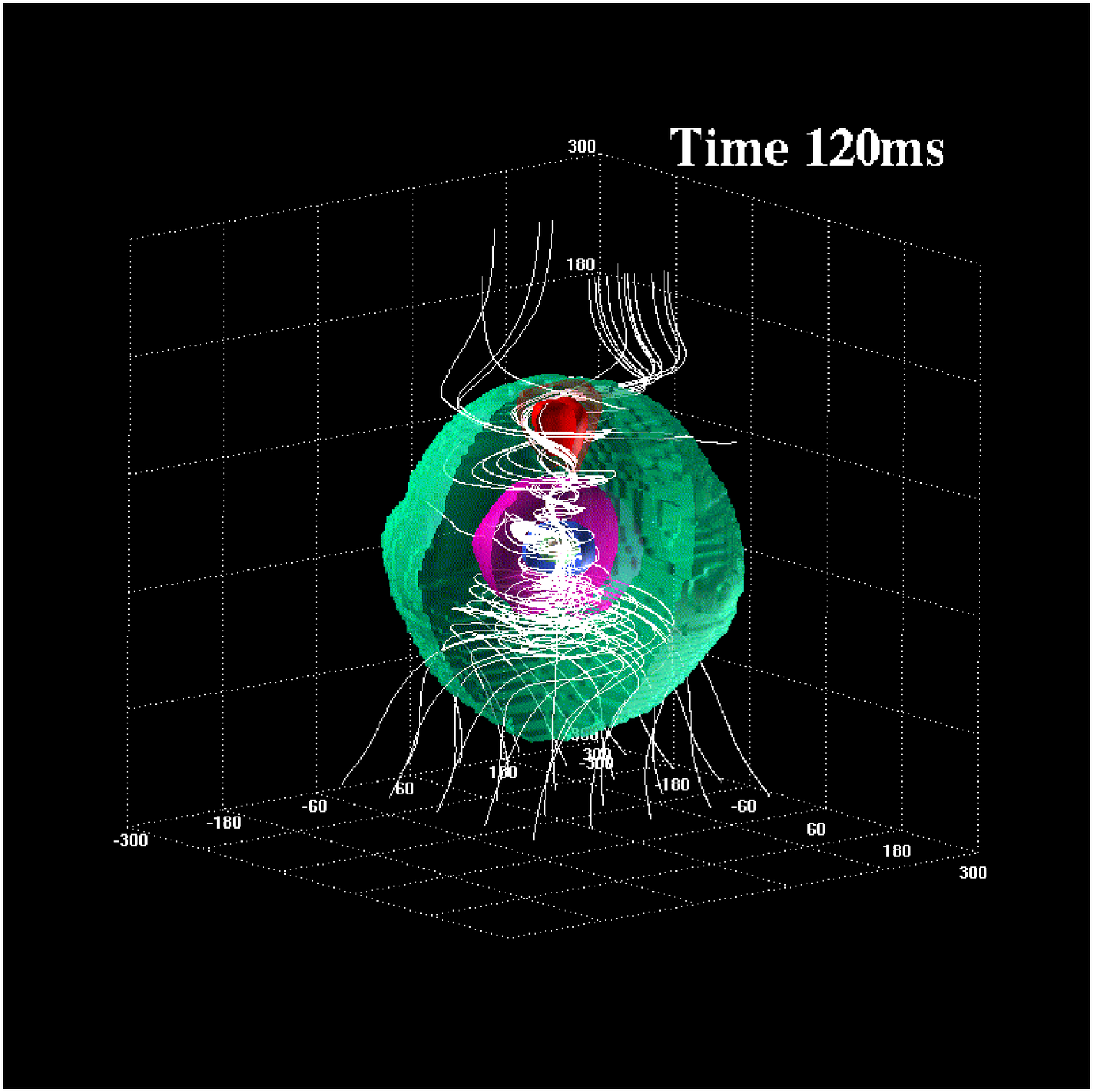}
\end{flushright}
\end{minipage}
\begin{minipage}{0.5\hsize}
\begin{flushleft}
\includegraphics[width=50mm]{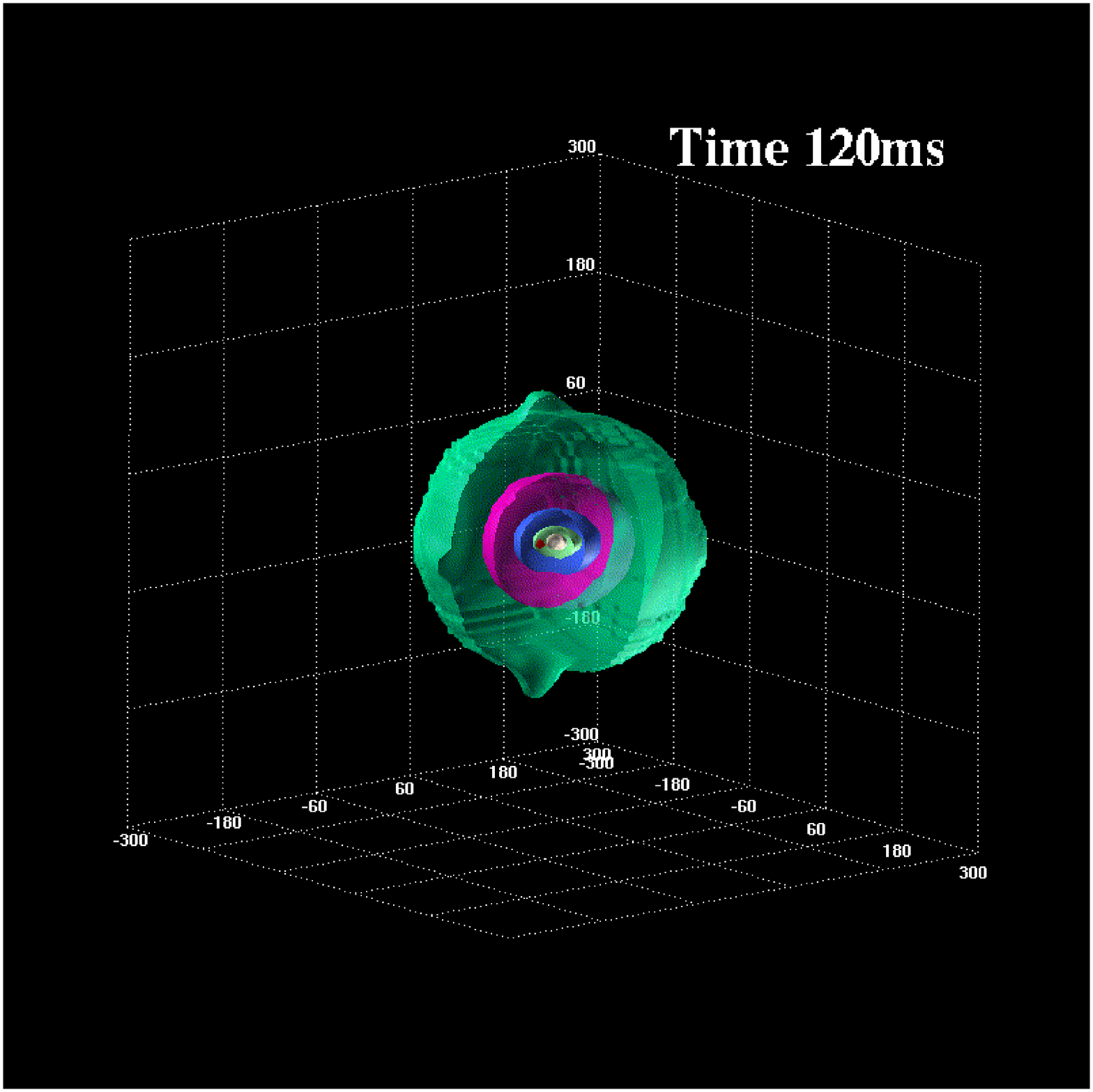}
\end{flushleft}
\end{minipage}
\end{tabular}
\begin{tabular}{cc}
\begin{minipage}{0.5\hsize}
\begin{flushright}
\includegraphics[width=50mm]{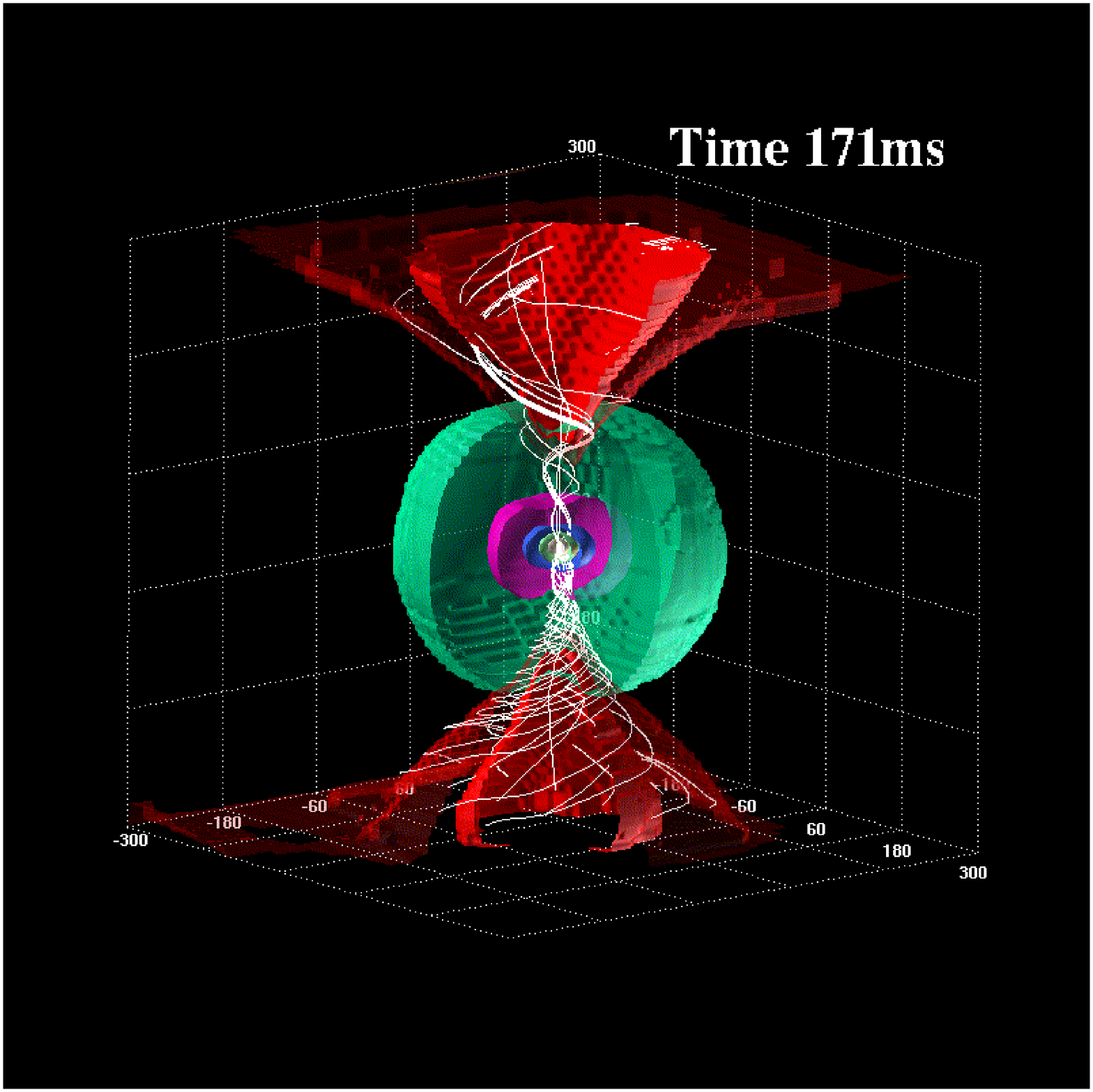}
\end{flushright}
\end{minipage}
\begin{minipage}{0.5\hsize}
\begin{flushleft}
\includegraphics[width=50mm]{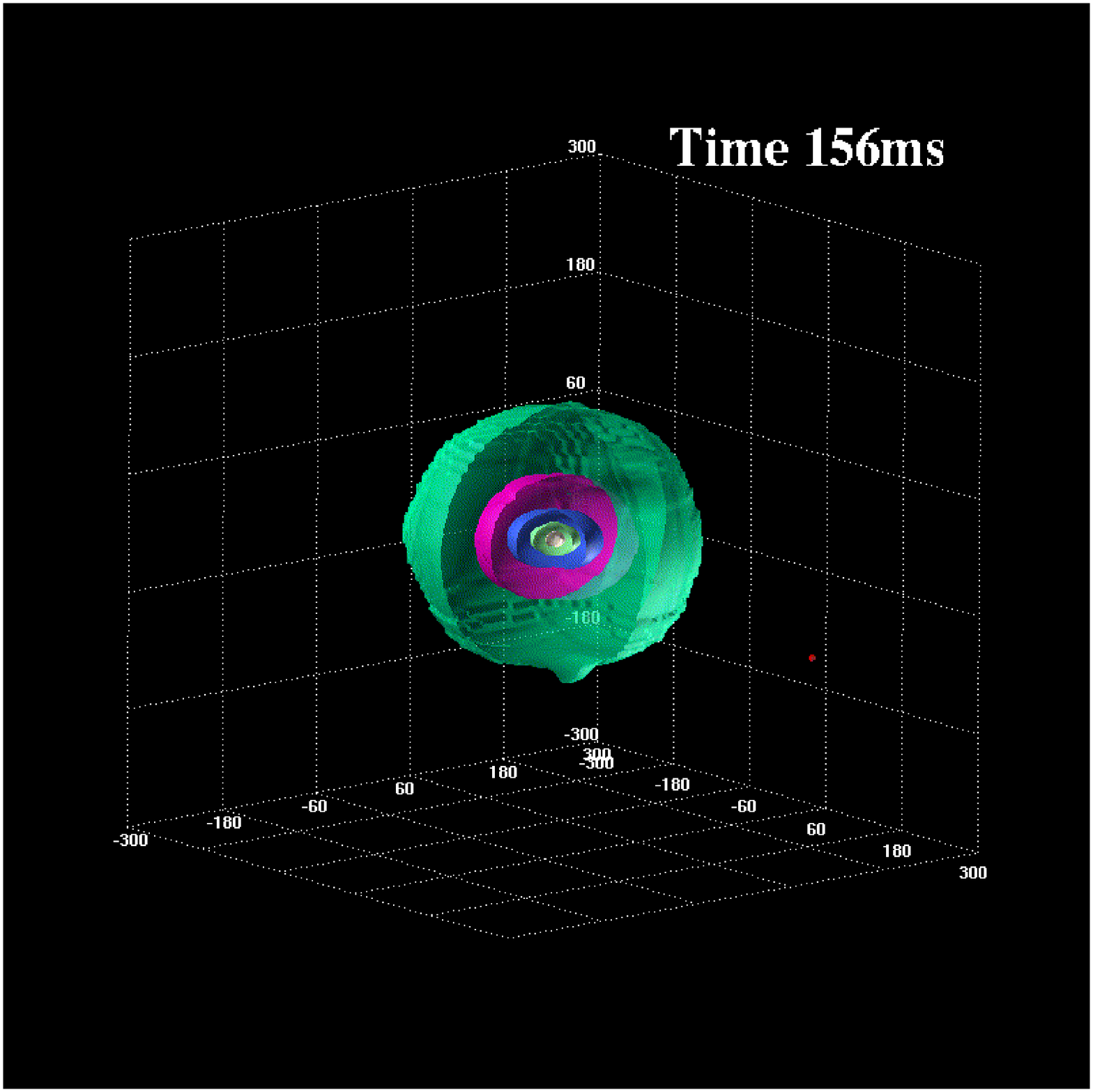}
\end{flushleft}
\end{minipage}
\end{tabular}
\caption{Three dimensional pictures of bipolar-outflow.
$Left$ two panels are "NB12R020Sf" and $right$ two are "NB00R020Sf" and $|x,y,z|\le300$km is drawn.
The central spherical-like shells represent iso-density surfaces and the outermost shell ($green$) corresponds to $10^9$g $\rm cm^{-3}$.
$White$ lines are the magnetic field lines. Translucent and opaque $red$ surfaces are iso-radialvelocity surfaces and each corresponds to
$v_{\rm r}=10^9$cm $\rm s^{-1}$ (translucent) and $v_{\rm r}=2\times10^9$cm $\rm s^{-1}$ (opaque).
In non-magnetized model "NB00R020Sf", high velocity bipolar outflow did not appear.
}
\label{pic:f10f13}
\end{figure}

\begin{figure}[htbp]
\begin{tabular}{cc}
\begin{minipage}{0.5\hsize}
\begin{flushright}
\includegraphics[width=50mm]{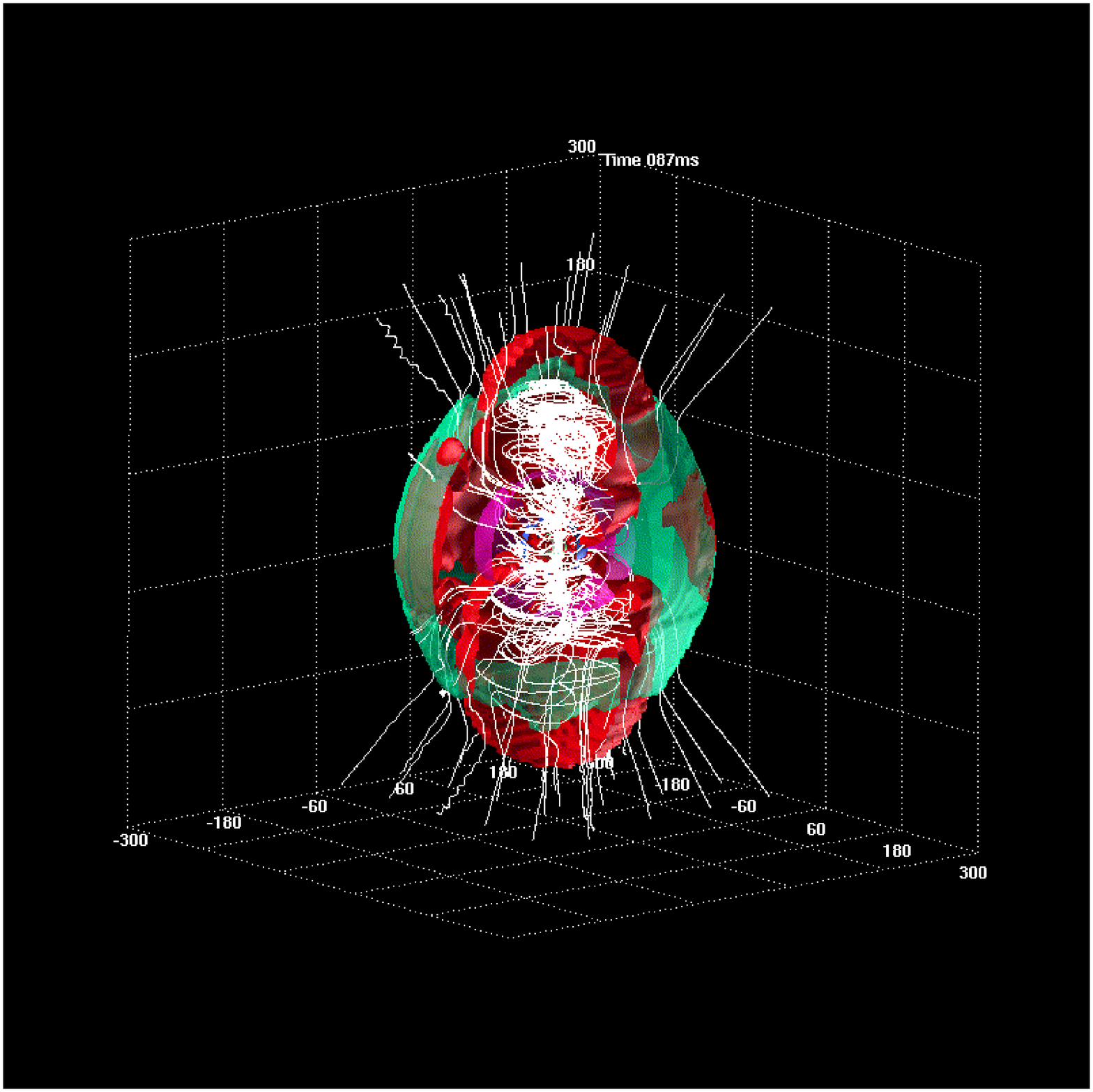}
\end{flushright}
\end{minipage}
\begin{minipage}{0.5\hsize}
\begin{flushleft}
\includegraphics[width=50mm]{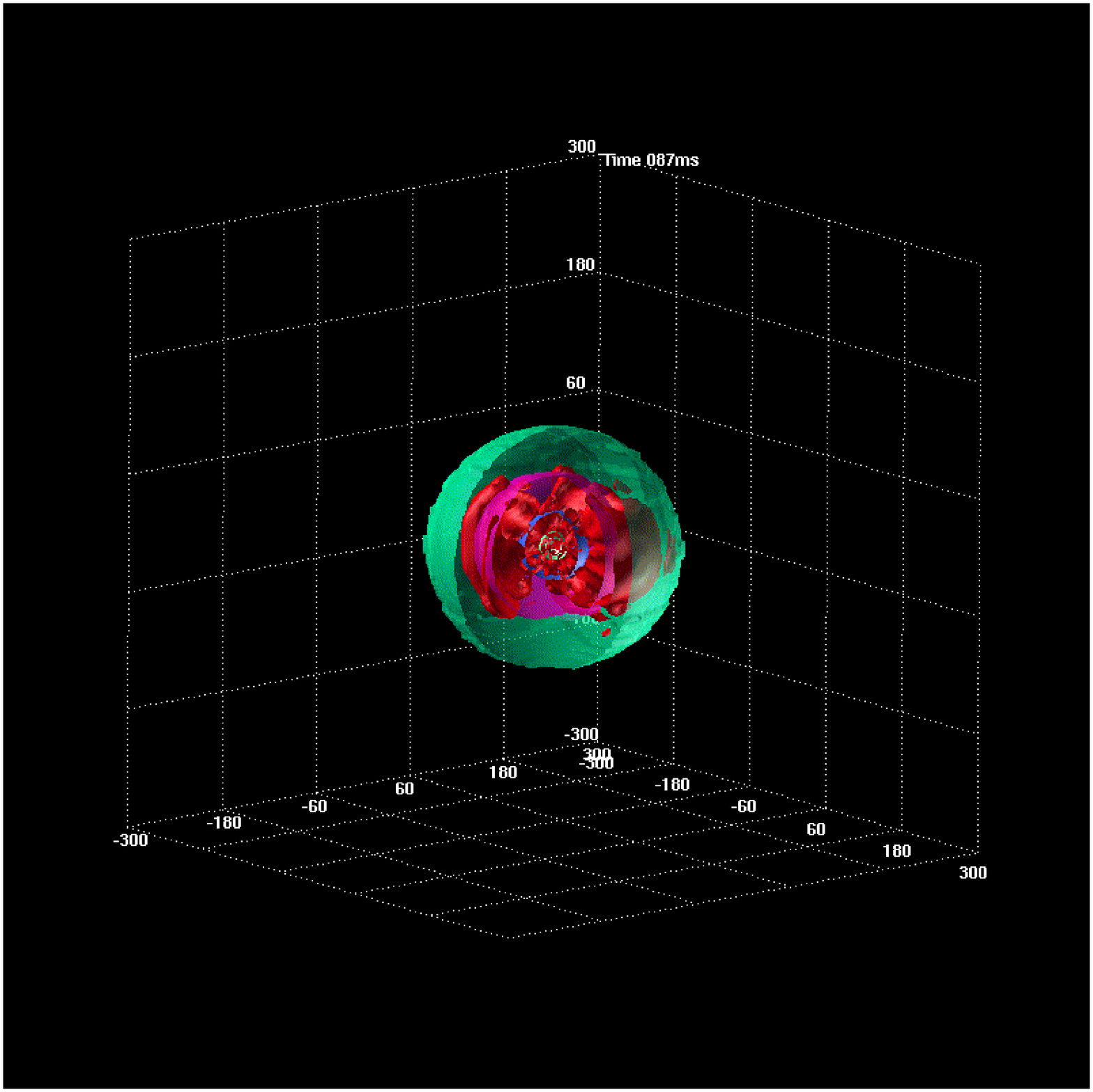}
\end{flushleft}
\end{minipage}
\end{tabular}
\caption{Same as Fig.\ref{pic:f10f13} but of GRMHD models ($left$; "GB12R020Sf'' and $right$; "GB00R020Sf'').
Both snapshots are taken at t=87 ms.
}
\label{pic:f21f22}
\end{figure}

Next, we describe the formation of bipolar outflow.
In all of our strongly magnetized models, bipolar outflow is formed in a similar manner and we thus present
mainly one representative model "NB12R020Sf" in this subsection.
Fig.\ref{pic:rho5rho7} and Fig.\ref{pic:rho8rho9} show the density contour in model "NB12R020Sf" at different time slices.
Fig.\ref{pic:f5f7} and Fig.\ref{pic:f8f9} are the same as Fig.\ref{pic:rho5rho7} but are with the color coded contour of plasma beta ($\beta_{\rm p}\equiv{\rm P_{gas}/P_{mag}}$) in logarithmic scale and the flow velocity in $white$ arrows.
$Black$ curves represent the iso-density contour.
The depicted region is $(x,y,z)=[-150,150]$km and, in each panel, $bottom$-$left$, $top$-$left$ and $top$-
$right$ part represents $xy$ (equatorial), $xz$ and $yz$ plane, respectively.
As shown in these figures, the strongly magnetized regions where $\log\beta_{\rm p}$ reaches $\sim0$ appear
along the rotational axis, which means the magnetic pressure is comparable to the matter pressure.
Then high velocity outflow is launched along the rotational axis (see, Fig.\ref{pic:f10f13} for NMHD and
Fig.\ref{pic:f21f22} for GRMHD) while inflow appears along the equatorial plane.
On the other hand, non-magnetized models and weakly magnetized model "NB09R020Sf"
do not form any bipolar outflow as seen from two $right$ panels of Fig.\ref{pic:f10f13}.

\begin{figure}[htbp]
  \begin{center}
    \includegraphics[width=40mm,angle=-90]{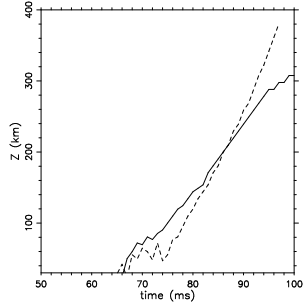}
    \caption{Time evolutions of the bipolar shock front along the rotational axis (Only those of north hemisphere
    are shown).
    $Solid$ and $dashed$ lines are results of NB12R020Sf and GB12R020Sf, respectively.
    }
    \label{pic:shockP.eps}
  \end{center}
\end{figure}

When we compare GRMHD(GB12R020Sf) and NMHD(NB12R020Sf) models, we see that the shock front of
bipolar outflow moves faster, approximately a factor of 2, in GRMHD model as can be seen in Fig.
\ref{pic:shockP.eps}.
Since the bipolar outflow is driven by the angular momentum transfer (described in the next section),
higher velocity outflow reflects that GB12R020Sf extracts larger angular momentum compared to
NB12R020Sf.
Then the magnetic energy is increased while the rotational energy is decreased as seen in $right$-$bottom$ panel of Fig. \ref{pic:f3}.

\subsubsection{Driving Mechanisms of Outflow}
\label{sec:DrivingMechanismsofOutflow}
\begin{figure}[htbp]
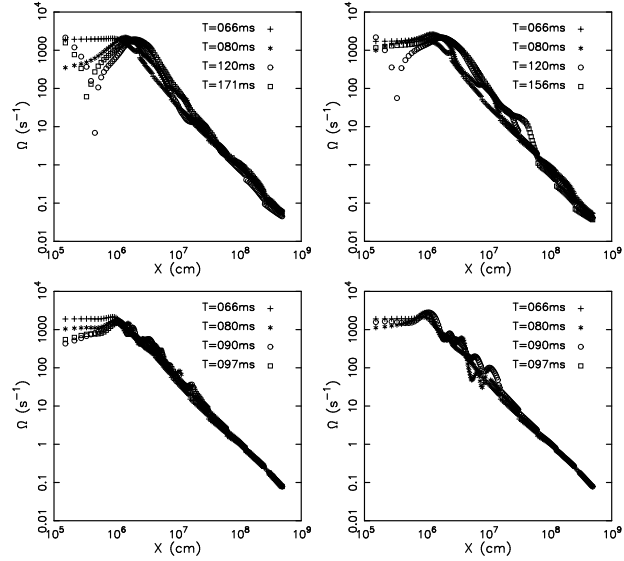

\begin{tabular}{cc}
\begin{minipage}{0.5\hsize}
\begin{flushright}
\includegraphics[angle=-90.,width=40mm]{f41.eps}
\end{flushright}
\end{minipage}
\begin{minipage}{0.5\hsize}
\begin{flushleft}
\includegraphics[angle=-90.,width=40mm]{f42.eps}
\end{flushleft}
\end{minipage}
\end{tabular}
\begin{tabular}{cc}
\begin{minipage}{0.5\hsize}
\begin{flushright}
\includegraphics[angle=-90.,width=40mm]{f43.eps}
\end{flushright}
\end{minipage}
\begin{minipage}{0.5\hsize}
\begin{flushleft}
\includegraphics[angle=-90.,width=40mm]{f44.eps}
\end{flushleft}
\end{minipage}
\end{tabular}
\caption{The angular velocity profiles along $x$ axis at different time. From $top$-$left$ to clock wise direction, "NB12R020Sf'', "NB00R020Sf''
, "GB00R020Sf'' and "GB12R020Sf'' are displayed.
Note that in all four models shown in this figure, the time of core bounce is approximately 66ms.
Kinks, e.g., shown around $x\sim3$km in the $top$-$left$ panel at $T=120$ms, represent the retrograde of the angular velocity.}
\label{pic:f14f17}
\end{figure}

In this section, we describe the driving mechanisms of the outflow.
As mentioned in previous subsection, all strongly magnetized models exhibit high velocity outflow along the rotational axis.
The ultimate energy source of outflow is the angular momentum transfer from the central object which can be
seen from Fig.\ref{pic:f14f17}.
In this figure, the angular velocities along the $x$ axis of different models at different time slices are shown.
For instance, from left two panels, we can see the angular velocity within $x\la10$km decreases rapidly, a
factor of several, from $t=66$ms to $t\sim80$-90ms.
On the other hand, non-magnetized models (right two panels) do not show any deceleration (note that
kink in the angular velocity profile at $x\sim3$km shown in $t=120$ms of "NB00R020Sf" is originated from
such as meridional circulation or displacement of the mass center).
From this fact, the angular momentum is extracted from the central object by the magnetic field.
The extracted angular momentum is first converted to the magnetic field mainly via the magnetic field lapping.
Then, there are two types of driving mechanisms of the outflow, the magneto-spring and the
magneto-centrifugally supported mechanisms.
We consider, from $left$ two panels of Fig.\ref{pic:f10f13}, that initial mechanism is the magneto-spring
effect and then it transitions to the magneto-centrifugally supported mechanism.
This is because from Fig.\ref{pic:f10f13}, we see the magnetic field lines are highly twisted inside the shell at the onset of launching the outflow (t=120ms),
however this twisted configuration is stretched eventually, as if the compressed spring would do, toward the north-south direction as seen in t=171ms panel.
Final magnetic field configuration is less toroidal dominant compared to that of t=120ms and the matters stream away along these helical magnetic field lines.
\begin{figure}[htbp]
\begin{tabular}{cc}
\begin{minipage}{0.5\hsize}
\begin{flushright}
\includegraphics[angle=-90.,width=45mm]{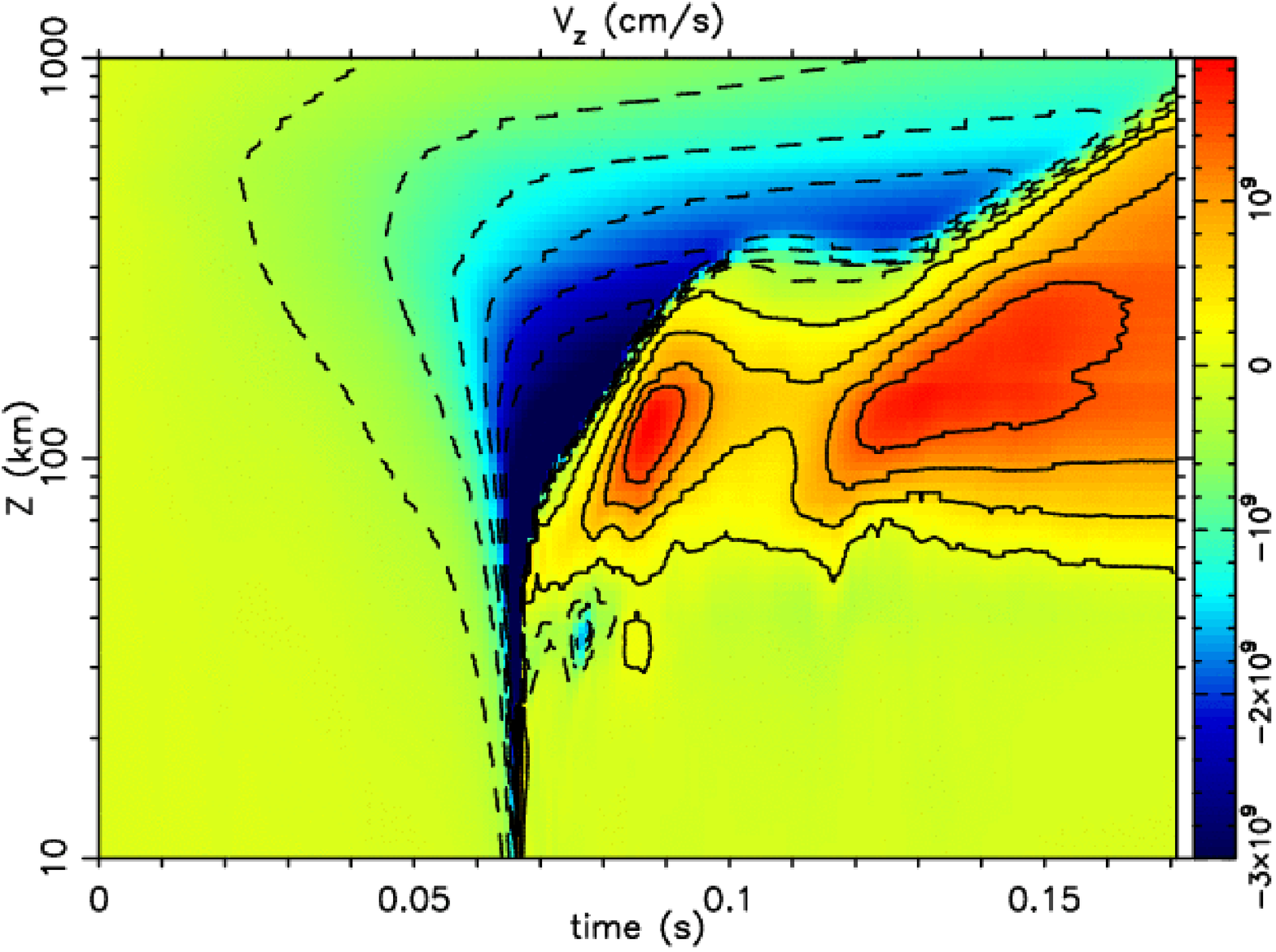}
\end{flushright}
\end{minipage}
\begin{minipage}{0.5\hsize}
\begin{flushleft}
\includegraphics[angle=-90.,width=45mm]{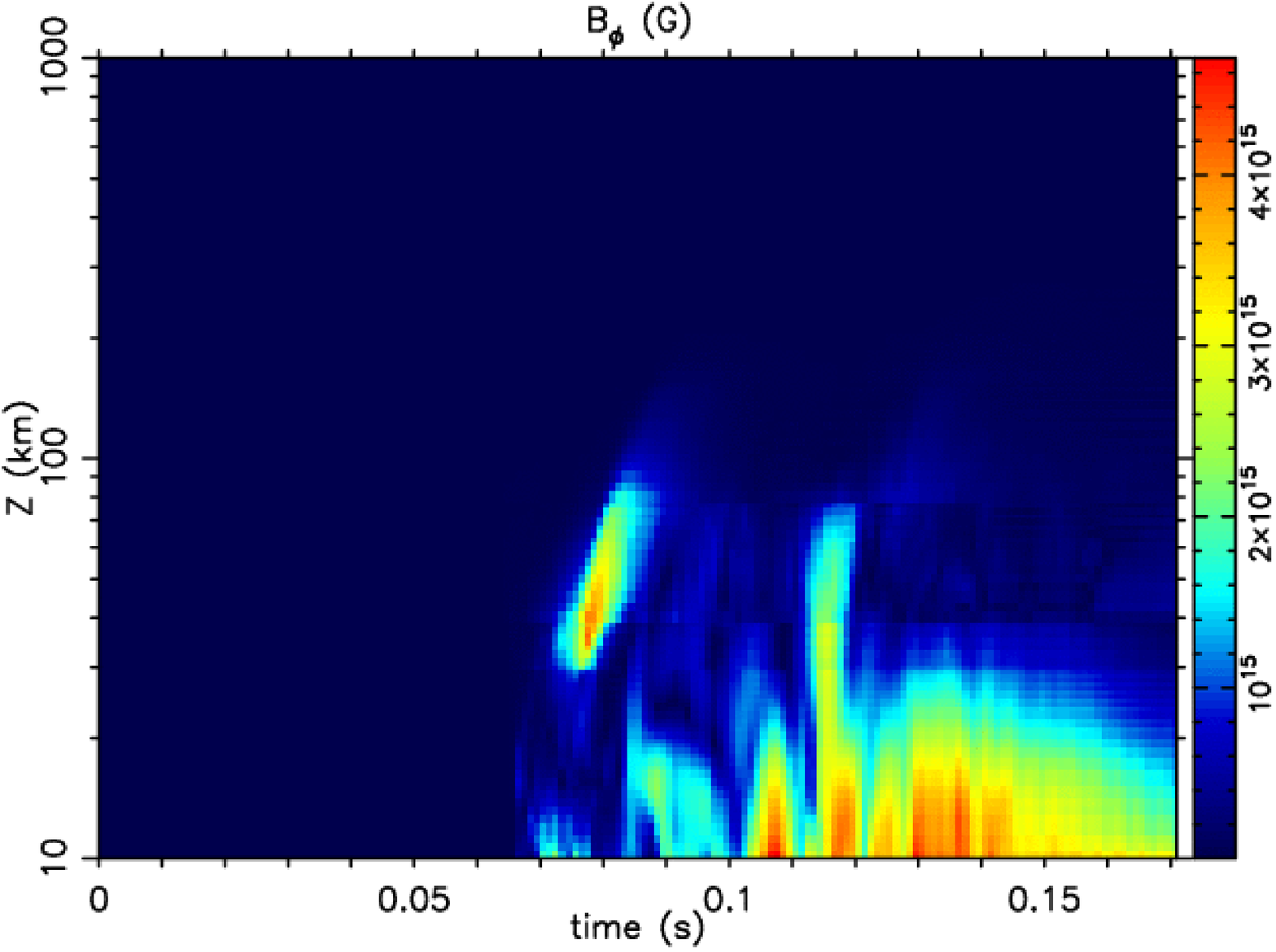}
\end{flushleft}
\end{minipage}
\end{tabular}
\begin{center}
\includegraphics[angle=-90.,width=45mm]{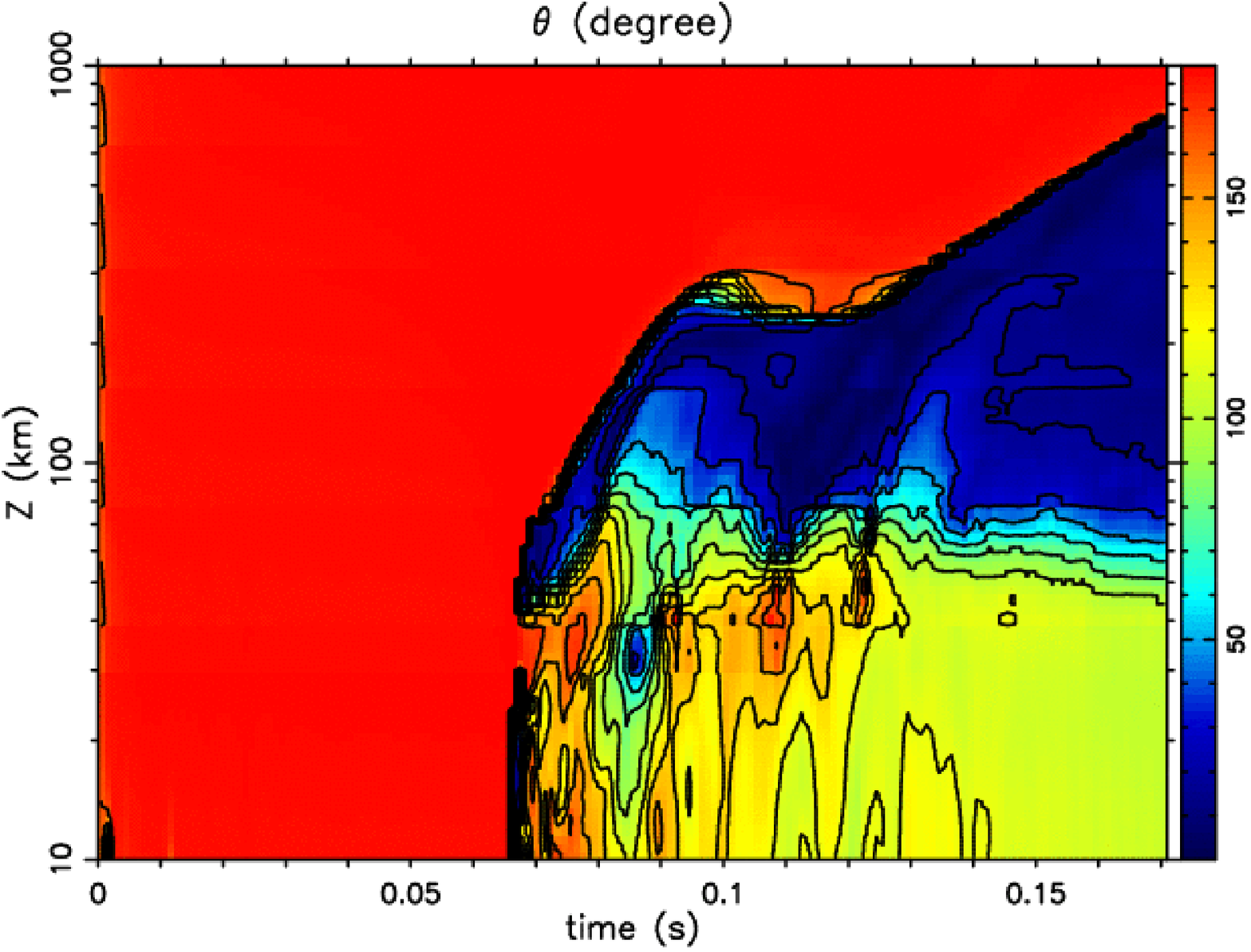}
\end{center}
\caption{Time evolutions of the velocity component ($V_z,\ upper$-$left$), the toroidal magnetic field ($B_\phi,\ upper$-$right$) and angle $\theta$ ($bottom$) between the magnetic field and the velocity 
vector, along the rotational axis of model "NB12R020Sf".}
\label{pic:f18f20}
\end{figure}
This is also seen in Fig.\ref{pic:f18f20} which shows time evolutions of the outflow velocity ($V_z$), the toroidal magnetic field ($B_\phi$) and angle $\theta\tbond\cos^{-1}(({\bf B\cdot V})/BV)$ between
the magnetic field and the velocity vector, along the rotational axis of model "NB12R020Sf".
From $upper$-$right$ panel, we find the strongly amplified $B_\phi$ at $t\sim120$ms and $z\la80$km which shortly disappears.
At the same time, the high velocity region appears in the $upper$-$left$ panel.
The angle $\theta$ is nearly $\sim90\degr$ at $z\la50$km and $t\ga120$ms which indicates that the outflow is driven by the gradient of magnetic pressure (i.e., the magneto spring effect).
Then $V_{\rm z}$ is accelerated up to $z\sim110$km along the magnetic field ($\theta\sim0\degr$, i.e., the magneto centrifugal effect).

In \citet{Shibata06}, they calculated 2D axisymmetric GRMHD simulations and reported MHD outflow is first driven by the magneto-spring effect and 
eventually by the magneto-centrifugally supported mechanism.
Thus our results of launching processes of MHD outflow are qualitatively similar to theirs.

\subsection{Non-Axisymmetric Motion}
\label{sec:Non-Axisymmetric motion}
\begin{figure}[htpb]
\begin{center}
\includegraphics[width=50mm,angle=-90.]{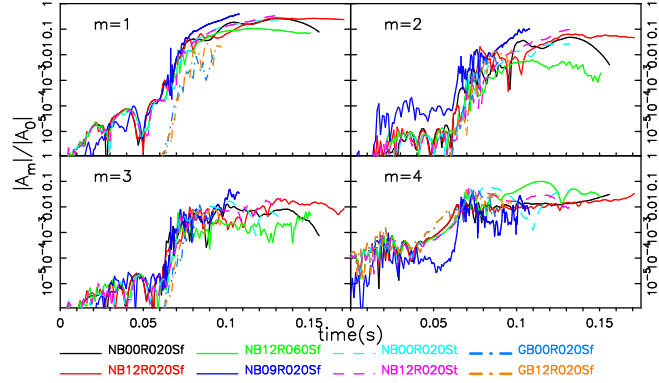}
\caption{Time evolutions of the normalized mode amplitude $|{\rm A_m}|/|{\rm A_0}|$.
Model names are listed in the bottom part.
}
\label{pic:f23}
\end{center}
\end{figure}

As described in the previous subsection, our results seen in dynamical evolutions are qualitatively the same as
those reported in previous 2D axisymmetric MHD works (for NMHD see, e.g.,
\citet{Kotake04,Sawai05,Burrows07} and for GRMHD see, e.g., \citet{Obergaulinger06,Shibata06}) in the way
that the equatorial inflow and the bipolar outflow along the rotational axis due to the magnetic field.
We also calculated several models with tilted magnetic field axis against the rotational axis to induce larger
non-axisymmetry and found the outflow is driven along the rotational axis similar to \citet{Mikami08}.
According to many previous studies, both in GR and the Newtonian limit \citep[e.g.,][]{Ott07,Scheidegger09}, the nascent neutronstar is
sensitive to the rotational instability predominated by the "m=1" non-axisymmetric mode within our initial rotational parameter
range.
To confirm our code can actually reproduce some nonaxisymmetric modes characteristic to rotational collapse
of massive stars, we monitor the non-axisymmetry with the same approach as \citet{Scheidegger09}.
We decompose the density into the Fourier components along the equatorial ring of radius 40km which is
is beyond the central rigidly rotating region and inside the prompt shock (see, Fig.\ref{pic:RhoX}).
Fourier amplitude of mode "m" is defined as the following equations.
\begin{eqnarray}
\label{eq:Fourier1}
\rho(\varpi,z,\phi)&=&\sum^{\infty}_{{\rm m}=-\infty}A_{\rm m}(\varpi,z)e^{i{\rm m}\phi}\\
\label{eq:Fourier2}
A_{\rm m}(\varpi,z)&=&\frac{1}{2\pi}\int^{2\pi}_{0}\rho(\varpi,z,\phi)e^{-i{\rm m}\phi}d\phi
\end{eqnarray}
In Fig.\ref{pic:f23}, we plot the time evolutions of normalized mode amplitude $|{\rm A_m}|/|{\rm A_0}|$.
From this figure, we see that m=4 mode is the most dominant mode before core bounce ($t\sim66$ms) since our cartesian grid
induces quadrupole numerical noise at initial.
However, the linear amplification phase starts immediately after core bounce and several ms later it reaches the non-linear phase
($t\ga75$ms).
During the non-linear phase, dominant mode becomes m=1 and their normalized amplitude exceed $\ga0.1$.
This is consistent with the structure, a so-called one-armed spiral structure seen in \citet{Ott05}.
Since $\beta_{\rm rot}$ at core bounce reaches $\sim3$\% and also keeps $\ga1$\% after core bounce, we
consider that the low-$|{\rm T/W}|$ instability causes this non-axisymmetric configuration.
Like these, even though the outflow structure is almost axisymmetric from broader perspective,
non-axisymmetry develops and show significantly large mode amplitude in the vicinity of center in the self gravitating system.
This non-axisymmetry may alter the gravitational wave form \citep{Scheidegger09}.

Next, we describe about the amplification of initially weak magnetic field ($B_0=10^9$G) in model
"NB09R020Sf".
As mentioned above in Sec.\ref{sec:FormationofOutflow}, only through the field-wrapping and the compression
mechanism, the magnetic field cannot be amplified strongly enough to drive the outflow soon after the core
bounce as seen in other models with initially strong magnetic field.
However, there might be several magnetic field amplification mechanisms to be occurred after core 
bounce such as the MRI or the dynamo mechanism \citep{Akiyama03,Cerda07,Obergaulinger09}
in addition to the aforementioned linear mechanisms.
If some of these mechanisms operate within dynamical time scale, the saturated magnetic field is considered
to possess enough capability to affect the explosion dynamics.
The largest difference between 3D and 2D(axisymmetric) in the amplification of the magnetic field is
conversion from toroidal to poloidal magnetic field, since the toroidal to poloidal conversion can never happen in 2D axisymmetric motion.
This is because, from the Faraday's law, time evolution of a poloidal component $B^{\rm pol}$ becomes
\begin{eqnarray}
\partial_t B^{\rm pol}&+&\partial_{\rm pol\bot}(B^{\rm pol}v^{\rm pol\bot}-v^{\rm pol}B^{\rm pol\bot})\nonumber\\
&+&\partial_{\rm tor}(B^{\rm pol}v^{\rm tor}-v^{\rm pol}B^{\rm tor})=0
\end{eqnarray}
Here, "pol" and "tor" represent a poloidal and toroidal component, respectively, and "pol$\bot$" is a
perpendicular one to both the "pol" and "tor" components.
In axisymmetry, $\partial_{\rm tor}=0$ and thus $B^{\rm tor}$ cannot be converted to $B^{\rm pol}$,
however in full 3D, several non-axisymmetric fluid motions (e.g., the Parker or the Tayler or the convective instabilities)
let $\partial_{\rm tor}\ne0$ and close the conversion cycle (i.e., from poloidal to toroidal and toroidal to poloidal).
Through our weakly magnetized model "NB09R020Sf", we examine how the magnetic field is amplified after
core bounce and whether the amplified magnetic field affect the explosion dynamics or not.

\begin{figure}[htbp]
\begin{tabular}{cc}
\begin{minipage}{0.5\hsize}
\begin{flushright}
\includegraphics[angle=-90.,width=60mm]{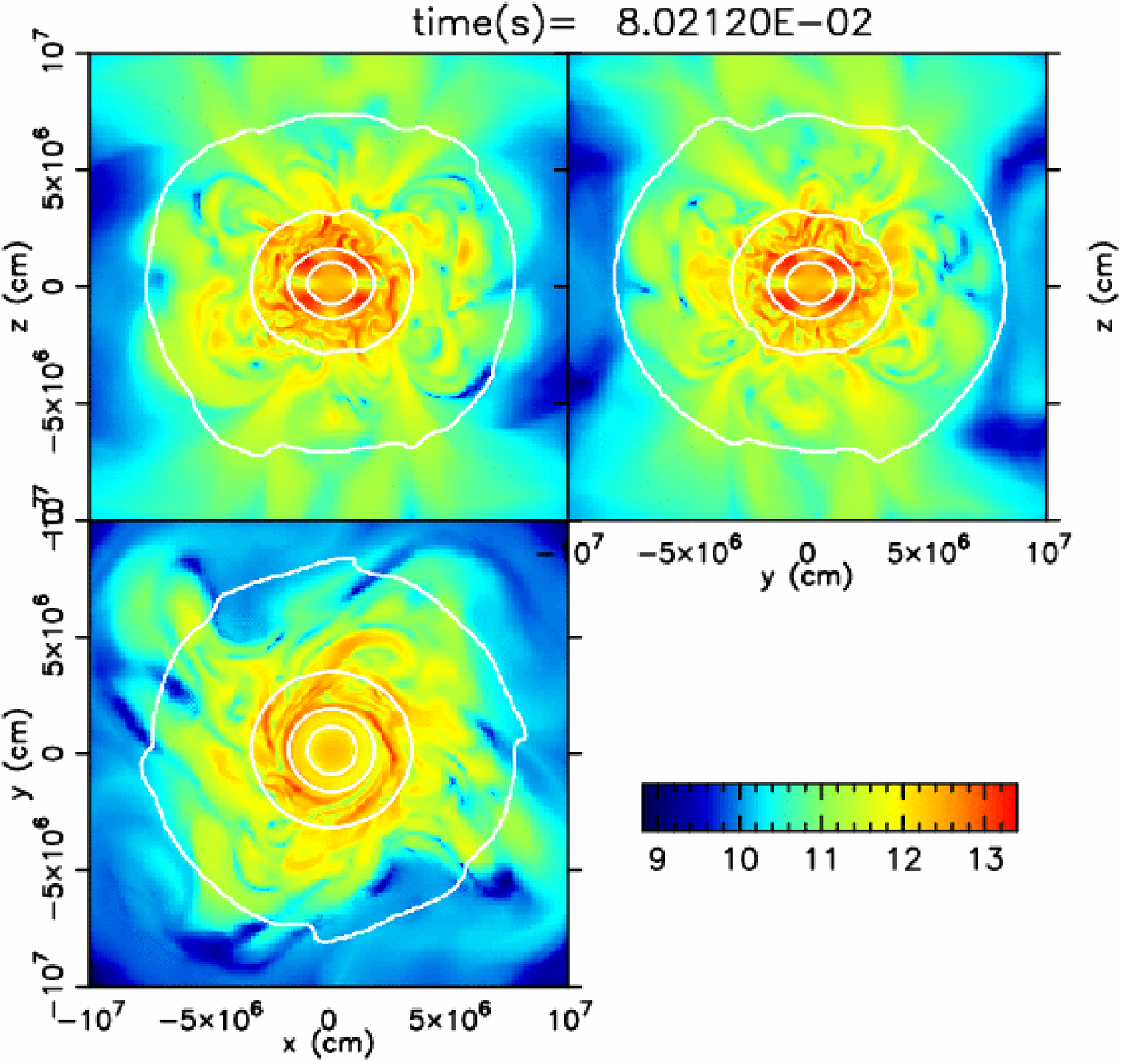}
\end{flushright}
\end{minipage}
\begin{minipage}{0.5\hsize}
\begin{flushleft}
\includegraphics[angle=-90.,width=60mm]{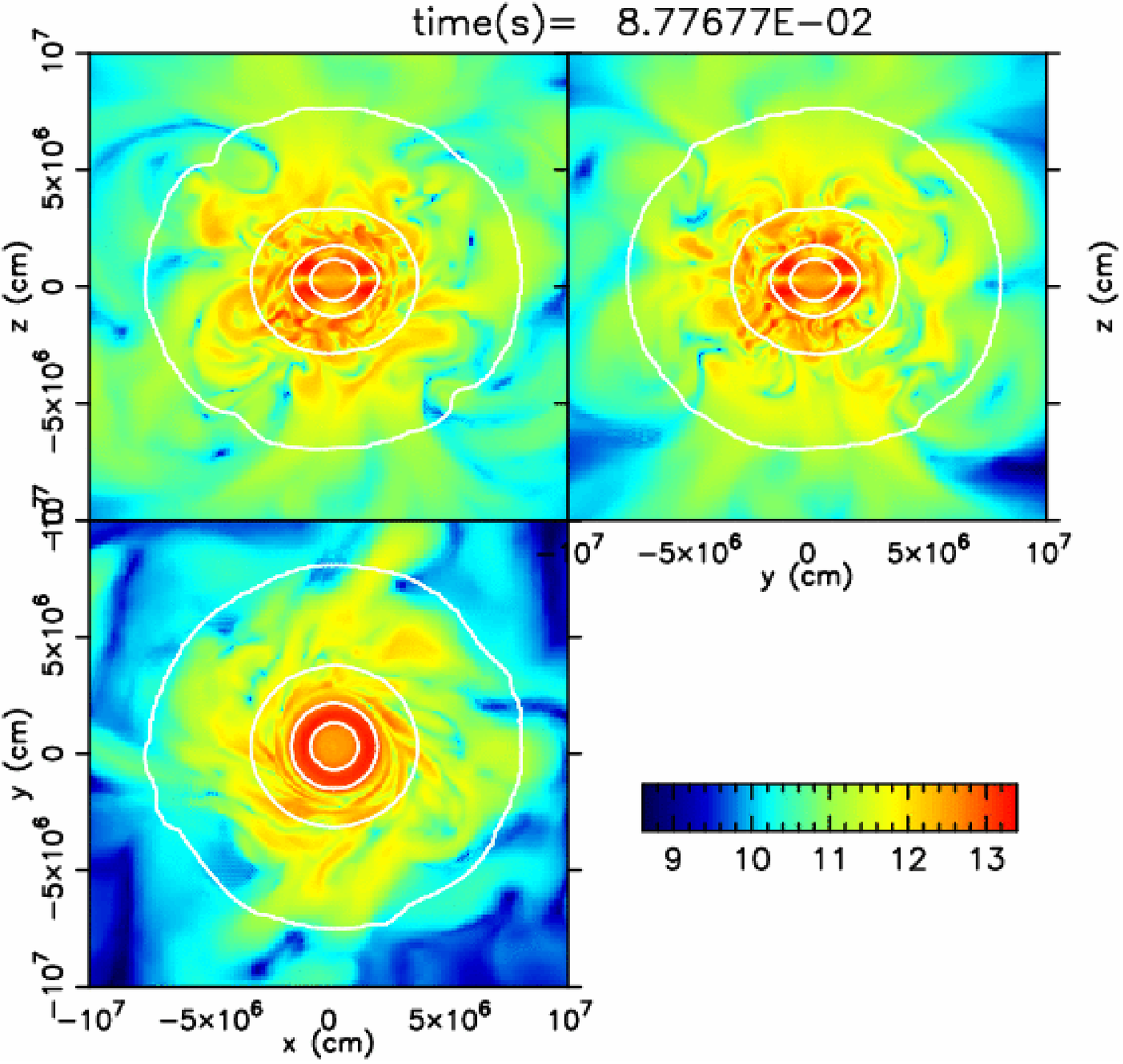}
\end{flushleft}
\end{minipage}
\end{tabular}
\begin{tabular}{cc}
\begin{minipage}{0.5\hsize}
\begin{flushright}
\includegraphics[angle=-90.,width=60mm]{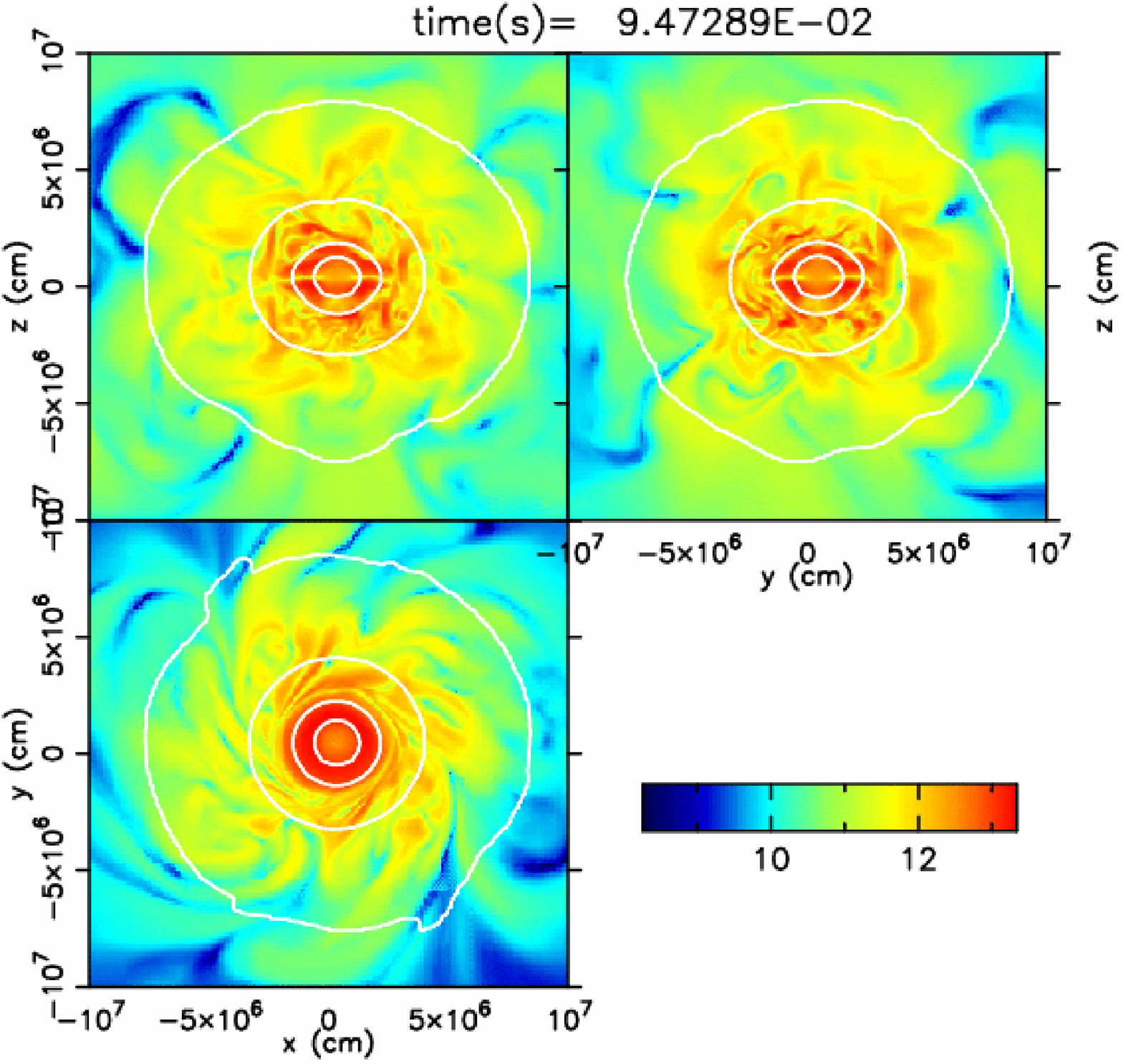}
\end{flushright}
\end{minipage}
\begin{minipage}{0.5\hsize}
\begin{flushleft}
\includegraphics[angle=-90.,width=60mm]{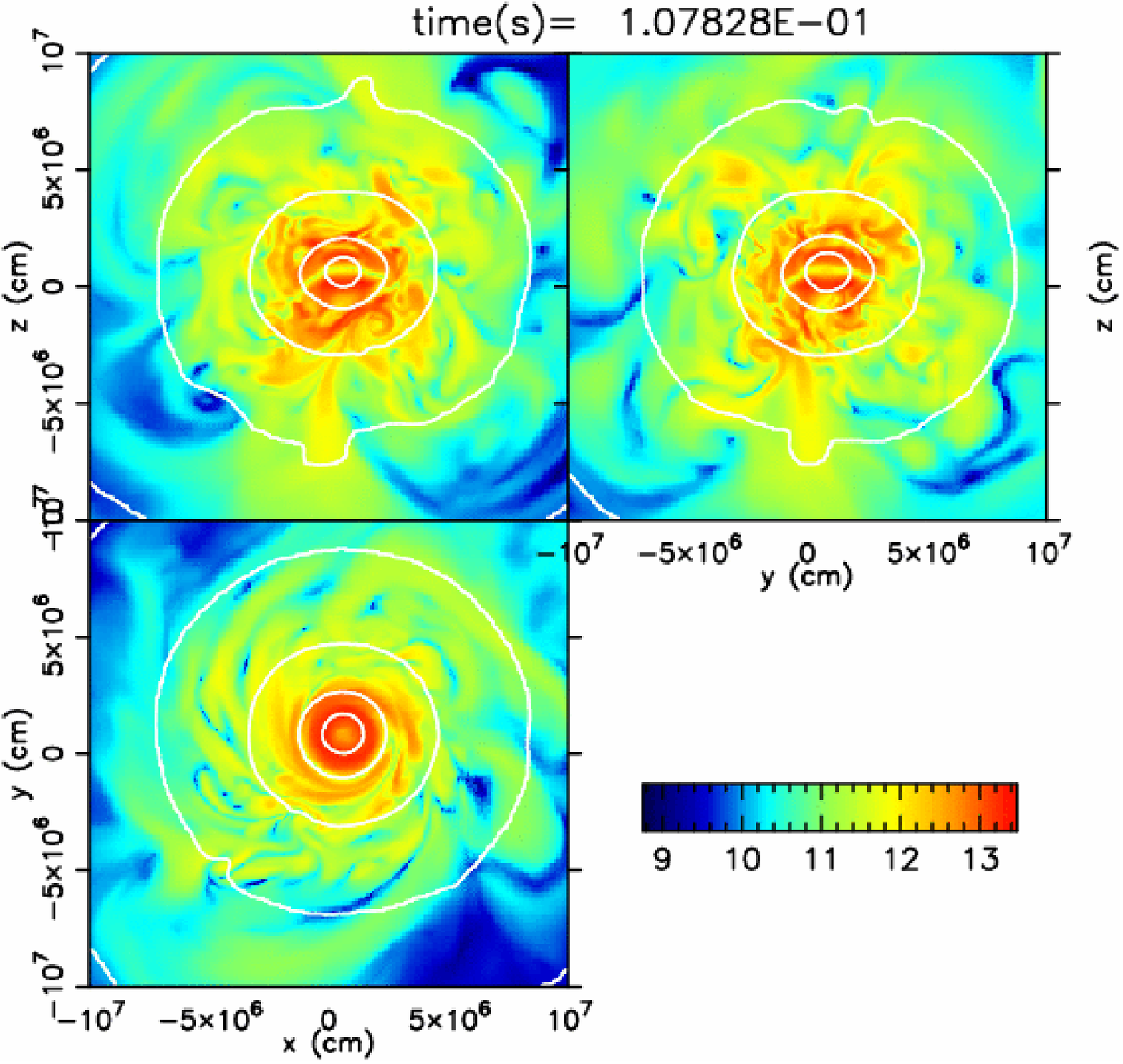}
\end{flushleft}
\end{minipage}
\end{tabular}
\caption{Time evolution of the magnetic field in logarithmic scale of model "NB09R020Sf".}
\label{pic:f24f27}
\end{figure}

In Fig.\ref{pic:f24f27}, we display time evolution of the magnetic field strength in logarithmic scale of model
"NB09R020Sf" which is high resolution run.
Within $r\la20$km, strength of the magnetic field shows stratified configuration compared to $r\ga20$km and is amplified
strongest ($\sim2\times10^{13}$G) among the numerical domain.
Since, within $10\la r\la20$km, matters rotate differentially with the steepest angular velocity gradient
(see, Fig.\ref{pic:f14f17}), the magnetic field strength is higher than the value of central region $r\la10$km
where the rotation is almost rigid.
Magnitude of the amplification is $\sim10^{13}/10^{9}=10^{4}$ which is close to the value predicted by
compression and we thus consider the dominant field amplification mechanisms within $10\la r\la20$km are
compression and the rotational winding effect.

\begin{figure}[htbp]
\begin{tabular}{cc}
\begin{minipage}{0.5\hsize}
\begin{flushright}
\includegraphics[angle=-90.,width=60mm]{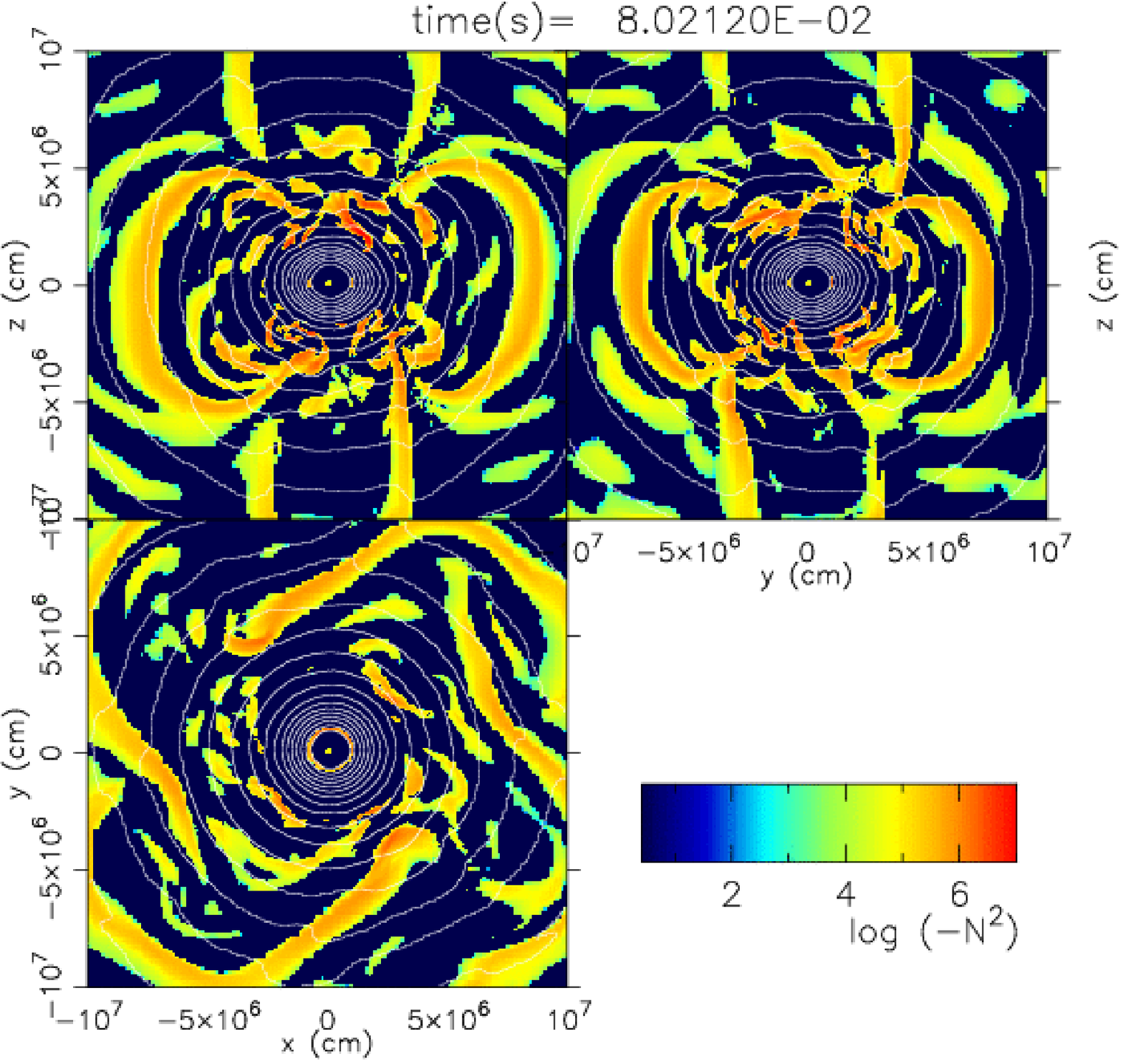}
\end{flushright}
\end{minipage}
\begin{minipage}{0.5\hsize}
\begin{flushleft}
\includegraphics[angle=-90.,width=60mm]{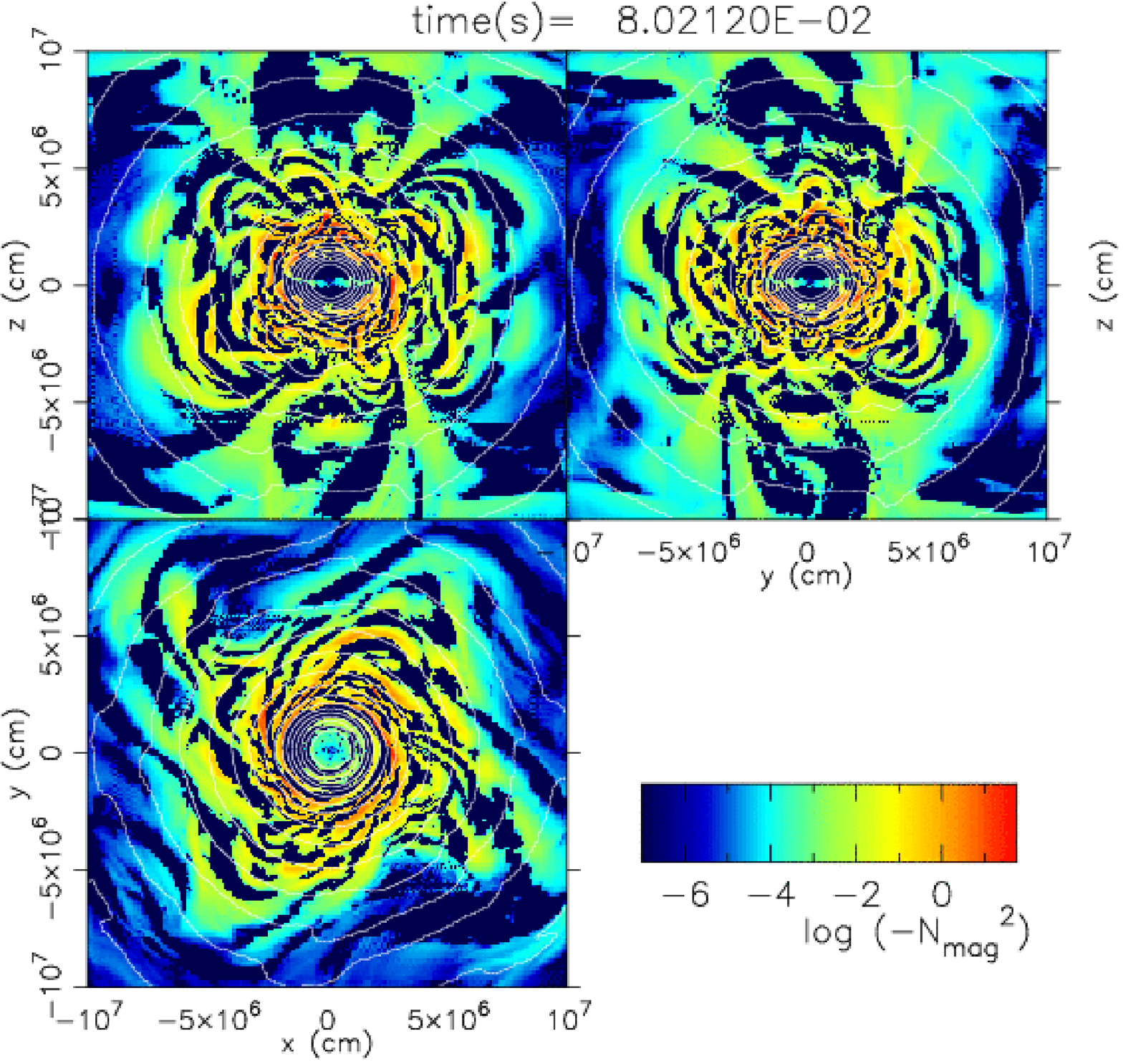}
\end{flushleft}
\end{minipage}
\end{tabular}
\caption{Logarithmic scale of the Brunt-$\rm V\ddot{a}is\ddot{a}l\ddot{a}$ frequency $-N^2$ ($left$)
and the magnetic buoyant frequency $-N_{\rm mag}^2$ ($right$) of model "NB09R020Sf".
Color-less area is where $N^2/N_{\rm mag}^2$ has positive value (i.e., stable region)
and $white$ curves are the iso-density contour.}
\label{pic:BVfreq}
\end{figure}

In contrast, we see the region $r\ga20$km is highly non-homologous and the magnetic field configuration
pattern changes momentarily, however the maximum strength keeps the same
level of the order of $\la10^{13}$G throughout our calculation ($\sim40$ms after core bounce).
The stochastic configuration pattern beyond $r\sim20$km is mainly triggered by the entropy driven convection.
The reason is like this.
There are several candidates to configure such flow pattern such as the convective motion, the MRI, the
Parker instability (or the magnetic buoyancy) or the Tayler instability.
However, among them, the growth time scales of the Tayler and the Parker instabilities are too long and they
are inefficient during our calculation time.
For instance, the growth time scale of the Tayler instability is the order of the $Alfv\acute en$ crossing time of
the system and $\sim\mathcal{O}(1)$ h according to \citet{Cerda07} in a weakly magnetized limit.
As for the Parker instability, the growth time scale ($\tau_{\rm mag}$) can be estimated by using the frequency
of the magnetic buoyancy $N_{\rm mag}$ as $\tau_{\rm mag}\sim 2\pi N_{\rm mag}^{-1}$.
According to \citet{Acheson79}, $N_{\rm mag}$ is defined by using the the $Alfv\acute en$ velocity
$c_{\mathcal A}$, the speed of sound $a$, the toroidal magnetic field $B_\phi$ and the rest mass density
$\rho$ as
\begin{equation}
N_{\rm mag}^{2}=\frac{c_\mathcal{A}^2}{\gamma a^2}\bf{g}\cdot \nabla\left(\rm{ln}\frac{B_\phi}{\rho}\right)
\end{equation}
Where, $\gamma\tbond d{\rm ln}p/d{\rm ln}\rho|_S$, $S$ is the entropy and $\bf g$ is the gravitational
acceleration.
In what follows, we adopt a pseudo-entropy defined by $S\tbond P_{\rm t}/P_{\rm c}$.
In Fig. \ref{pic:BVfreq}, we display color coded contour of $N_{\rm mag}^2$ ($right$) in addition to the
Brunt-$\rm V\ddot{a}is\ddot{a}l\ddot{a}$ frequency $N^2$ ($left$) defined by
\begin{eqnarray}
N^2=\frac{1}{p\gamma}\frac{\partial p}{\partial S}\biggr|_\rho {\bf g}\cdot \nabla S
\end{eqnarray}
in log scale of model "NB09R020Sf".
In each panel, color-less area is where $N^2/N_{\rm mag}^2$ has positive value and is thus stable region against each mechanism.
The specific angular momentum with positive gradient with respect to $r$ has stabilizing effect on the
convective motion, however it has negative gradient in our models and the unstable region becomes larger when we
consider the contribution from it (i.e., the Solberg-H$\o$iland criterion).
From Fig. \ref{pic:BVfreq}, we see the PNS is convectively and magnetic buoyantly unstable.
However, the growth time scale of each mechanism differs widely, $\tau\sim 2\pi N^{-1}\sim1-10$ms for convection
and $\tau_{\rm mag}\sim 2\pi N_{\rm mag}^{-1}\ga 0.1-10$s for magnetic buoyancy.
Since our calculation time is $\sim40$ms after core bounce, the Parker instability does not grow
while the convection can grow sufficiently within our simulation times.
The convective-dynamo is thus considered to contribute to the magnetic field amplification mechanism.

\begin{figure}[htpb]
\begin{tabular}{cc}
\begin{minipage}{0.5\hsize}
\begin{flushright}
\includegraphics[angle=-90.,width=60mm]{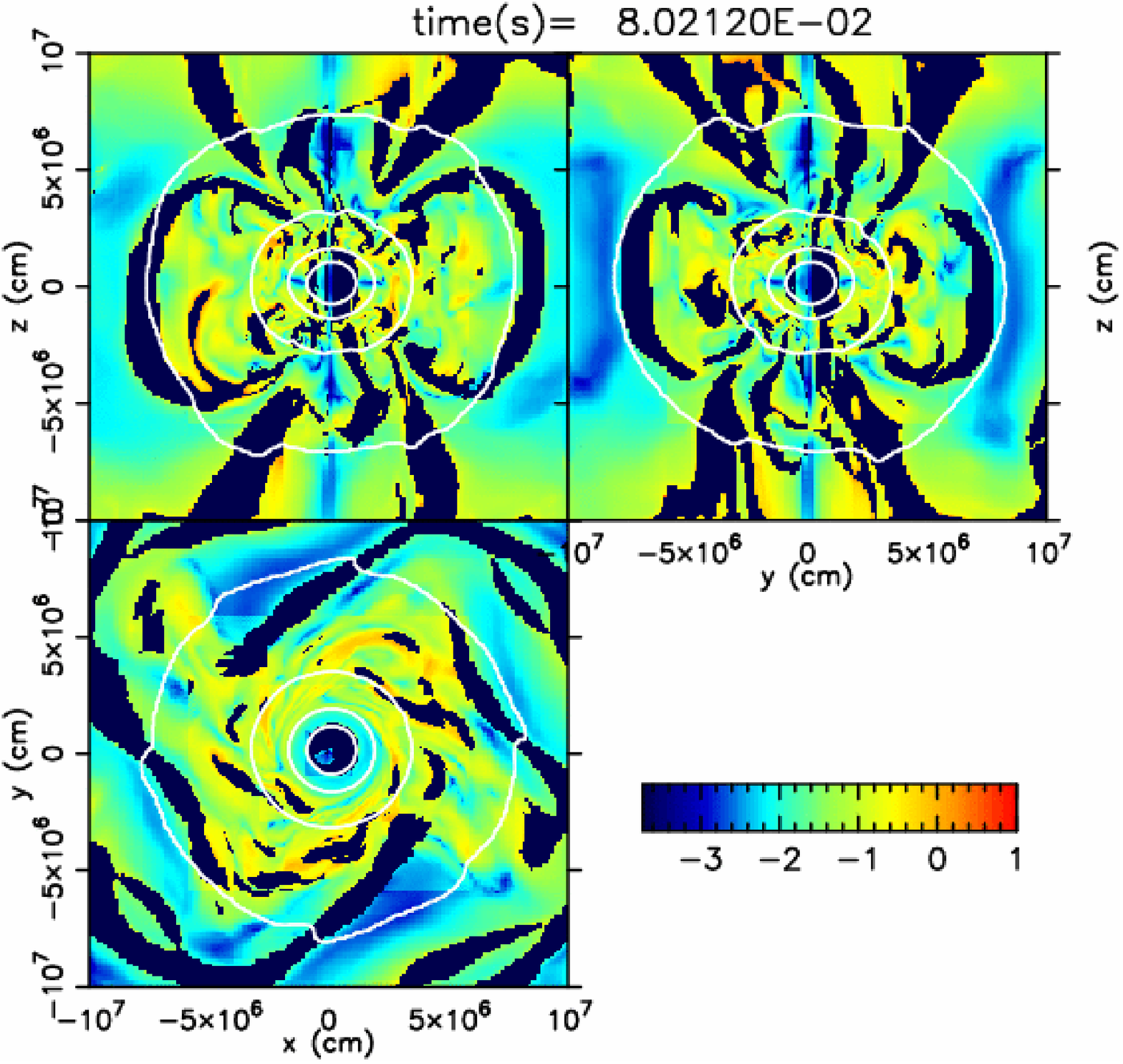}
\end{flushright}
\end{minipage}
\begin{minipage}{0.5\hsize}
\begin{flushleft}
\includegraphics[angle=-90.,width=60mm]{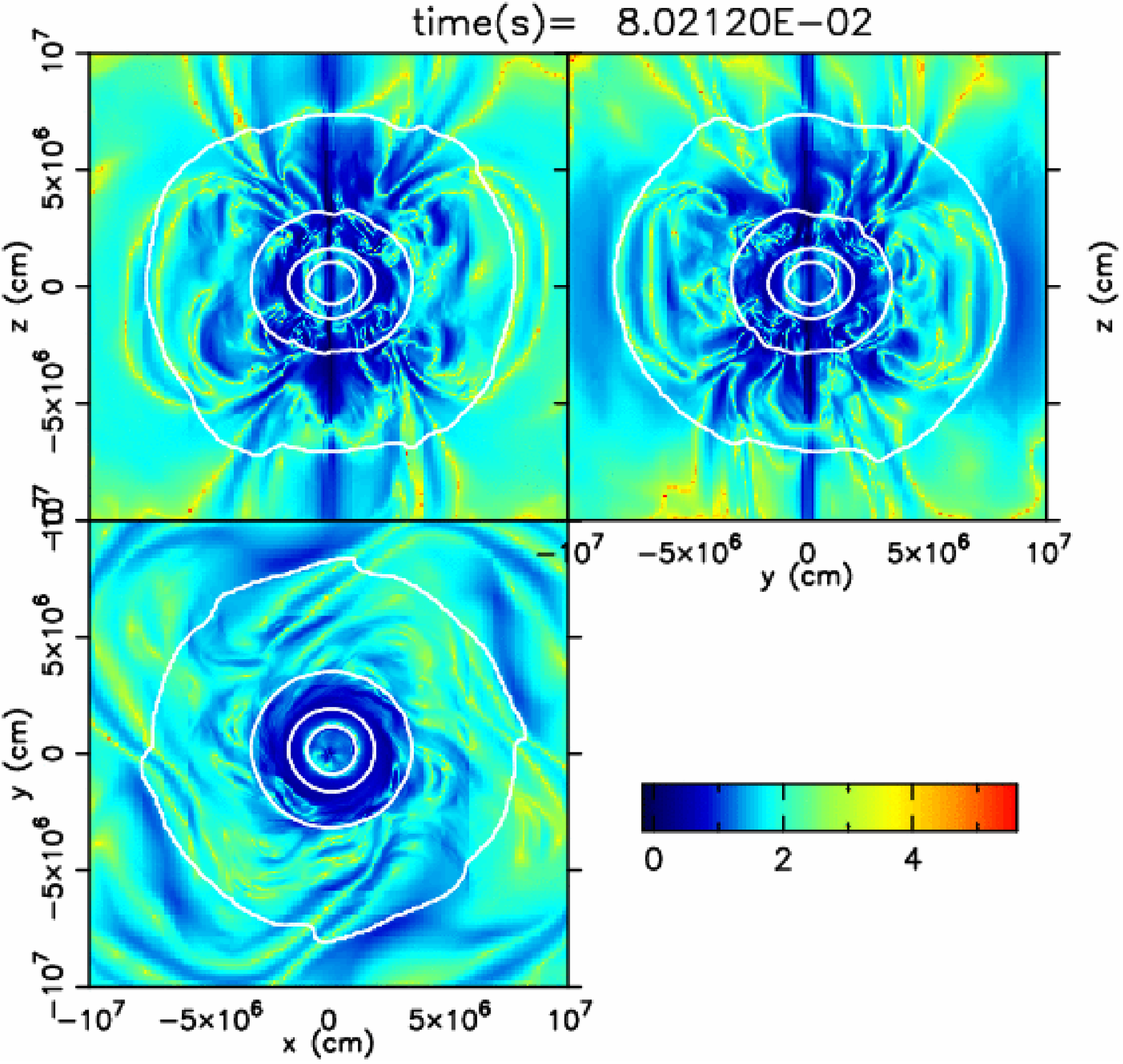}
\end{flushleft}
\end{minipage}
\end{tabular}
\caption{Contour of $\lambda_{\rm MRI}/\Delta x$ ($left$) and $\tau_{\rm MRI}$ (ms) ($right$) in log scale from model "NB09R020Sf".
Colorless areas in $left$ panel are where the MRI stable regions (i.e., positive angular velocity gradient).
}
\label{pic:f28f29}
\end{figure}

We also examine the possibility of the MRI.
Since, from local linear analysis, the MRI would occur when the rotation is differential with negative
angular velocity gradient \citep{Balbus91}.
From Fig.\ref{pic:f14f17}, we see inside the shock ($r\la100$km) has negative angular velocity gradient and is thus unstable against the MRI.
However, to follow the MRI by numerical simulation, the critical wave length of the MRI
"$\lambda_{\rm MRI}$" has to be resolved at least $\sim10$ numerical grids \citep{Obergaulinger09}.
Here, $\lambda_{\rm MRI}$ is defined by
\begin{eqnarray}
\label{eq:lambdaMRI}
\lambda_{\rm MRI}&=&4\pi c_{\mathcal A} \left(-\varpi\frac{\partial\Omega^2}{\partial\varpi}\right)^{-1/2}\\
\end{eqnarray}
In Fig.\ref{pic:f28f29}, we display the ratio of the critical wave length $\lambda_{\rm MRI}$ to the local numerical
grid width in $left$ panel and rough estimation of the growth time scale of the fastest growing MRI mode
"$\tau_{\rm MRI}$(ms)" in log scale in $right$ panel.
$\tau_{\rm MRI}$ is defined by \citep[see,][]{Balbus91}.
\begin{eqnarray}
\label{eq:tauMRI}
\tau_{\rm MRI}&=&4\pi  \left(\Bigl|\varpi\frac{\partial\Omega}{\partial\varpi}\Bigr|\right)^{-1}
\end{eqnarray}

From $left$ panel of this figure, we see the most part has negative value and therefore the MRI cannot be resolved.
We have to employ $\ga10$ times higher resolution ($\Delta x\la30-60$m) to resolve the MRI or,
if we adopt $\sim10$ times larger initial magnetic field ($\sim10^{10}$G), we possibly manage to do it with our
high resolution run marginally, since the wave length $\lambda_{\rm MRI}$ is proportional to $|B|$.
Here, we comment about our strongly magnetized models ($B_0\sim10^{12}$G).
If we extend what we mentioned just above (i.e., to adopt stronger initial magnetic field), we may easily
resolve the MRI since the critical wave length $\lambda_{\rm MRI}$ is approximately $10^3$ times larger
compared to weakly magnetized model.
From Fig. \ref{pic:f28f29}, we can estimate the wave length as $\lambda_{\rm MRI}\sim60$m ($yellow$ region)
and $\lambda_{\rm MRI}\sim600$m ($orange$ region) for $30\la r\la60$km where the resolution is $\Delta x\sim600$m.
Then in the strongly magnetized models, $\lambda_{\rm MRI}$ multiplied $10^3$ becomes $\lambda_{\rm MRI}\sim60-600$km.
However, the system scale (i.e., inside the prompt shock) is $\sim200$km and therefore those modes which have larger wave length
than the system cannot last.
In case an MRI mode lasts and if we can resolve it,
the magnetic field soon reaches saturation strength in several times of the growth time scale.
\citet{Akiyama03} derived the saturated magnetic field strength $B_{\rm sat,MRI}$ as
\begin{equation}
B_{\rm sat,MRI}^2\sim4\pi\rho v_\phi^2
\end{equation}
and it becomes $\sim10^{15}$ G with our initial rotational parameter.
All of our strongly magnetized models exhibit saturated magnetic field strength of the order of
$\sim10^{15-16}$ G after core bounce (the value is consistent with those reported by previous many studies,
e.g., for NMHD \citet{Kotake04,Sawai05,Burrows07},
for GRMHD, \citet{Obergaulinger06,Shibata06} and for 3D works, \citet{Mikami08,Scheidegger09}).
The value is comparable to $B_{\rm sat,MRI}$ and we thus consider the initially strong magnetic field is first amplified by the compression and the winding effect with the amplification magnitude of the order of $\sim 10^3$ and then the MRI operates to amplify the magnetic field up to the saturation strength
\citep[in][they reported the MRI operates with adopting similar initial magnetic field strength $\sim10^{12}$G]{Obergaulinger06,Shibata06}.
However, since just only the linear amplification mechanisms amplify the magnetic field up to $\sim10^{15}$G which is close to the MRI saturation level,
to see the effects of the MRI amplification more clearly, we have to adopt sufficiently weak magnetic field (e.g., $B_0\sim10^{10}$G)
which does not reach $B_{\rm sat,MRI}$ only through the linear amplification mechanisms but sufficiently strong
to resolve by numerical simulation.

As for the calculation time, the length of 40 ms after core bounce is marginally sufficient for the inner region
($r\la40$km) from the $right$ panel of Fig. \ref{pic:f28f29}, however beyond that region we have to evolve more than several
hundreds ms.
If we capture the linear amplification, it may reaches the saturation phase within the several rotational periods.
At this saturation phase, whether the magnetic field is sufficiently strong to affect the explosion dynamics and
how the magnetic field configuration is cannot be clarified without numerical simulation and this would be our future work.

\section{Summary and Discussions}
\label{sec:Summary and Discussions}
The explosion mechanisms of the core-collapse supernovae have been unknown and fascinating problems for several decades.
Recent observations show several common features seen in the CCSNe that some types of them are bipolar
like and sometimes non-axisymmetric explosions.
Motivated by these, we now have to take into account the effects of asymmetry into numerical works to
uncover the explosion dynamics.
Fortunately, recent development of computational resources enable us to handle the numerical simulations in
the context of three dimension.
We therefore have developed two types of three dimensional magneto hydrodynamical codes.
One is in the Newtonian approximation (NMHD) and the other is in the full general relativity (GRMHD).
The features of our codes are; 
(1) Adoptive Mesh Refinement to cover the wide dynamical ranges;
(2) high resolution shock capturing scheme with Roe-like (in NMHD) and HLL (in GRMHD) flux;
(3) several reconstruction schemes to maintain high spatial resolution;
(4) time update of the matters and the metric is done by the iterative Crank-Nicholson scheme;
(5) the constrained transport to evolve the magnetic field;
(6) any types of the EOS can be adopted;
(7) the Poisson solver with BiConjugate Gradient Stabilized Method under our AMR structure to solve the self gravity (in NMHD) and
the non-linear Poisson like equations for the Hamiltonian and the momentum constrains (in GRMHD).

In this paper, we described our numerical methods in detail and did several tests to confirm their abilities
through the simple shock tube tests; the Poisson solver for the spherically distributed matters;
conservation of the mass, the energy and the local/global angular momentum; the quadrupole linearized gravitational wave; the rotating neutronstar in
equilibrium states and so on.
Through these tests, we confirmed that our codes reproduce the numerical error convergence predicted
by our adopted reconstruction schemes and also confirmed that the accuracy of our code is sufficient to follow
the dynamical evolution of CCSNe.
And as for the first test of CCSN simulation, we calculated collapse of a $15M_\odot$ progenitor with varying the initial magnetic field strength,
the angular velocity and the stiffness of the Polytropic EOS by our GRMHD and NMHD codes.
Our main results and some discussions are as the following.

(1) After a short while ($\sim20$ms) from the time of core bounce, high velocity ($V_{\rm r}\sim2\times10^9$cm $\rm s^{-1}$) bipolar outflow is driven from surface of the proto-neutronstar
($|z|\sim30$km) and moves through along the rotational axis.
The bipolar outflow does not appear in the non-magnetized and the initially weak magnetized models which indicate the outflow is magnetically driven outflow.
The energy source of this outflow is the extracted angular momentum of the central proto-neutronstar which is
transfered by the magnetic torque.
The driving mechanisms are first by the magneto spring effect and then we consider the magneto centrifugally
supported outflow.

(2) In our self gravitating system, the non-axisymmetry develops immediately after the core bounce with the linear amplification at first and soon reaches the non-linear phase.
The dominant non-axisymmetric mode is the m=1 mode during the non-linear phase and the one-armed spiral structure is also can be seen.
Since our initial rotational velocity ($\Omega_{\rm c}\ge2$ rad/s) satisfies the low-$|T/W|$ instability criterion ($\beta\ga1$\%) after core bounce,
we consider that these non-axisymmetric, spiral mode is originated from the low-$|T/W|$ instability.
However, these non-axisymmetric motions are confined in the vicinity of the center and, in general terms, the global structure of bipolar outflow
is qualitatively the same as those reported in previous 2D axisymmetric MHD works in the way that the equatorial inflow and the bipolar outflow
along the rotational axis.

(3) In weakly magnetized model in which the initial central, poloidal magnetic field is $10^9$G, the convective over turn highly deforms the magnetic field
configuration.
However, with our resolution of $\Delta x\sim 300-600$m and limited computational time ($\sim40$ms after core bounce), we did not find
the exponential growth of the magnetic field which can be seen if the magneto-rotational instability works.
If we employ 10 times higher resolution or 10 times stronger initial magnetic field, we possibly capture the MRI marginally.
Time scale of the MRI is $\tau_{\rm MRI}\sim\mathcal O$(1) ms inside $r\la30$km which is comparable to the dynamical time scale and is sufficiently short to follow by the numerical simulation.
Even if the MRI operates in a weakly magnetized model, whether the MRI amplifies the magnetic field strongly
enough to affect the explosion dynamics or not and whether the amplified magnetic field contributes to launch
the outflow are big issues.
Especially, since the the amplified magnetic field through the MRI would be less directional 
(i.e., the magnetic field is not amplified intensively along the rotational axis as seen in
Fig.\ref{pic:f10f13} \& \ref{pic:f21f22}), we have to examine whether the amplified magnetic field affect
the mangeto-rotational explosion scenario.

(4) By comparing GRMHD and NMHD models, we found that the gravitational effect works a little bit stronger in GRMHD models which can be seen in the increase of the central density as $\sim30$\%.
However, the global dynamical evolutions are similar such as the time of core bounce and formation of the bipolar outflow.
Therefore we consider that the Newtonian approximation for the low mass range ($\le15M_\odot$) is acceptable at least for several ten ms after core bounce.
In this study, our progenitor is a $15M_\odot$ star which is small among the mass range of CCSNe progenitors and thus, if we adopt much larger mass such as
$\sim 40-100M_\odot$, the general relativistic effects become stronger and the qualitative differences may be appeared even in a similar time scale
as those used in this report.
Such high mass range calculations are now in progress and will be reported in near future.

To confirm the validation of numerical results, we have to connect them to the observations.
One important observed object is the gravitational wave.
Since we cannot observe directly in the vicinity of the center of CCSNe by the electro-magetic wave, the gravitational wave is one of the limited ways
with which we can observe directly.
In this report, though we do not evaluate the gravitational wave forms, the non-axisymmetric motion and the bipolar configuration appear.
These motions alter the gravitational wave forms as reported by \citet{Scheidegger09} and we will examine the
effects of, e.g., the progenitor mass or the non-axisymmetric motions or the magnetic field in the context of full
general relativity.
This will be our future work.
Another object to confirm is the ejected elements accompanied with the explosions.
CCSNe eject abundant heavy elements which are synthesized during the progenitor's main sequence age to their final fate and also during the
explosion via, such as, $r$-process nucleosynthesis \citep{Wanajo02,Fujimoto07,Kuroda08}.
However, the ejected chemical compositions and abundances depend on the detailed properties of the explosion dynamics and
we can still not explain the observed chemical abundances, one reason is due to the lack of comprehension about the explosion dynamics.
By comparing our numerical results with the observations, we can feed back the observational studies to our
numerical models and input physics.

\acknowledgements
Numerical computations were carried out on Cray XT4 at Center for Computational Astrophysics, CfCA, of National Astronomical Observatory of Japan.
This work was partly supported by Grants-in-Aid for JSPS Fellows and for the Scientific Research from the Ministry of Education, Science and Culture of Japan (20105004).
We are grateful to an anonymous referee for his/her valuable and constructive comments.

\end{document}